%% file: _paper-springer.tex
\RequirePackage{tikz}
\documentclass[sn-basic]{sn-jnl}

\usepackage[authoryear]{natbib}
\usepackage{tikz}
\usepackage{pgfplots}
\pgfplotsset{compat=newest}
\usetikzlibrary{arrows, chains, positioning,fit,calc, shapes, shapes.geometric, external}
\usepgfplotslibrary{statistics}%

\usepackage{subcaption}
\usepackage[printonlyused, nohyperlinks]{acronym}

\newacro{AAS}[AAS]{Asset Administration Shell}
\newacro{cd}[CD]{Complexity Degree}
\newacroplural{cd}[CDs]{Complexity Degrees}
\newacro{PRISMA}[PRISMA]{Preferred Reporting Items for Systematic Reviews and Meta-Analyses}
\newacro{DT}[DT]{Digital Twin}
\newacroplural{DT}[DTs]{Digital Twins}
\newacro{cps}[CPS]{Cyber Physical System}
\newacroplural{cps}[CPS]{Cyber Physical Systems}
\newacro{cpps}[CPPS]{Cyber Physical Production System}
\newacroplural{cpps}[CPPS]{Cyber Physical Production Systems}
\newacro{MAS}[MAS]{Multi-Agent System}
\newacroplural{MAS}[MAS]{Multi-Agent Systems}
\newacro{rami}[RAMI 4.0]{Reference Architecture Model Industry 4.0}
\usepackage{tabularx}
\newcolumntype{Y}{>{\centering\arraybackslash}X} 
\newcolumntype{Z}{>{\raggedleft\arraybackslash}X} 
\usepackage{longtable}
\usepackage{xltabular}
\usepackage{ltxtable}
\newcommand*\rot{\rotatebox{270}}
\newcommand{\scopus}{3,063}

\newcommand{\recexclphone}{981}
\newcommand{\recexclphtwo}{344}

\newcommand{\repexcl}{90}
\newcommand{\recexclone}{41}
\newcommand{\recexcltwo}{27}
\newcommand{\recexclthree}{22}

\newcommand{\recincl}{145}

\usepackage{sparklines}
\definecolor{hsurot}	{cmyk}{.00 1 .59 .26}
\definecolor{hsugrau}	{cmyk}{.38 .37 .39 .15}

\definecolor{hsugelb}	{cmyk}{0 .16 .80 0}
\definecolor{hsublau}	{cmyk}{1 .40 0 .82}
\definecolor{hsuturkis}	{cmyk}{1 .14 .60 .49}
\definecolor{hsugruen}	{cmyk}{.16 .16 .91 .28}
\definecolor{hsubraun}	{cmyk}{.00 .57 1 .17}
\definecolor{hsuorange}	{cmyk}{.01 .87 .77 .13}

\usepackage{hyperref}
\usepackage[english]{babel} 	
\addto\extrasenglish{}
\addto\extrasenglish{}
\addto\extrasenglish{}
\addto\extrasenglish{}
\addto\extrasenglish{}
\addto\extrasenglish{}

\Newlabel{1}{1}
\jyear{2023}%

\theoremstyle{thmstyleone}%
\theoremstyle{thmstyletwo}%

\theoremstyle{thmstylethree}%

\begin{document}

\title[Systematic Comparison of Software Agents and Digital~Twins]{Systematic Comparison of Software Agents and Digital~Twins: Differences, Similarities, and Synergies in Industrial Production}


\author*[1]{\fnm{Lasse M.} \sur{Reinpold}}\email{lasse.reinpold@hsu-hh.de}
\author[1]{\fnm{Lukas P.} \sur{Wagner}}
\author[1]{\fnm{Felix} \sur{Gehlhoff}}
\author[1]{\fnm{Malte} \sur{Ramonat}}
\author[1]{\fnm{Maximilian} \sur{Kilthau}}
\author[1]{\fnm{Milapji S.} \sur{Gill}}
\author[1]{\fnm{Jonathan T.} \sur{Reif}}
\author[1]{\fnm{Vincent} \sur{Henkel}}
\author[1]{\fnm{Lena} \sur{Scholz}}
\author[1]{\fnm{Alexander} \sur{Fay}}
\affil[1]{\orgdiv{Institute of Automation Technology}, \orgname{Helmut Schmidt University / University of the Federal Armed Forces} \city{Hamburg}, \country{Germany}}

\abstract{To achieve a highly agile and flexible production, a transformational shift is envisioned whereby industrial production systems evolve to be more decentralized, interconnected, and intelligent. Within this vision, production assets collaborate with each other, exhibiting a high degree of autonomy. Furthermore, information about individual production assets is accessible throughout their entire life-cycles. To realize this vision, the use of advanced information technology is required. Two commonly applied software paradigms in this context are Software Agents (referred to as Agents) and \acp{DT}. 

This work presents a systematic comparison of Agents and \acp{DT} in industrial applications. The goal of the study is to determine the differences, similarities, and potential synergies between the two paradigms. The comparison is based on the purposes for which Agents and \acp{DT} are applied, the properties and capabilities exhibited by these software paradigms, and how they can be allocated within the Reference Architecture Model Industry 4.0.

The comparison reveals that Agents are commonly employed in the collaborative planning and execution of production processes, while \acp{DT} are generally more applied to monitor production resources and process information. Although these observations imply characteristic sets of capabilities and properties for both Agents and \acp{DT}, a clear and definitive distinction between the two paradigms cannot be made. Instead, the analysis indicates that production assets utilizing a combination of Agents and \acp{DT} would demonstrate high degrees of intelligence, autonomy, sociability, and fidelity. To achieve this, further standardization is required, particularly in the field of \acp{DT}. 
}
\acresetall

\keywords{Digital Twin, Industry 4.0,  RAMI 4.0, Software Agent, Production, PRISMA, Systematic comparison}


\maketitle

\section*{Abbreviations}
The following abbreviations are used in this manuscript: 
\begin{description}
    \item[\textbf{AAS}]{Asset Administration Shell}
      \item[\textbf{CD}]{Complexity Degree}
        \item[\textbf{DT}]{Digital Twin}
                \item[\textbf{MAS}]{Multi-Agent System}
                   \item[\textbf{PRISMA}]{Preferred Reporting Items for Systematic Reviews and Meta-Analyses}
                   \item[\textbf{RAMI 4.0}]{Reference Architecture Model Industry 4.0}
\end{description}
\input{1introduction}
\input{2method}
\input{3results}
\input{4discussion}

\backmatter

\bmhead{Supplementary information}
The dataset related to this article can be found at \url{https://doi.org/10.5281/zenodo.8120624               } \citep{dataset}.


\section*{Declarations}

\subsection*{Funding}
This research is partly funded by the projects 'OptiFlex', 'iMOD', and 'ProMoDi' within dtec.bw – Digitalization and Technology Research Center of the Bundeswehr which we gratefully acknowledge. dtec.bw is funded by the European Union – NextGenerationEU. This work was also supported by the Federal Ministry for Economic Affairs and Climate Action (Projektträger Jülich GmbH, FKZ: 03EI6035A and VDI/VDE Innovation + Technik GmbH, FKZ: 16KN102724) as well as the Federal Ministry of Education and Research (Projektträger Jülich GmbH, FKZ 03HY116). 

\subsection*{Competing interests}
The authors have no competing interests to declare that are relevant to the content of this article.

\subsection*{Data availability statement}
All data generated or analysed during this study are included in this published article (\autoref{secA1}).

\subsection*{Author's contributions}
Conceptualization - L.R., L.W., F.G., M.K., M.R., J.R., M.G., A.F., 	Data curation - L.R., L.W., F.G., M.K., M.R., J.R., M.G., Formal analysis - L.R., L.W., F.G., M.K., M.R., J.R., M.G., A.F., 	Funding acquisition - A.F., 	Investigation - L.R., L.W., F.G., M.K., M.R., J.R., M.G., L.S, V.H., 	Methodology - L.R., L.W., F.G., M.K., M.R., J.R., M.G., A.F., 	Project Administration - L.R., L.W., F.G., 			Supervision - L.R., L.W., F.G., A.F., 	Validation - L.R., L.W., F.G., M.K., M.R., J.R., M.G., 	Visualization - L.R., L.W., 	Writing - original draft - L.R., L.W., F.G., M.K., M.R., J.R., M.G., A.F., 	Writing- review \& editing - L.R., L.W., F.G., M.K., M.R., J.R., M.G., L.S, V.H., A.F.

\begin{itemize}
\item Ethics approval: Not applicable
\item Consent to participate: Not applicable
\item Consent for publication: Not applicable
\item Availability of data and materials: see 'Supplementary information' and \autoref{secA1}
\item Code availability: Not applicable
\end{itemize}


\begin{appendices}

\section{Comprehensive Overview of Results}\label{secA1}
\input{abbr_table_results}
\clearpage
\input{tab_properties.tex}




\end{appendices}



\end{document}

%% file: 1introduction.tex
\section{Introduction} \label{sec:introduction}
Industrial enterprises face demanding challenges concerning the competitiveness and sustainability of their operations \citep{AHX+22}. Global competition, resource scarcity, and increasing customer demands for individualized products necessitate shorter product development cycles and more agile production environments~\citep{AJL+19}. 
To tackle these challenges, it is envisioned that production systems become increasingly interconnected, cooperative, and autonomous. Assets such as production machines and products will exhibit a high degree of intelligence, supported by enhanced data collection and information processing capabilities \citep{Monostori.2014}. This vision is commonly referred to as Industry~4.0 \citep{SaLe20}. The realization of this vision is expected to yield several benefits, including the optimization of production processes \citep{Monostori.2014}, increased product individualization \cite{Monostori.2014}, improved robustness of production systems \citep{TQC+23}, enhanced system transparency \citep{ZSS+23}, and increased resource efficiency \citep{Cardin.2019}. To realize these benefits and implement autonomous, cooperative, and interconnected production systems, suitable use of information and communication technology is required.

The \ac{DT} and its standardized implementation, the \ac{AAS}, are often seen as one of the most prominent paradigms to enable the process of digitization within Industry 4.0 \citep{SLL22}. Software Agents, particularly \acp{MAS}, are often viewed as a key enabler for autonomous and self acting \acp{DT} \citep{SLL22, VSC+20, SaLe20} as well as an enabler for realizing the full potential of Industry~4.0 \citep{KRL+19}.

Therefore, this work investigates two software paradigms commonly viewed as key enablers of Industry 4.0: Software Agents (referred to as Agents) and Digital Twins \citep{KLR+20}. 

\subsection{Agent Paradigm} \label{sec:agents}
An Agent is "[…]~a computer system, situated in some environment, that is capable of flexible autonomous action in order to meet its design objectives"~\citep{JSW98}. Agents are commonly associated with properties and capabilities such as autonomy, reactivity, proactivity, deliberativeness, persistence, encapsulation, and communicative abilities  \citep{vdi12345}, which are required for interacting with one another in so called \acp{MAS}. There are a number of advantages that distinguish Agent-based automation solutions from conventional, centralized concepts \citep{JeBu03,vdi12345}: 
\begin{itemize}
\item Conceptual advantages arise from inherent problem decomposition during engineering
\item No single point of failure or communication bottleneck
\item Interaction with the environment and utilization of local knowledge, reducing communication traffic. 
\item Flexible communication paths and organizational relationships
\item The distribution of computing power enables high scalability, facilitating flexible and efficient adaptation to changing production requirements and workloads. 
\item Adaptive and flexible behavior, allowing for dynamic reconfiguration and coordination of Agents in response to changing conditions and requirements in the industrial environment
\item Independent operation of Agents, enabling MAS to compensate failures of other Agents
\item By coordinating efforts among Agents, decentralized optimization of production processes can be achieved, improving resource allocation, task allocation, and scheduling
\end{itemize}

\subsection{Digital Twin Paradigm}\label{sec:dt}
The concept of the \ac{DT} as a virtual representation of an asset has first been described by \citet{Gri14}. Additionally,the literature offers a multitude of definitions for the \ac{DT} \citep{Kritzinger.2018}. In this work, the definitions introduced by \cite{Tao.2018} and \cite{Kritzinger.2018} are referred to and used as a guideline in order to establish a common understanding of this concept. \citet{Tao.2018} define the \ac{DT} as a digital representation of an asset throughout its life cycle, capable of reflecting its static properties, its dynamic behavior as well as its current condition. In this context, a \ac{DT} can mirror the real world object or system in real time, thus being beneficial for use cases like  monitoring, forecasting, analysis and optimization purposes as well as control \citep{Kritzinger.2018, Tao.2018}. Interaction between the digital representation, consisting of relevant digital models, and the physical asset is achieved by establishing a bidirectional and automated data flow  \citep{Kritzinger.2018}. This enables continuous model improvement and synchronization with the real world asset or system \citep{Tao.2018}. The \ac{DT} includes static data such as technical documentation, as well as dynamic data such as mathematical or simulation models that reflect the asset's behavior \citep{Kritzinger.2018}. 

In this work the \ac{DT} is considered to take both, a passive as well as an active role. In the passive role, similar to the digital shadow concept proposed by \cite{Kritzinger.2018}, synchronization of digital models with the physical asset occurs by establishing a unidirectional data flow. During this process, data is collected, stored, and used for condition monitoring, analysis, or optimization using the existing digital models. However, no immediate or simultaneous influence is exerted on the physical asset. In contrast, in the active role, the  \ac{DT} is fully integrated by establishing a bidirectional data flow and has control over the physical asset. Both roles are taken into account by the authors of this paper for the systematic comparison of the paradigms.

The advantages of using \acp{DT} are the following \citep{Kritzinger.2018, Tao.2018, Tao.2019, Pires.2021b, Mashaly.2021}: 
\begin{itemize}
\item Efficient design and development of Industry~4.0 conform production systems by means of virtual modeling and simulation of products, processes, and systems in the early stages of development or during reconfiguration. 
\item Facilitating cooperation among designers, engineers, operators, and manufacturers during the engineering phase, thanks to a digital system model. 
\item Improving operational efficiency by providing real time information about the condition, performance, and behavior of physical assets. 
\item Enabling condition monitoring using real time operational data and historical data
\item Supporting operational decisions by visualizing operational data.
\end{itemize}

\subsection{Motivation} \label{sec:motivation}
While previous conceptual descriptions of Agents and \acp{DT} suggest certain differences between the two paradigms, they are sometimes applied for similar purposes. For instance, both Agents and \acp{DT} have been used for production scheduling \citep{KLJ+18, VNB+21}, fault diagnosis \citep{XSL+19, SeSr10}, and the control of production resources at the control device level \citep{LLY+20, SSF+11}. 

On the other hand, some publications apply Agents and \acp{DT} synergistically. For example \cite{VOS21} utilize \acp{DT} as a standardized knowledge base for Agents, while \cite{AJL+19} present an architecture for an intelligent \ac{DT} and employ Agents to facilitate a co-simulation between different \acp{DT}. Another example of the concurrent and synergistic use of Agents and \acp{DT}, is the application of design patterns from the Agent paradigm to develop "proactive" \acp{AAS}, resulting in \acp{DT} exhibiting Agent-like capabilities such as proactivity~\citep{LCE+23}. 

To summarize the previous paragraphs, three observations should be highlighted: 
  \begin{itemize}
         \item Agents and \acp{DT} are sometimes applied for similar purposes.
         \item There are use cases in which Agents and \acp{DT} are used synergistically. 
         \item The formal definitions of Agents and \acp{DT} indicate significant differences between the paradigms.
    \end{itemize}
    
This work further explores these observations to determine the similarities and differences of Agents and \acp{DT} within actual implementations in industrial automated production systems. By focusing on actual implementations of Agents and \acp{DT}, purely conceptual reports are disregarded in this work. This is done to obtain insights into the capabilities required for industrial applications and evaluate their feasibility. The exclusion of purely theoretical reports is based on the premise that concepts can be proposed regardless of their applicability in or suitability for industrial practice.

Therefore, the research questions pursued in this work can be summarized as follows: 
\begin{enumerate}
    \item What are the capabilities and properties exhibited by Agents and \acp{DT}?
    \item What purposes do Agents and \acp{DT} serve within industrial production systems?
    \item What are differences, similarities, and potential synergies between Agents and \acp{DT} in industrial practice? 
\end{enumerate}

To address these research questions, a systematic analysis of reports describing applications of Agents and  \acp{DT} is conducted. Based on the analysis, the properties and capabilities exhibited by Agents and  \acp{DT} are identified. Additionally, each application of Agents and  \acp{DT} is allocated to the elements of the \emph{Reference Architecture Model Industry 4.0} (\acsu{rami}), i.e., the layers, hierarchy levels and life cycle phases, which the respective report primarily addresses. Finally, the purposes fulfilled by the Agents and \acp{DT} are determined for each report. The precise method for conducting these analyses is described in \autoref{sec:method}.

The insights obtained from the conducted analyses can serve as a basis for practitioners, looking for the most suitable software paradigm to solve specific problems. 
The focus on actual implementations thereby allows to move beyond conceptual comparisons and investigate what is actually implemented in practice. This in turn can serve as a base for discussions on the conceptual definitions of the paradigms' definitions. In fact, it has been suggested by other authors that \emph{"the digital twin would benefit from a more detailed comparison and review in the context of similar and connected fields"}~\citep{JSN+20} and that \emph{"some authors use the term [digital twin] merely as a catchphrase"}~\citep{SLF+20}, indicating the lack of a clear and common definition and standardization, which is investigated within this work.  
Finally, the results of this analysis provide insights into potential synergies between the two paradigms, thereby identifying future research opportunities. 

\subsection{Related Works}\label{sec:relatedworks}
Several reviews have been published aiming to describe characteristics of Industry~4.0-related paradigms and/or a comparison thereof. \Citet{PAS20} review simulation in Industry 4.0 and derive "design principles" to classify simulation paradigms, including Agents and \acp{DT}. However, they do not conduct a detailed comparison between Agents and \acp{DT} \citep{PAS20}. \citet{MMA+21}  present a list of abilities and characteristics to describe industrial autonomous systems, which is then used to exemplify a single use case  \citep{MMA+21}. \citet{JSN+20} aim at characterizing the \ac{DT} paradigm in depth, providing a description of its characteristics, but do not apply the list to compare different \ac{DT} applications \citep{JSN+20}. \citet{CAR23} conduct a review of artificial intelligence applications in \acp{DT} and offer recommendations on which artificial intelligence techniques are suitable for different purposes of \ac{DT} applications. Agents are not within the scope of \citet{CAR23}. \citet{LLH+23} explore the symbiotic relationship between intelligent \acp{DT} and \acp{MAS}. \citet{LLH+23} highlight the similarities in terms of specific characteristics of \acp{MAS}, such as being active, online, goal seeking, and anticipatory. However, their comparison between \acp{DT} and \acp{MAS} is qualitative, and no systematic literature review has been conducted. 

Furthermore, research has been devoted to analyzing the alignment of Agents and \acp{DT} with \ac{rami}. \citet{VeBa20} propose an approach to implement an autonomous Industry 4.0 system using Agents and classify the  mandatory and optional capabilities of Agents base on the "layers" and "hierarchy levels" axes of \ac{rami}. \citet{MLL23} align the \ac{DT} concept with the different dimensions of \ac{rami} and consider the capabilities of \acp{MAS} to support the development of digital twin ecosystems.

As mentioned earlier, numerous literature reviews exist, but none conduct a systematic comparison of Agents and \acp{DT} to identify differences, similarities, and synergies as well as common purposes.

\subsection{Contributions of this Systematic Comparison}
The analysis of related works reveals that the research questions stated in \autoref{sec:motivation} have not been addressed in previous studies. Therefore, this systematic comparison of Agents and \acp{DT} offers the following contributions:
\begin{enumerate}
\item Systematic literature review following the most recent \ac{PRISMA} statement to identify relevant reports concerning the application of Agents and \acp{DT} in industrial automated production systems.
\item Comparison of Agents and \acp{DT} based on their capabilities and properties as well as their allocation within \ac{rami}
\item Analysis of use cases of Agents and \acp{DT} including the determination of the purposes for which they are applied.
\item Identification of differences, similarities, and synergies between Agents and \acp{DT}.
\end{enumerate}

The remainder of this work is structured as follows: \autoref{sec:method} describes the methodology employed to (1) identify relevant reports regarding industrial applications of Agents and \acp{DT} and (2) systematically compare them.  \autoref{sec:results} presents the results of the systematic literature review, along with the determination of system properties, capabilities, and purposes. In  \autoref{sec:discussion}, a discussion is conducted on the applied methods and the presented results.

%% file: 2method.tex
\section{Methodology} \label{sec:method}
The methodology applied in this work consists of two main parts: a systematic review process to identify relevant reports from the existing literature (\autoref{sec:method-comp}) and a method to facilitate a systematic comparison of the reports identified within the systematic review process  (\autoref{sec:mbse}ff).

\subsection{Method of the Systematic Literature Review} \label{sec:method-comp}
The systematic literature review adheres to the  \ac{PRISMA} 2020 statement \citep{PMB+21}. The following sections outline the search query, the reviewing process, and the criteria for excluding or including records. In this context, the term "record" refers to publications obtained from the search engine based on the search query, while "report" refers to those records that met the inclusion criteria and were subsequently analyzed concerning the application of Agents or \acp{DT}.

To identify potentially eligible records, a systematic search was conducted on June 15, 2023, using the Scopus \small{\copyright} \ database \citep{Els22}. The search query is presented in  \autoref{tab:searchquery}.

\begin{table}[h]
\centering
\caption{Search query used to identify relevant records.}
\label{tab:searchquery}
\begin{tabularx}{\linewidth}{lX}
\toprule
Field & Title, Abstract, Keywords \\
\midrule
\multirow{4}{*}{Search term} & (agent* OR mas OR digital-twin OR "digital twin" OR twin OR "administration shell" OR aas) \\
& AND (industr* OR produc* OR manufact* OR process) \\
& AND automa* \\
& AND ("case study" OR application OR implement*)  \\
\midrule
Excluded & blockchain OR health* OR human OR bio* OR city OR home OR vehicle\\
\bottomrule
\end{tabularx}
\end{table}

Results were limited to English and German publications. Since the Industry 4.0 framework was introduced at the Hanover fair in 2011 \citep{GCK+11}, records dating back to 2011 were included. Additionally, citation thresholds based on the year of publication were applied to ensure the inclusion of only relevant literature (see \autoref{tab:thresholds}). 

\begin{table}[h]
\centering
\caption{Minimum number of citations by year of publication.}
\label{tab:thresholds}
\begin{tabularx}{.65\linewidth}{XY}
\toprule
Year of publication & Minimum number of citations \\
\midrule
2022, 2023	&	0	\\
2021	&	1	\\
2020	&	2	\\
2019	&	3	\\
2018	&	4	\\
2017	&	5	\\
older than 2017	&	10	\\
\bottomrule
\end{tabularx}
\end{table}

The reviewing process was structured into three phases, with the same inclusion and exclusion criteria applied throughout the entire process. For a record to be included, it had to fulfill all of the following criteria; otherwise, it was excluded. These three inclusion criteria were evaluated simultaneously:
\begin{enumerate}
    \item The record must primarily focus on either Agents or \acp{DT}.  
    \item The area of application must be industrial automated production systems.
    \item The record must include an application, implementation, or case study. Purely conceptual or review papers were excluded. The implementation described should provide sufficient detail to allow characterization using the method presented in \autoref{sec:mbse}.
\end{enumerate}

The records underwent analysis in three consecutive phases. In phase 1, only the title of each record was analyzed to eliminate those that were clearly out of scope. During phase 2, the title, abstract, and keywords of the remaining records were screened. Possible ratings included 0 (clearly out of scope), 1 (unclear), and 2 (criteria for inclusion fulfilled). If a record received a rating of 1, it was independently screened again by a different reviewer. For all records rated 1 or 2 in phase 2, the full text was obtained and analyzed in phase 3. The analysis followed the method outlined in \autoref{sec:mbse}ff. If the analysis of the full text revealed that any of the inclusion criteria were not met, the respective record was excluded during the full text screening in phase 3.

To prevent bias during the first two phases, information regarding authors, journals/conferences, and citation counts was omitted. Additionally, different reviewers were assigned to screen each record in every phase. To ensure consistent decisions regarding inclusion and exclusion, a joint analysis of at least 15 reports was conducted in each phase. Furthermore, 10~\% of the included reports were independently screened by two reviewers in phase 3.

\subsection{Methodology for the Systematic Determination of Capabilities, Properties, and Purposes} \label{sec:mbse}
To compare Agents and \acp{DT}, the purposes, properties, and capabilities of both paradigms were used as comparative aspects. \autoref{tab:mbse-def} presents the definitions of the terms purpose, property, capability as interpreted in this work. 
The terms have been adapted from model-based systems engineering  \citep{Wei14} serving to characterize complex systems. In this context, the term property is utilized to characterize the system in its target state, offering an overview or abstract representation of the system's potential in achieving its intended objectives. On the other hand, capability, synonymous with functionality, delineates the specific skills or abilities that the system must possess, and thus, is more granular and precise compared to properties.
\begin{table}[h]
\centering
\caption{Definition of purpose, properties, and capabilities}
\label{tab:mbse-def}
\begin{tabularx}{\linewidth}{lX}
\toprule 
\textbf{Aspect} & \textbf{Definition}\\
\toprule 
System purpose  & The system purpose describes the future state the system should achieve. \citep{Wei14}  \\
\midrule  
System properties  & System properties characterize the system in its target state. They represent and group the system's capabilities on an abstract level. \citep{FrKl19}\\
\midrule
System capabilities  & System capabilities extend a system's properties and describe the power or skill of the system to fulfill the purpose of the system \citep{FGS+19}. It encompasses the expression ability as well. \\
\bottomrule
\end{tabularx}
\end{table}

The sets of capabilities, properties, and purposes which were attributed to the applications of Agents and \acp{DT} were derived from a literature analysis and iteratively adjusted in the initial screening. In total 26 capabilities were identified as well as four properties and nine categories of purposes. The process of identifying the capabilities, properties, and purpose of the Agents and \acp{DT} in a given report is described in \autoref{sec:capabilities},  \autoref{sec:capandprop}, and \autoref{sec:method-goals}, respectively.

\subsubsection{Identification of Capabilities} \label{sec:capabilities}

The capabilities of Agents and \acp{DT} were identified based on text passages describing the implementation of the paradigms in the corresponding reports. This analysis focused on the description of the actual implementation of Agents and \acp{DT}, usually found in the methodology section of a report. This means that capabilities which were addressed in the introduction of a report but not in the section describing the actual implementation were not considered. 

The set of capabilities exhibited by Agents and \acp{DT} is presented in \autoref{tab:capabilites}. Initially, the set of capabilities was iteratively adjusted to encompass all the functions and behaviors of Agents and \acp{DT}. The set of capabilities was created irrespective of the paradigm, i.e., no capability is considered as exclusive to one paradigm a priori. The exact definitions of the capabilities as applied in this work are stated in \autoref{tab:properties-autonomy}-\ref{tab:properties-fidelity}.

The following text passage provides an example of identifying the capabilities exhibited by a \ac{MAS}. In the report, Agents are utilized as a decision support for human operators in planning the adaptation of production resources. The passage describes how different options for adapting the production resources to fulfill a given product order are presented to a human operator: \emph{"To support the decision maker in the last phase of the adaptation process, the previously found adaptation options ('\textbf{flexibility}') are compared by their KPIs and only a preselection ('\textbf{prioritization}') of adaptation options is given to the user ('\textbf{human interaction}') who is then capable of selecting the most appropriate one depending on the context. The preselection is done via negotiation ('\textbf{negotiation}') between the different debater roles ('\textbf{internal system interaction}') in the Agent system. Suitable approaches for that can be found in the field of multi criteria decision making ('\textbf{decision making}'). In most cases, the selected adaptation option by the user requires the adaptation ('\textbf{adaptability}') of the underlying model of the manufacturing machines."} \citep{MWH+17} It is worth noting that this text passage contains a notably high number of indicators of capabilities and was chosen for illustrative purposes.

It was assumed that the \ac{MAS} described above exhibits the capabilities highlighted in the quoted text passage. No distinction was made regarding the level of detail or the number of indicators for each capability. Thus, an application of Agents or \acp{DT} either exhibits a given capability (1) or does not exhibit it (0). 

\begin{table}[ht]
\centering
\caption{Capabilities for the Comparison}
\label{tab:capabilites}
\begin{tabularx}{\textwidth}{X|X|X}
\toprule 
Adaptability&Condition Monitoring&Context-Awareness\\
\midrule Cooperation&Coordination&Decision-Making\\
\midrule Encapsulation&External-System-Interaction&Flexibility\\
\midrule Human-Interaction&Internal-System-Interaction&Learning\\
\midrule Mobility&Negotiation&Prediction\\
\midrule Prioritization&Proactivity&Reactivity and External-System-Control\\
\midrule Real-time Capability&Robustness and Resilience&Simulation\\
\midrule Synchronisation&Uncertainty-handling&Visualisation\\
\bottomrule
\end{tabularx}
\end{table}

\subsubsection{Identification of Properties} \label{sec:capandprop}
Four properties were evaluated: \emph{autonomy}, \emph{intelligence}, \emph{sociability}, and \emph{fidelity}, as defined in \autoref{tab:properties}. The decision to differentiate between \emph{intelligence} and \emph{autonomy} as separate categories was made due to the fact that there can be technical systems that exhibit \emph{intelligent} and partially \emph{autonomous} behavior while still being fully controllable by external sources (see also \citep{MMA+21} for this distinction). However, the fact that a system is controllable from the outside contradicts common definitions of \emph{autonomy}. Another example is an \emph{autonomous} system that is unable to cooperate with its peers. The system might be very \emph{autonomous} in the sense that it can handle unforeseen events etc. but it lacks capabilities in the area of \emph{sociability}. Thus, it makes sense to analyze \emph{sociability} and \emph{autonomy} separately. The different properties are not mutually exclusive and partially overlap with each other. The chosen properties are not exclusively associated with one paradigm, e.g., \emph{intelligence}, which is a key property of agents, is also increasingly exhibited by \acp{DT} \citep{AJL+19}.

\begin{table}[ht]
\centering
\caption{System properties and their descriptions}
\label{tab:properties}
\begin{tabularx}{\textwidth}{p{2.5cm}X}
\toprule 
\textbf{Property} &  \textbf{Description}\\
\toprule 
Fidelity	& \cite{JSN+20} define fidelity by the number of parameters exchanged between the physical and the virtual system as well as parameter accuracy and their level of abstraction. A high fidelity virtual model is designed to accurately replicate the physical entity and to realistically correspond to the system (and its behavior) it is created to emulate \citep{JSN+20, Maier.2017}. \\
\midrule
Intelligence & \cite{Kri} defines that intelligent systems emulate aspects of intelligence observed in nature. This includes the ability to process experiences and learn from them, the ability to store knowledge, and the ability to recognize new situations and react to them. Intelligent applications are usually associated with capabilities such as decision making, process control, recognition of patterns and their environment as well as a degree of autonomous behavior in,e.g. maneuvering unknown environments \citep{KMB+22}.\\
\midrule
Autonomy	& Autonomous systems can be defined as systems which achieve their objectives within an uncertain environment systematically and without external intervention. \citep{MMA+21}. \\
\midrule
Sociability	& The property to communicate, interact and coordinate among distributed systems is often called sociability. A paradigm exhibiting sociability is able to transfer data or information using communication protocols and moreover can negotiate with other systems \citep{Han12}. \\
\bottomrule
\end{tabularx}
\end{table}

To facilitate a comparison of the properties of Agents and \acp{DT} on an ordinal scale, a property fulfillment score is introduced and determined. The property fulfillment score is calculated based on the previously defined set of capabilities. 
For each capability, it was determined whether and to what extent it contributes to the fulfillment score of a given property. This determination is inspired by established measures of the degree of \emph{autonomy} in technical systems, which are described in the following paragraphs. Similarly, the relevant capabilities for the properties of \emph{intelligence}, \emph{sociability}, and \emph{fidelity} could be determined and their contribution to the property fulfillment score could be established. 

There are several attempts, mainly domain dependent, to determine the \emph{autonomy} of a system: 

In 2007, the "ALFUS framework" was published, which defined a framework for \emph{autonomous} capabilities of unmanned aerial systems. \cite{HPN+07}  propose measuring the \emph{autonomy} of the system based on the decreasing level of human intervention required. The levels of robot \emph{autonomy}, described by \cite{BFR14}, consider the degree of required human intervention in sensing, planning, and acting primitives. The authors propose nine levels, ranging from manual task execution to full \emph{autonomy} within each of the primitives \citep{BFR14}.

The SAE standard J3016 describes different levels of \emph{autonomy} for \emph{autonomous} driving systems \citep{SGS+20}. 
These levels are based on the allocation of responsibility and task execution that are either executed by the driver or the automation system. The tasks include motion control, sensing of the environment and all necessary sub-tasks to fulfill a secure transportation mission. 
The increasing \emph{autonomy} mainly refers to the decreasing reliance on a driver as a fallback solution in case of unforeseen events and operating in unfamiliar domains. 
The levels range from 0 (no driving automation) to 5 (full driving automation). The "Industry 4.0 platform" has adapted these levels to industrial production applications, focusing on the degree of required assistance by the user as well \citep{Fed19}. Therein, the "Industry 4.0 platform" provides an indication of which capabilities a system requires to achieve a given level of \emph{autonomy}. 

The above mentioned frameworks associate the level of \emph{autonomy} that a system exhibits with the degree to which the respective system is capable of executing autonomous behavior. Thus, as a higher level of \emph{autonomy} requires more complex capabilities, a system with a higher level of \emph{autonomy} reflects the possession of more complex capabilities. To avoid the pitfall of associating a high level of autonomy to a system that possesses a number of complex capabilities but lacks other, more basic ones, the methodology of this contribution directly maps the capabilities to so called \ac{cd}. The \acp{cd} are integers from 0 to 5. They rate how complex the implementation of a given capability is and how strongly it contributes to the property fulfillment score of each of the four properties. Thus, each capability is analyzed in the context of each property. It should be noted though that each of the four subsets of the different capabilities and associated \acp{cd} does not contain all 26 capabilities, since not every capability is relevant to the respective property. 

The fulfillment score of a property is calculated using \autoref{propertyscore}. In \autoref{propertyscore},  $x$ is the number of capabilities which contribute to the property, $n_i$ represents the complexity degree of the $i$-th capability, and $cap_{p,i}$ takes the value 1 if the $i$-th capability is exhibited by the Agents or \acp{DT} in report $p$ and takes the value 0 otherwise. This means that a property fulfillment score of 100\% is achieved if all capabilities contributing to the property are exhibited by the Agents or \acp{DT} in a given report. 

\begin{align}
    \frac{1}{\sum \limits_{i=0}^{x} n_i} \cdot \sum  \limits_{i=0}^{x} \text{cap}_{p, i} \cdot n_i \label{propertyscore}
\end{align} 

In the following, each property and the underlying capabilities that contribute to property fulfillment scores of all four properties are described and assigned a \ac{cd} based on a short literature analysis.   

\paragraph{Autonomy and its Associated Capabilities} \label{sec:autonomy}
The set of capabilities which contribute to the \emph{autonomy} of a system is based on the definitions reviewed by \cite{MMA+21} who argue that the absence of human intervention, self determination, perception of the environment, decision making capabilities, goal orientation and the ability to adapt as well as proactive planning and execution of actions are important aspects of a system's \emph{autonomy}. This is in line with other definitions, for example in \cite{VSC+20} or \cite{MFH09}. Therefore, the  capabilities listed in \autoref{tab:properties-autonomy}, sorted by their associated \ac{cd}, have been extracted from the literature as contributors to \emph{autonomy}.

\captionsetup[longtable]{labelfont=bf}
\renewcommand*{\arraystretch}{1.2}
\tablebodyfont{
\begin{longtable}{p{.15cm}p{1.5cm}p{8.6cm}} 
\caption{Capabilities within autonomy} \label{tab:properties-autonomy} \\
\toprule
\!\!\!CD & Capability & Definition  \\
\midrule
\endhead
1 & External-System Interaction & The system is able to perceive \citep{MMA+21} and exchange information with external systems (applications, services,   users, resources, systems, ...) \citep{BCC+16}.   \\
\midrule1 & Encapsu\-lation & The system’s inner states, functions, goals, and strategies are hidden from the outside and cannot be accessed by external   entities \citep{VSC+20, Woo02}.  \\
\midrule2 & Reactivity and External System Control & The system is able to react in a timely and   suitable manner to the data and information it perceives \citep{MMA+21}. The reaction can have a physical effect within the system’s domain, i.e., it can use physical actors if necessary. This is in contrast to actions   that solely change, for example, the inner states of a controller \citep{BJW04}.  \\
\midrule3 & Decision making & A decision is the choice between different   alternatives (incl. the option not to act at all) \citep{LGS18}. There   must be at least two alternatives between the autonomous system can choose.   At least two of these alternatives must be decided by the fact that one   alternative achieves a certain goal better or worse \citep{LGS18}. In   other words, the alternatives must have different effects \citep{BJW04} or achieve the same effect with a different degree of efficiency  \citep{MMA+21}.  \\
\midrule3 & Mobility & An autonomous (software) system is able to   transmit itself, i.e., its program and states, across a computer network and   recommencing execution at a remote site \citep{Woo02}. \\
\midrule 4 & Flexibility & The system is capable of using multiple ways   of achieving its goals \citep{PaWi04} as well as deliberating   about which goals to pursue \citep{BPM+05}. \\
\midrule 4 & Proactivity & The autonomous system chooses its actions   according to its goal \citep{vdi12345}. Such an explicit goal orientation often   requires some kind of strategy \citep{WoJe95} and is more   complex than the implicit design objective of, for example, a simple if-then   rule-based controller. This also requires the capability to plan and select   the most appropriate actions according to a certain goal \citep{MMA+21}. As every system is designed to meet certain design objectives, the   autonomous system needs to be able to (pro)actively pursue these goals   without being triggered by external events or entities \citep{MMA+21, VSC+20}. \\
\midrule 4 & Adaptability & An autonomous system is adaptable if it can   change its own structure, shape or behavior \citep{MMA+21}. \\
\midrule 4 & Robustness and resilience & The autonomous system can maintain its system   states and functionality under the influence of internal and external   disturbances \citep{Gor12} and withstands high impact disruptive events   \citep{DeNe18}. \\ 
\bottomrule
\end{longtable}
}
\small

\paragraph{Sociability  and its Associated Capabilities} \label{sec:sociability}
Distributed systems should possess the ability to optimize their interactions and communicate with other systems. According to \cite{Woo02}, \emph{sociability} can be classified into different degrees, which determine the complexity of the capabilities described in this work. \autoref{tab:properties-sociability} defines the set of capabilities of \emph{sociability}.

\tablebodyfont{
\begin{longtable}{p{.5cm}p{1.75cm}p{8cm}}
\caption{Capabilities within sociability} \label{tab:properties-sociability} \\
\toprule
\!\!\!CD& Capability & Definition  \\
\midrule
\endhead
1 & External-System Interaction & see \autoref{tab:properties-autonomy}   \\
\midrule 1 & Human-Interaction & Human interaction refers to systems that   offer visual representations of a system’s behavior or personalized   recommendations to users \citep{Pires.2021b}.   \\
\midrule 1 & Internal-System Interaction & A system which is able to interact with other   systems of the same type is capable of the internal-system-interaction \citep{BCC+16}.   \\
\midrule 2 & Coordination & The ability to organizing and aligning the   actions of multiple entities towards a common objective while minimizing the   use of resources. \citep{GLM+18} \\
\midrule 3 & Cooperation & The ability to collaborate with other systems   to achieve goals that may be beyond the capabilities of individuals alone  \citep{Woo02}.  \\
\midrule 3 & Negotiation & The system is capable of interacting with   other systems that may have conflicting interests, in order to negotiate and   reach mutually acceptable agreements among all parties involved \citep{ECR+20}.   \\  
\bottomrule
\end{longtable}
}
\small
\paragraph{Intelligence  and its Associated Capabilities} \label{sec:intelligence}
\cite{MFH09} provide an overview of intelligent products, define \emph{intelligence} and present levels for the classification of intelligent products. In this work, these findings are generalized to intelligent production resources. The minimum requirement to be classified as \emph{intelligent} is the ability to handle information sent by external systems, such as sensors. Products that can derive problems from the received information are classified at the second level of \emph{intelligence}. At the third level, resources possess the degree of \emph{intelligence} to manage themselves and make their own decisions. These levels of product \emph{intelligence} serve as a basis for determining the \ac{cd} of the capabilities associated with \emph{intelligence}. \autoref{tab:properties-intelligence} defines which capabilities constitute \emph{intelligence}.

\tablebodyfont{
\begin{longtable}{p{.15cm}p{1.5cm}p{8.6cm}}
\caption{Capabilities within intelligence} \label{tab:properties-intelligence} \\
\toprule
\!\!\!CD& Capability & Definition  \\
\midrule
\endhead
1 & Condition monitoring & Capability of dynamic representation by an   automated, at least one-way data connection to the physical system \citep{Kritzinger.2018}.  \\
\midrule
2 & Context awareness & A paradigm is context-aware when it utilizes   contextual data to provide the user with significant information or services   that are dependent on the user’s task \citep{Dey01}. In addition, the system   should be capable of detecting changes in environmental conditions and   effectively utilizing them in accordance with varying requirements, thus   contextualizing them appropriately \citep{RoGe18}.   \\
\midrule
3 & Decision making & see \autoref{tab:properties-autonomy}   \\
\midrule
3 & Sociability \ac{cd} 1 & External-system-, Internal-system-, and   Human-Interaction  (see \autoref{tab:properties-sociability}) \\
\midrule
4 & Prioritization & A system providing prioritization should be   able to classify incoming messages and tasks based on their relevance in   achieving its own goals \citep{SBE04}. \\
\midrule
4 & Uncertainty handling & Uncertainty handling refers to the ability of   a paradigm to effectively manage imprecise, contradictory, uninterpretable,   and even missing contextual information \citep{BBK+21}. Even though different uncertainties can be faced by a paradigm in production settings, all kinds of uncertainties are considered under this capability.  \\
\midrule
4 & Flexibility & see \autoref{tab:properties-autonomy}    \\
\midrule
4 & Sociability \ac{cd} 2 \& 3 & Coordination, Cooperation, Negotiation (see \autoref{tab:properties-sociability})  \\
\midrule 4 & Prediction & Capability to forecast future system data output based on historical data or external input \citep{Birk.2022}.   \\
\midrule 5 & Learning & The system possesses the capability to learn   from known situations/experiments and adapt its behavior accordingly \citep{BCC+16}.  \\
\midrule 5 & Autonomy \ac{cd} 4 & Proactivity, Adaptability, and Robustness (see \autoref{tab:properties-autonomy})  \\ 
\bottomrule
\end{longtable}
}
\small

\paragraph{Fidelity  and its Associated Capabilities} \label{sec:fidelity}
Both Agents and \acp{DT} should possess the ability to accurately analyze, describe, and reflect the physical system(s) they are associated with. This often requires a high \emph{fidelity} of models used to describe the system. 
\autoref{tab:properties-fidelity} lists the capabilities of which \emph{fidelity} consists.

\tablebodyfont{
\begin{longtable}{p{.15cm}p{1.5cm}p{8.6cm}}
\caption{Capabilities within fidelity} \label{tab:properties-fidelity} \\
\toprule
\!\!\!CD & Capability & Definition  \\
\midrule
\endhead
1 & Visualization & Capability to graphically represent the system or its behavior \citep{Marcus.06112003}.  \\
\midrule
2 & Condition monitoring & see \autoref{tab:properties-intelligence} \\
\midrule
3 & Real-time capability & Capability to react under tight temporal restrictions \citep{JuBo04} which are predefined by the physical system.  \\
\midrule
3 & Synchro\-nization &  Capability to recognize and react to changed data in the real system in order to uphold its consistency to it \citep{Talkhestani.2018}. \\
\midrule
4 & Prediction & see \autoref{tab:properties-intelligence}   \\
\midrule
5 & Simulation & Capability to describe the physical system via mathematical models or a simulation environment in order to enable a reflection of dynamic system behavior \citep{Kritzinger.2018, Sahlab.}.   \\  
\bottomrule
\end{longtable}
}
\small

\subsubsection{Identification of Purposes} \label{sec:method-goals}
The utilization of Agents or \acp{DT} in the context of production always serves a specific purpose. Therefore, based on the literature analysis presented, several categories of purposes were identified. These categories are the main focus when applying Agents and \acp{DT} in the context of industrial automated production systems. The analysis conducted for the first research question (\autoref{sec:motivation}) aims to identify whether certain purpose categories are more commonly associated with one paradigm or the other. This will subsequently address possible correlations between required capabilities and specific purpose categories. The analysis primarily focuses on the life-cycle phases of commissioning and operation of existing applications of Agents and \acp{DT}. To identify these purpose categories, an initial list was compiled based on the existing literature, and this list was further extended and updated during the initial screening process.

For the initial definition of purpose categories, the work of \cite{Lei09} and \cite{Colombo.2020} was consulted. These works introduced an architecture for manufacturing control systems, and the individual elements of this architecture were interpreted as purposes in this work. The major purpose categories in manufacturing control systems include  \emph{planning, scheduling, dispatching} and \emph{control}. Typically, these tasks are executed sequentially in the mentioned order. \emph{Monitoring} is used to track essential information about processes, products, and resources from the shop floor or manufacturing system. This information is then used as a feedback loop to improve \emph{planning, scheduling, dispatching,} and \emph{control} activities \citep{Lei09}. The results of \emph{monitoring} can also be utilized for diagnosis, error identification and correction, and predictive maintenance which are grouped under the category \emph{diagnosis and fault management}. Other purpose categories frequently associated with Agents and \acp{DT}, but not covered by \cite{Lei09}, include \emph{user assistance}, \emph{virtual commissioning}, and general \emph{process optimization}. 
\autoref{tab:goals} provides a summary of the individual purpose categories along with a brief description. The purpose of an application involving Agents or \acp{DT} was determined by the reviewer after thoroughly reading a report and assigned to one of the nine purpose categories listed in \autoref{tab:goals}.

\begin{table}[!ht]
\renewcommand{\arraystretch}{1.3} 
\centering
\caption{System purposes and their descriptions}
\label{tab:goals}
\begin{tabularx}{\textwidth}{p{2.5cm}X}
\toprule 
\textbf{Goal Category} &  \textbf{Description}\\
\toprule 
Planning	& Planning includes the decision of when and in what quantities a product is to be produced and by when the products are due. In Planning, possible operations to produce the products are derived \citep{Gro16} and resources are assigned to operations \citep{Colombo.2020}. In this work, the planning of reconfiguration actions are also included in planning.  \\
\midrule
Scheduling & 	Scheduling deals with the issue of what action a given resource performs at what time. Scheduling is performed well ahead of the actual manufacturing (days or even weeks). The constraints of the resulting schedule are predefined by the planning. \citep{Colombo.2020}\\
\midrule
Dispatching	&In dispatching, production resources are coordinated to fulfil a job that has already been released onto the shop floor. Therein it is decided what job a resource will perform next, taking into account the current status of available resources. \citep{Lei09}\\
\midrule
Control	&Control is considered to be the process of receiving sensor signals and applying a given control logic in order to send the required actor signals. Control happens in real time. \citep{Colombo.2020} \\
\midrule
Diagnosis and fault management (error recovery / predictive maintenance)	&Diagnosis and fault management deal with the evaluation of abnormal behavior of resources to identify its source. Adequate reactions to abnormal behavior are also determined in this category \citep{Colombo.2020}. In this work, we consider predictive maintenance to be included in diagnosis and fault management.  \\
\midrule
User assistance 	&User assistance includes use cases where a computer program is used to proactively assist a user working on a given problem \citep{Woo02}. The problems can fall within one of the other categories presented here, so the focus is on the interaction of the user with a computer program. \\
\midrule
Monitoring	&Monitoring provides other use cases like planning or dispatching with the necessary information. This includes the state of production resources and the progress of operations \citep{Colombo.2020}. The health of a resource or a set of process parameters can be the captured in monitoring. \\
\midrule
Virtual commissioning	&Virtual commissioning aims at the verification of manufacturing systems by connecting a virtual model of a production resource with a real controller or a suitable controller model, thus allowing to simulate and verify both resource and controller behavior. \citep{Lee.2014} \\
\midrule
Process optimization	& Process optimization has the goal of improving the design of a new resource or the performance of a given resource. The improvements can lead to increased resource efficiency or production volume \citep{Terwiesch.2001}. The focus of process optimization as applied in this work lies on electrical, mechanical, chemical or biological processes. Therefore, the adjustment of process parameters or the replacement of components can be the result of process optimization. \\
\bottomrule
\end{tabularx}
\end{table}
\clearpage

\subsection{Allocation of Applications in the Reference Architecture Model Industry 4.0} \label{sec:method-cap-rami}
\begin{figure}[h]
    \centering
    \includegraphics[width=.9\linewidth]{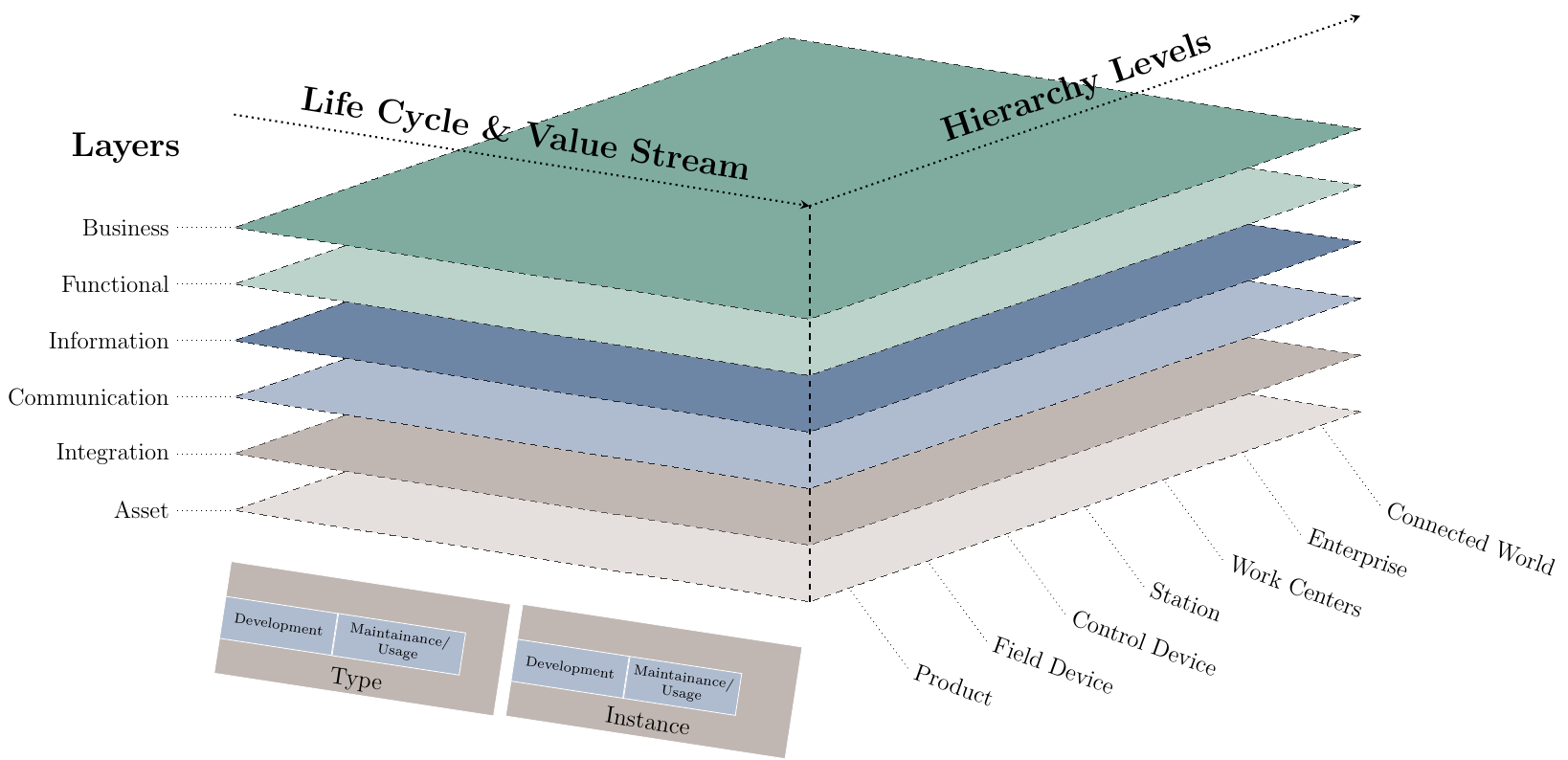}
    \caption{\ac{rami} \citep{91345}}
    \label{fig:rami}
\end{figure}

The use cases described in the reports were further classified based on the individual elements of the axes of \ac{rami} \citep{91345}  (see \autoref{fig:rami}). This classification was performed to enable both the categorization and aggregated comparison of all use cases for each paradigm.Since both Agents and \acp{DT} are widely applied in Industry 4.0 use cases, \ac{rami} serves as a suitable framework for classifying these paradigms. \ac{rami} allows for the description of an Industry 4.0 component throughout its lifecycle within the context of Industry 4.0  \citep{91345}.
The classification within \ac{rami} aims to reveal insights about frequently mentioned layers, hierarchy levels, and life cycle phases in applications of each paradigm. A comprehensive description of the six layers, seven hierarchy levels, and four life cycle phases can be found in \citep{91345}.

To allocate reports describing the application of Agents and \acp{DT} within \ac{rami}, an examination was conducted to determine whether a specific element of one of RAMI 4.0's three axes was relevant to the application described in the respective report. For example, a single element could be the  \emph{functional} layer, at the \emph{work center} level in the usage phase of an \emph{instance}. If a particular element was identified as being of significant importance for the use case, it was marked as the "focus element". Elements that were addressed within a report but were not considered the focus of the report were still extracted to analyze the application of the paradigms across all elements.

As the focus of this work is the analysis of implemented Agents and \acp{DT}, all implementations are considered to be \emph{instances}. Therefore, the elements of the "Lifecycle \& Value Stream" axis are  always considered as an \emph{instance} in the usage phase. Consequently, the \emph{lifecycle and value stream} axis was not explicitly extracted from the reports and will not be discussed in the results section (\autoref{sec:results}).
\clearpage

%% file: 3results.tex
\section{Results} \label{sec:results}
This section presents the results of the analysis conducted using the information extracted from the reports included in the systematic literature review. \autoref{sec:review-results} provides a brief overview of the results obtained from the literature review.  Subsequently, \autoref{sec:capmasdt} describes the capabilities and properties associated with the applications of Agents and \acp{DT} within the reports. \autoref{sec:placementrami} illustrates the placement of Agents' and \acp{DT}' applications within \ac{rami} and offers an interpretation of the findings.  \autoref{sec:systemgoals} compares system purposes of applications of both paradigms. A comprehensive overview of all the results can be found in \autoref{tab-resultsoverview}.

\subsection{Results of the Systematic Literature Review} \label{sec:review-results}
\begin{figure}[h]	
\includegraphics[width = .65\linewidth]{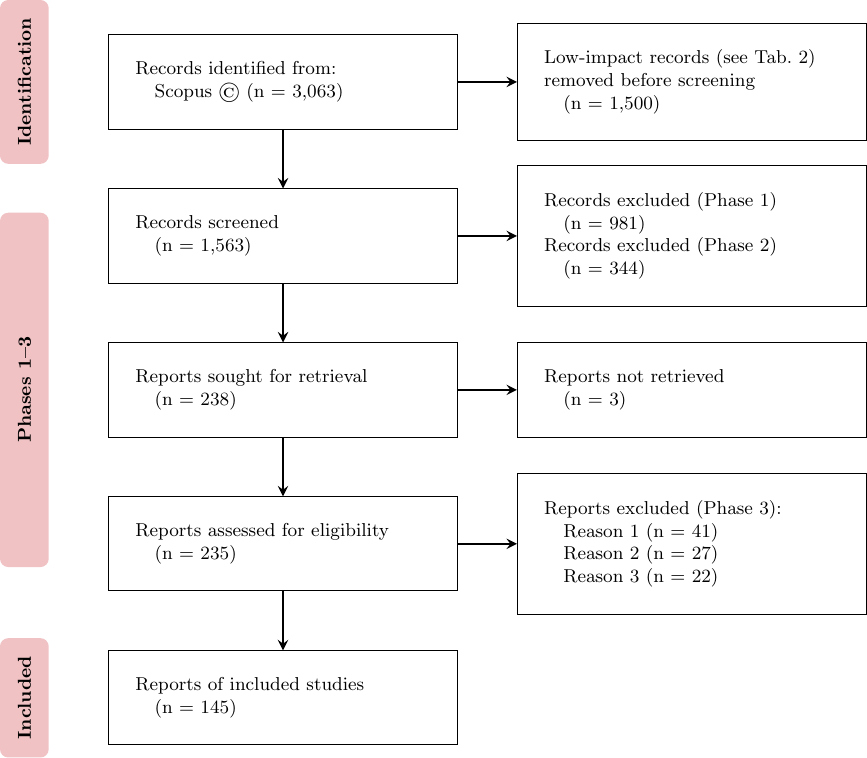}
\centering
\caption{PRISMA 2020 flow diagram according to \citet{PMB+21}.} 
\label{fig-flowchart}
\end{figure}

As shown in \autoref{fig-flowchart}, the three-phase approach outlined in \autoref{sec:method-comp},  led to the exclusion of \recexclphone \ records in phase~1 and \recexclphtwo \ records in phase 2. The full text review in phase 3 resulted in the exclusion of \repexcl \ reports (R1: \recexclone, R2: \recexcltwo, R3: \recexclthree).  Consequently, a total of \recincl \ reports were included for the identification of system properties, capabilities, and purposes (as explained in \autoref{sec:mbse}) and for the subsequent systematic comparison. Out of these  \recincl \ reports, 58 reports focus on Agents, 86 reports focus on \acp{DT}, and one report focuses on both Agents and \ac{DT} \citep{XSK+21}. The analysis of the latter report was divided into separate analyses of the \ac{DT} implementation and the Agent implementation, with the results being incorporated into the overall results of the respective paradigm (see also \autoref{tab-resultsoverview}). 

\subsection{Capabilities and Properties of Agents and Digital Twins} \label{sec:capmasdt}
The following sections describe Agent's and \ac{DT}'s capabilities (\autoref{sec:capagentsdt}) and properties (\autoref{sec:capdt}).
\subsubsection{Agents' and Digital Twins' Capabilities} \label{sec:capagentsdt}
\begin{figure}[h]	
 \centering
   \includegraphics[width = \linewidth]{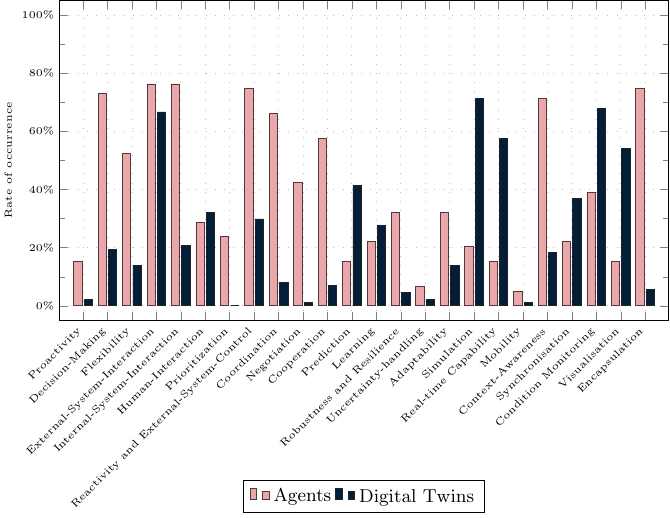}
 \caption{Rate of occurrence of capabilities in Agents and Digital Twins}
 \label{fig-results-capabilities}
 \end{figure}

\autoref{fig-results-capabilities} shows the rate of occurrence of the investigated set of capabilities in Agents and \acp{DT}. The rate of occurrence is calculated by dividing the number of implementations of one paradigm exhibiting a specific capability by the total number of reports focusing on that paradigm. For example, out of 59 reports presenting Agents, 12 implementations exhibit the capability of \emph{learning}, which amounts to 21~\% as shown in \autoref{fig-results-capabilities}.

The analysis reveals that some capabilities are rarely found in both Agents and \acp{DT}. \emph{Proactivity}, \emph{uncertainty handling}, and \emph{mobility} are capabilities that were scarcely addressed in the investigated reports. It should be noted that although Agents and \acp{DT} may possess these capabilities, the respective reports did not discuss them, suggesting that they may not be essential for the specific use-cases.

There are also sets of capabilities that are predominantly found in either Agents or \acp{DT}. A capability is considered "Agent-specific" or "digital twin-specific" if it occurs at a rate of 50~\% or higher in one paradigm but not in the other. Therefore, the Agent-specific capabilities include \emph{encapsulation}, \emph{internal system interaction}, \emph{context awareness}, \emph{decision making}, \emph{external system control}, \emph{coordination}, \emph{cooperation}, and \emph{flexibility} (in descending order). It is expected that Agents possess these capabilities, as they are commonly applied as MAS, thus requiring capabilities such as \emph{internal system interaction}, \emph{cooperation}, and \emph{encapsulation}. However, the higher occurrence rate of the capability of \emph{external system control} in Agents compared to \acp{DT} is surprising, considering that a two-way connection between an asset and a \ac{DT} is often described as a requirement in DT definitions   \citep{Kritzinger.2018, Tao.2018}. 

The \ac{DT}-specific capabilities include \emph{simulation}, \emph{condition monitoring}, \emph{real time capability}, and \emph{visualization}. This observation aligns with common definitions of \acp{DT}. However, it is surprising that the capability of \emph{simulation} occurs in less than 70~\% of \ac{DT} use cases, considering that \emph{simulation} models are often regarded as a core feature of \acp{DT}. Additionally, the capability of \emph{adaptability} occurs less frequently than expected, with a rate of less than 20~\%. This indicates the challenge of detecting asset changes and adjusting the \ac{DT} accordingly to maintain accuracy. The same applies to the capability of \emph{synchronization}, which occurs in less than 40~\% of cases.
\begin{figure}[h]	
 \centering
 \includegraphics[width = \linewidth]{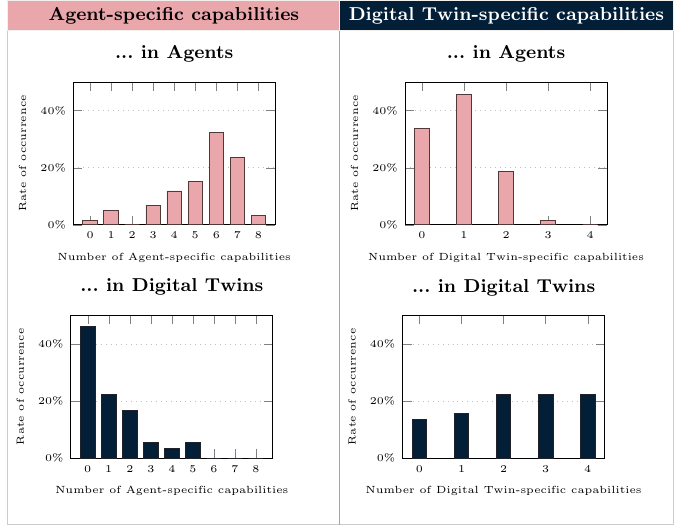}
 \caption{Comparison of characteristic capabilities of Agents and Digital Twins}
 \label{fig-characteristic-capabilities}
 \end{figure}
 
This initial analysis reveals substantial differences between the paradigms of Agents and \acp{DT}. However, it is also observed that almost no capability is exclusive to one paradigm (with the exception of \emph{mobility}, which was identified in only three applications of Agents, as seen in \autoref{fig-results-capabilities}). This observation is further supported by \autoref{fig-characteristic-capabilities}, which illustrates the occurrence of Agent-specific and \ac{DT}-specific capabilities in both paradigms.

In the case of Agents, it is observed that the majority possess at least five out of the seven Agent-specific capabilities. Two-thirds of Agents possess at least one \ac{DT}-specific capability. Similarly, approximately half of the analyzed \acp{DT} possess at least one Agent-specific capability. Five \ac{DT} applications possess as many as five Agent-specific capabilities. The distribution of \ac{DT}-specific capabilities among \acp{DT} is relatively uniform. Twelve \ac{DT} applications possess no \ac{DT}-specific capabilities, while 20 applications possess all four capabilities.

\subsubsection{Agents' and Digital Twins' Properties} \label{sec:capdt}
\begin{figure}[h]
    \centering
    \begin{subfigure}{0.48\textwidth}
    \includegraphics[width = .95\linewidth]{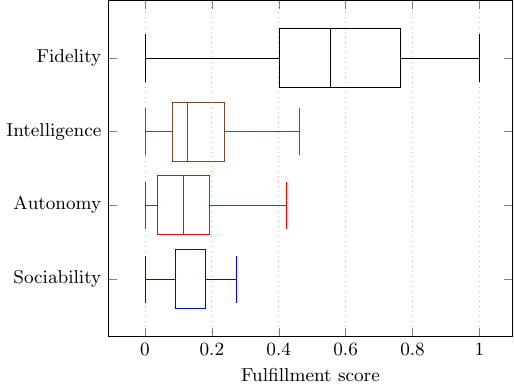}
\centering
\caption{Properties of \acp{DT} and Their Fulfillment Score}
\label{fig-properties-boxplot-dt}
    \end{subfigure}
    \begin{subfigure}{0.48\textwidth}
       \includegraphics[width = .95\linewidth]{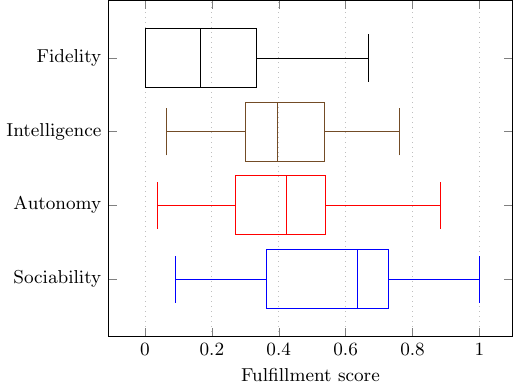}
\centering
\caption{Properties of Agents and Their Fulfillment Score}
\label{fig-properties-boxplot-agents}
    \end{subfigure}
    \caption{Properties of \acp{DT} and  Agents and their fulfillment score}
    \label{fig-properties-agentsdt}
\end{figure}

\autoref{fig-properties-boxplot-dt} illustrates the property fulfillment score of \acp{DT} for the mentioned properties based on the screened capabilities. It is evident that \emph{fidelity} is the most distinctive property of \acp{DT}, as indicated by its median score of 56~\% and the top quartile at 77~\% fulfillment. 

This aligns with the objective of \ac{DT} applications, which aim to accurately represent physical assets. On the other hand, \emph{sociability} has a low property fulfillment score, with a median of 9~\% fulfillment. This is consistent with the definition provided in \autoref{sec:dt}, where social behaviors such as \emph{cooperation} or \emph{communication} are not emphasized. However, it is surprising that \emph{autonomy} also has a low property fulfillment score. Definitions of \acp{DT} often emphasize their capability to independently interact with physical assets, but the screening results indicate that this is rarely implemented in practice.

Additionally, \emph{intelligence} has a low fulfillment score, with its median being at 13~\%. Also in this case, definitions like \citep{Tao.2018,Reiche.2021, Jazdi.2021} claiming that \emph{intelligent} components exist within the \ac{DT}, differ from actual applications.

In contrast, Agents exhibit a different property profile, as shown in \autoref{fig-properties-boxplot-agents}. The majority of the screened applications only demonstrate limited \emph{fidelity}, suggesting that Agents may not be capable of representing physical assets in great detail. However, in terms of \emph{intelligence}, \emph{autonomy}, and \emph{sociability}, Agents exhibit significantly higher scores than \acp{DT}.

\begin{table}[ht]
    \centering
     \caption{Average fulfillment score of properties by complexity degree of capabilities}
    \label{tab:propertiesbylevel}
    \begin{tabular}{ll*{6}{c}}
    \toprule
  \textbf{  Property}	& \textbf{ Paradigm	}	& & \textbf{  CD 1} & 	\textbf{ CD 2 }& \textbf{ CD 3} & 	\textbf{ CD 4} & 	\textbf{ CD 5} \\
 \cmidrule(lr){1-8}
\multirow{2}{*}{Fidelity} &	Agents&	\begin{sparkline}{5}\sparkspike 0.1 0.163934426229508 \sparkspike 0.3 0.409836065573771 \sparkspike 0.5 0.19672131147541 \sparkspike 0.7 0.147540983606557 \sparkspike 0.9 0.180327868852459 \end{sparkline} &	16 \% &	41 \% &	20 \% &	15 \% &	18 \% \\
&	DT&	\begin{sparkline}{5}\sparkspike 0.1 0.552941176470588 \sparkspike 0.3 0.682352941176471 \sparkspike 0.5 0.470588235294118 \sparkspike 0.7 0.411764705882353 \sparkspike 0.9 0.705882352941177 \end{sparkline} &	55 \% &	68 \% &	47 \% &	41 \% &	71 \% \\
 \cmidrule(lr){1-8}
\multirow{2}{*}{Intelligence} &	Agents&	\begin{sparkline}{5}\sparkspike 0.1 0.409836065573771 \sparkspike 0.3 0.704918032786885 \sparkspike 0.5 0.639344262295082 \sparkspike 0.7 0.381733021077283 \sparkspike 0.9 0.254098360655738 \end{sparkline} &	41 \% &	70 \% &	64 \% &	38 \% &	25 \% \\
&	DT&	\begin{sparkline}{5}\sparkspike 0.1 0.682352941176471 \sparkspike 0.3 0.176470588235294 \sparkspike 0.5 0.347058823529412 \sparkspike 0.7 0.102521008403361 \sparkspike 0.9 0.108823529411765 \end{sparkline} &	68 \% &	18 \% &	35 \% &	10 \% &	11 \% \\
 \cmidrule(lr){1-8}
\multirow{2}{*}{Autonomy} &	Agents&	\begin{sparkline}{5}\sparkspike 0.1 0.745901639344262 \sparkspike 0.3 0.786885245901639 \sparkspike 0.5 0.39344262295082 \sparkspike 0.7 0.331967213114754 \end{sparkline} &	75 \% &	79 \% &	39 \% &	33 \% \\	
&	DT&	\begin{sparkline}{5}\sparkspike 0.1 0.364705882352941 \sparkspike 0.3 0.317647058823529 \sparkspike 0.5 0.105882352941176 \sparkspike 0.7 0.0764705882352941 \end{sparkline} &	36 \% &	32 \% &	11 \% &	8 \% \\	
 \cmidrule(lr){1-8}
\multirow{2}{*}{Sociability} &	Agents&	\begin{sparkline}{5}\sparkspike 0.1 0.60655737704918 \sparkspike 0.3 0.672131147540984 \sparkspike 0.5 0.5 \end{sparkline} &	61 \% &	67 \% &	50 \% \\		
&	DT&	\begin{sparkline}{5}\sparkspike 0.1 0.396078431372549 \sparkspike 0.3 0.0823529411764706 \sparkspike 0.5 0.0411764705882353 \end{sparkline} &	40 \% &	8 \% &	4 \% \\		
         \bottomrule
    \end{tabular}
\end{table}
The data displayed in \autoref{tab:propertiesbylevel} shows for each property and paradigm to what extent the capabilities of each \ac{cd} have been achieved. For example, a score of 16~\% for \emph{fidelity} capabilities of \ac{cd}1 means that on average an Agent-based implementation exhibited 16~\% of the \ac{cd}1 \emph{fidelity} capability subset. As there is only the capability \emph{visualization} for \ac{cd}1 \emph{fidelity}, this also means that 16~\% of the Agent-based implementations exhibited this capability. However, in case of, for example, the two \ac{cd}3 capabilities for the property \emph{autonomy}, \emph{decision making} and \emph{mobility}, the average score of 39~\% for Agents only refers to the average fulfillment and does not necessarily mean that 39~\% of the implementations exhibited both capabilities. This is also visible in \autoref{fig-results-capabilities}, showing that \emph{mobility} is less frequent in Agents than \emph{decision making}.  

\emph{Sociability}, which scored the highest with a median of 64~\% in reports featuring Agents, is distributed similarly across the \acp{cd} (see \autoref{tab:propertiesbylevel}). Therefore, most screened reports featuring Agents had the required basic capabilities for interaction, as well as more complex capabilities such as \emph{coordination} and \emph{cooperation}. This finding aligns with the definition provided in \autoref{sec:agents}.

The results for \emph{intelligence} in \autoref{tab:propertiesbylevel} show that Agents excel in medium complexity capabilities (\ac{cd} 2 and 3) such as \emph{decision making} and social aspects, which aligns with common definitions of Agents. However, higher \acp{cd} of \emph{intelligence}, involving \emph{learning} and a high degree of \emph{autonomy}, are less frequently achieved. Nevertheless, Agents still score significantly higher in these more complex degrees of \emph{intelligence} compared to \acp{DT}.

Autonomy, being one of the core properties of Agents, is also notably strong compared to \acp{DT}. However, it is important to note that most approaches exhibit low degrees of \emph{autonomy}, while the higher degrees, consisting of capabilities like \emph{flexibility}, \emph{proactivity}, and \emph{adaptability}, are rarely achieved (see \autoref{tab:propertiesbylevel}). This may be due to the divergence between the definitions of Agents and the requirements of production, which was the focus domain of this work. In production environments, high degrees of \emph{autonomy} commonly associated with Agent-based concepts are often neither desirable nor necessary. For example, in many applications, Agents are responsible for making decisions based on local information to control the routing of orders and workpieces through the system. Within such constrained environments, it is often enough to incorporate dynamic states of machines and the environment (e.g., \emph{context awareness}) in simple rule based decision processes. This is in contrast to environments where the Agents encounter situations that are fully unknown and thus no general applicable rules can be provided, corrective actions cannot be taken in time by humans due to latencies, or the environment exposes the Agent to significant threats. In such cases, higher degrees of \emph{autonomy} would be required. Generally, Agents exhibit significantly higher scores of \emph{autonomy} compared to \acp{DT}.

\subsection{Allocation of Paradigms' Applications in RAMI 4.0} \label{sec:placementrami}
As described in \autoref{sec:method-cap-rami}, part of the analysis is to investigate which elements within \ac{rami} are addressed in the analyzed reports. The results of this analysis are presented in the following paragraphs. 

\begin{figure}[h]
    \centering
    \begin{subfigure}{0.48\textwidth}
        \centering
            \includegraphics[width = \textwidth]{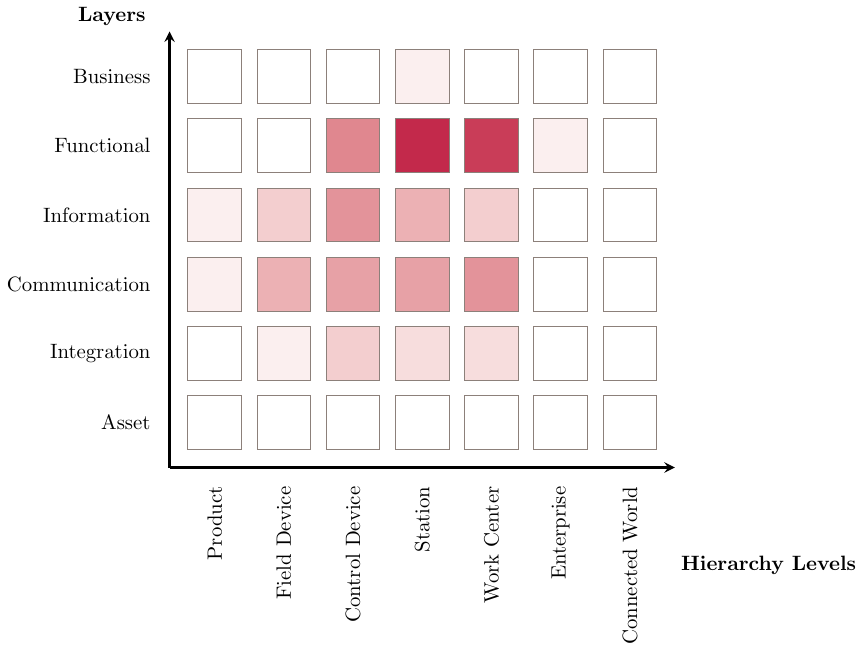}
        \caption{Focus Elements}
        \label{fig-placementramiagents-focus}
    \end{subfigure}
    \begin{subfigure}{0.48\textwidth}
        \centering
              \includegraphics[width = \textwidth]{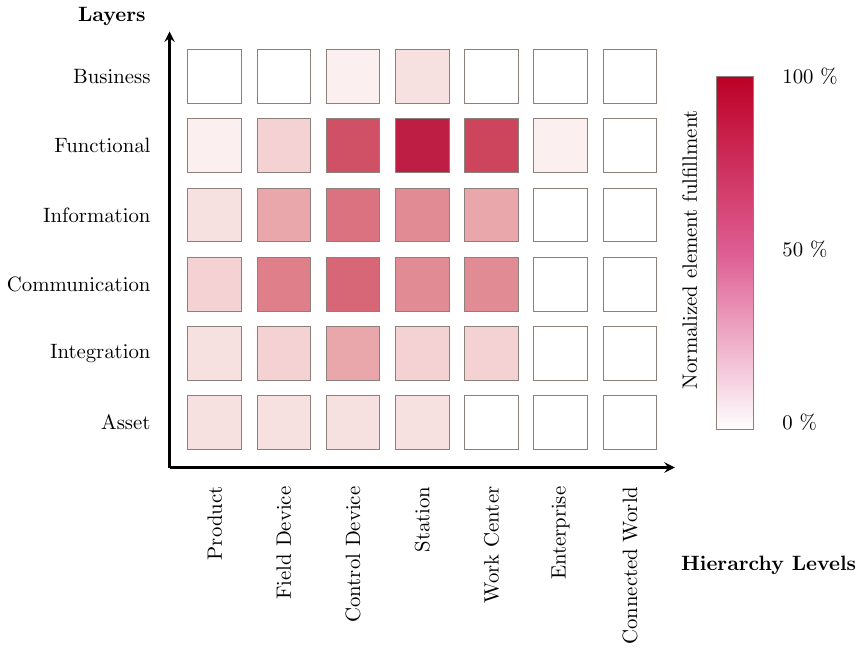}
        \caption{All Elements}
        \label{fig-placementramiagents-all}
    \end{subfigure}
    \caption{Placement of Agents' applications in \ac{rami}}
    \label{fig-placementramiagents}
\end{figure}

\begin{figure}[h]
     \centering
     \begin{subfigure}{0.48\textwidth}
         \centering
              \includegraphics[width = \textwidth]{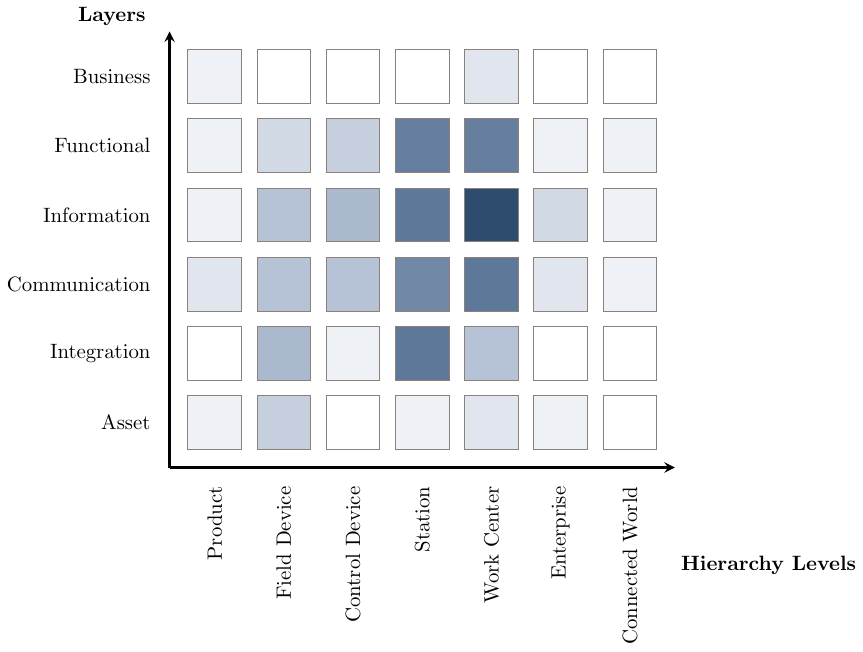}
         \caption{Focus Elements}
         \label{fig-placementramidt-focus}
     \end{subfigure}
     \begin{subfigure}{0.48\textwidth}
         \centering
              \includegraphics[width = \textwidth]{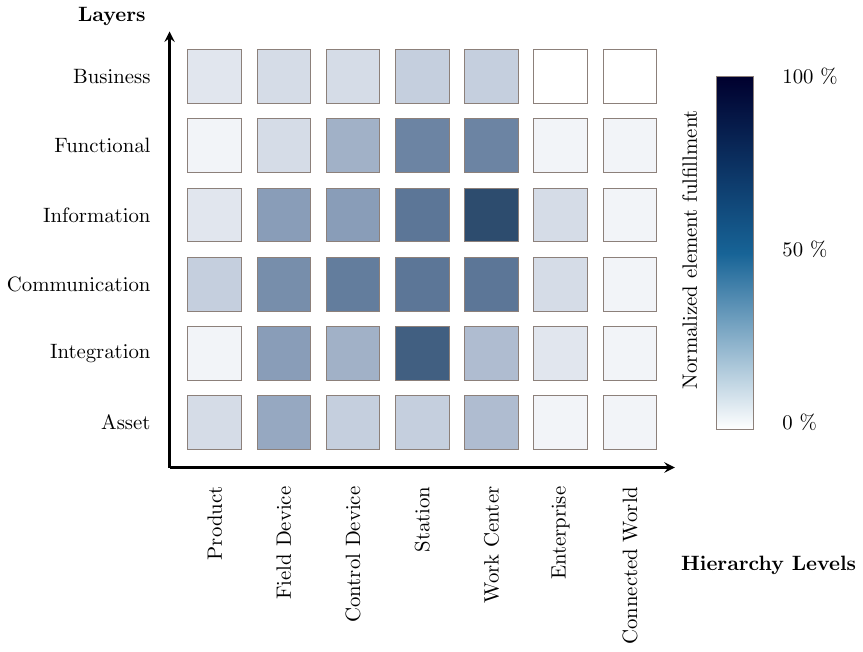}
         \caption{All Elements}
           \label{fig-placementramidt-all}
     \end{subfigure}
\caption{Placement of  Digital Twins' applications in \ac{rami}}
\label{fig-placementramidt}
\end{figure}

\autoref{fig-placementramiagents} and~\ref{fig-placementramidt} show for each element of \ac{rami} at what rate the respective elements occurred as the focus layer of the reports (\autoref{fig-placementramiagents-focus} and \ref{fig-placementramidt-focus}) and at what rate the respective elements occurred within the reports, in general (\autoref{fig-placementramiagents-all} and \ref{fig-placementramidt-all}). The value of the rate of occurrence (indicated by the tone of the color in each \ac{rami} element) is determined by dividing the number of occurrences in the respective \ac{rami} element by the total number of reports focusing on the respective paradigm. The visualization of the results is adapted from \citet{VeBa20}.

\autoref{fig-placementramiagents} presents the allocation of Agents' applications in \ac{rami}.  \autoref{fig-placementramiagents-focus} displays only the focus elements, showing that the focus of Agents' applications lies primarily on the \emph{integration} to \emph{business} layers as well as the  \emph{field device}, \emph{control device}, \emph{station}, and \emph{work center} hierarchy levels. The most frequently selected focus element is  \emph{functional} -- \emph{station}. The most frequently selected focus element is \emph{functional} - \emph{station}. The top three focus elements are found within the \emph{functional} layer, specifically at the \emph{control device} through \emph{work center} hierarchy levels. The \emph{asset} layer is never focused on, nor are the \emph{product} and \emph{enterprise} hierarchy levels, or the \emph{business} layer. This indicates that Agents are primarily applied within industrial production for establishing \emph{communication} between different units of a production system and executing \emph{functions} to facilitate the correct operation of the system. Interestingly, the processing of \emph{information} is rarely the focus of Agents, suggesting that Agents are commonly applied in industrial applications where \emph{information} processing is either a relatively simple task or performed by means other than Agents.
As Agents typically do not run on \emph{field devices} or \emph{control devices} (even though such applications exist \citep{USV12}), their frequent allocation to the \emph{station} and \emph{work center} levels are as expected.

\autoref{fig-placementramiagents-all} depicts the rate of occurrence of \ac{rami} elements without differentiation of focus element. It is evident that almost all Agents' applications (94~\%) feature the \ac{rami} element of \emph{functional} - \emph{station}, which is the most commonly selected element. The distribution of applications throughout \ac{rami} extends to the edges of the axes, indicating that the application of Agents requires consideration of many aspects within \ac{rami}, essentially up to the \emph{enterprise} level.

\autoref{fig-placementramidt} presents the categorization of \acp{DT}' applications. As seen in  \autoref{fig-placementramidt-focus}, \acp{DT} show a focus on the \emph{communication}, \emph{information}, and \emph{functional} layers, as well as the \emph{station} and \emph{work center} levels. The combination of the \emph{work center} level and the \emph{information} layer appears to be particularly focused in \acp{DT}' applications. However, the surrounding layers are also covered, and there is no layer or level that is not addressed.   

\autoref{fig-placementramidt-all} displays the categorization of \acp{DT}' applications, considering all elements. The results align with the focus element categorization, as the same layers and levels are most commonly found. However, there is a broader spread of observed \ac{rami} elements across both layers and levels, covering almost every combination of layer and level. This indicates a wide range of application areas for \acp{DT}.

Comparing the placement of Agents' applications (\autoref{fig-placementramiagents}) and \acp{DT}' applications (\autoref{fig-placementramidt}) it is evident that both paradigms mainly focus on the \emph{communication}, \emph{information}, and \emph{functional} layers, with a primary emphasis on the \emph{station} and \emph{work center} hierarchy levels. \acp{DT}' applications exhibit a stronger presence towards the edges of the axes compared to Agents' applications. Another interesting observation is that \ac{DT} applications have a strong focus on the processing of \emph{information}, whereas Agents' applications are primarily focused on executing \emph{functions} within the \emph{planning} and \emph{control} of production processes. This aligns with the definition presented in \autoref{sec:dt}. However, it is worth noting that for Agents, one might have expected their cooperative and social nature to also benefit use cases in the field of \emph{information} processing and management. Nevertheless, such use cases appear to be less common in practice.

\subsection{Purposes of Agents and Digital Twins} \label{sec:systemgoals}
This section presents the results of the analysis regarding the purposes connected to the applications of Agents and \acp{DT}. As described in \autoref{sec:method-goals}, nine categories of purposes have been identified that Agents and \acp{DT} are meant to fulfill: process optimization, user assistance, virtual commissioning, planning, scheduling, dispatching, control, monitoring, and diagnosis and fault management.

\begin{figure}[h]
\centering
              \includegraphics[width = .5\linewidth]{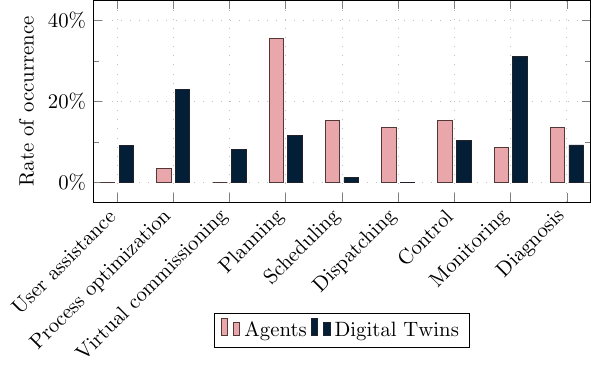}
\caption{Rate of occurrence of purpose categories in Agents and Digital Twins}
\label{fig:goalCategories}
\end{figure}

\autoref{fig:goalCategories} shows how often Agents and \acp{DT} are applied to serve a purpose each one of the nine categories. Five out of the nine purpose categories are predominantly addressed by one paradigm: \emph{user assistance}, \emph{process optimization}, and \emph{virtual commissioning} are specific to \acp{DT}, while \emph{scheduling} and \emph{dispatching} are mostly associated with Agents.

On the other hand, \emph{planning}, \emph{control}, \emph{monitoring}, and \emph{diagnosis and fault management} are areas where both Agents and \acp{DT} are applied. This indicates that Agents are primarily used during the operational phase of a production system's life cycle, whereas \acp{DT} are also utilized in the engineering and commissioning phases, demonstrating the wider range of use cases for \acp{DT} observed in \autoref{sec:placementrami}.

During the operational phase, Agents are employed for production \emph{planning}, \emph{scheduling}, and \emph{dispatching}. Production \emph{control} is a category that is evenly distributed between Agents and \acp{DT}. \emph{Monitoring} tasks are predominantly performed by \acp{DT}, while \emph{diagnosis and fault management} is carried out by both Agents and \acp{DT}, with Agents being the more commonly used paradigm. These observations align with the findings from  \autoref{sec:capmasdt}, which indicated that there are certain areas where Agents and \acp{DT} exhibit strong similarities, while other areas are more specific to each paradigm.

The observation that the \ac{DT} paradigm is applied to industrial use cases spanning from the engineering phase (\emph{process optimization}) to the operational phase (e.g., \emph{control}) confirms the observations presented in \autoref{sec:placementrami}, indicating that the \ac{DT} paradigm encompasses a wider range of elements within \ac{rami}. Similarly, as discussed in \autoref{sec:placementrami}, Agents are primarily utilized for executing functions within the \emph{station} and \emph{work center} levels. By referring to \autoref{fig:goalCategories}, these functions can now be attributed to the domains of production \emph{planning}, \emph{scheduling}, \emph{dispatching}, as well as \emph{diagnosis and fault management}.

%% file: 4discussion.tex
\section{Discussion} \label{sec:discussion}
This work presented a systematic comparison of applications of Software Agents and \acp{DT} in industrial automated production systems. The goal of this study was to identify similarities and differences between the two software paradigms, as well as potential synergies between them. The comparison was based on a systematic literature review conducted according to the \ac{PRISMA} 2020 statement. The focus of the review was on actual implementations of Agents and/or \acp{DT}. The two software paradigms were compared in terms of their purposes, properties, and capabilities. Additionally, the study investigated which elements of the \ac{rami} architecture are addressed when applying Agents and \acp{DT}.

The remainder of this section discusses the methods employed in this study. Subsequently, the results are discussed, and future research opportunities are identified.

\subsection{Discussion of Applied Methods} \label{sec:Discussion}
The set of reports analyzed in this study depended largely on the chosen search term and database. \emph{Scopus} was used as experience shows that it yields the highest number of reports. The search term described in \autoref{sec:method} was formulated to capture reports in the context of automated industrial production systems. Special care was taken to ensure a neutral phrasing of the search term to avoid any bias towards a specific software paradigm, application, use-case, or \ac{rami} element. The balance of 58 reports on Agents to 86 reports on \acp{DT}, with one report describing both an Agent's and a DT's application \citep{XSK+21}, indicates that this goal was satisfactorily achieved.

It is worth noting that the \scopus \ results obtained from Scopus represent only a fraction of all research conducted on Agents and \acp{DT} in the context of industrial automated production systems. However, it is estimated that increasing the number of reports would have a minimal impact on the quality of the results, as the sample size of 2,635 is already large and the presented results are clear, based on observations that are unlikely to be due to statistical variations. 

Another aspect of the literature review that can be discussed is the number of reviewers per report. Most reports were analyzed by one reviewer, with 15 reports being analyzed by all reviewers to align the review process before the major screening effort. Additionally, 10~\% of the reports were analyzed by one additional reviewer after the initial screening to ensure consistency. The results of the comparative screening remained consistent throughout all phases. Minor deviations were observed between individual reviewers, such as whether an implementation of the \ac{DT} had an individual capability or not. However, these deviations were uncommon, and the results consistently portrayed a clear picture across all reviewers. Having multiple reviewers with different specializations, as was the case in this study, helps prevent bias in a particular direction.

It was decided during the design of this study to only consider reports that presented actual implementations of Agents and \acp{DT} for industrial applications. This decision was made to obtain a sense of what kind of capabilities are required and useful in the industrial context and what capabilities are realistically implementable. Purely conceptual reports were omitted because it was deemed that concepts can be proposed irrespective of the need for a concept and its applicability in industrial practice.

To compare Agents and \acp{DT}, a set of capabilities, properties, and purposes was created. These sets were developed based on a review of existing literature.

The challenge in establishing a set of capabilities was to encompass all the tasks, features, and functions fulfilled by an Agent or \ac{DT}. The review process demonstrated that the established list of capabilities adequately met this requirement, although it should be acknowledged that the set of capabilities could be adjusted by adding new elements or combining existing ones. An issue encountered in identifying the capabilities of a given application was that reports often use different terminology or certain capabilities were only implicitly observed. In such cases, the interpretation of the reviewer became necessary, and it was in these instances that most divergences between reviewers occurred. A crucial aspect of the chosen approach was to provide clear definitions for each capability, along with corresponding examples, to establish a common understanding among the authors and readers.

The applied methodology for the calculation of the fulfillment score of the different properties enabled a quantitative comparison of the paradigms. It was also utilized to outline the fulfillment of the different \acp{cd} within each property. However, the fact that the \ac{cd} was also utilized to determine the weight of each capability for the property fulfillment score puts a strong emphasis on the more complex capabilities. This emphasis could also be put on the basic or fundamental capabilities, which would lead to different results. The argument for the latter approach would state that, for example, systems are not \emph{autonomous} if they do not possess the basic capability of \emph{encapsulation}. However, the analysis in this contribution followed established metrics like those described in \autoref{sec:autonomy} that emphasise more complex autonomous capabilities. 

The set of nine categories of purposes presented in \autoref{sec:method-goals} fully encompassed the purposes pursued by the analyzed reports. In some cases, multiple categories of purposes were assigned to a single application of Agents or \acp{DT}. However, this was only done in a few cases where the authors explicitly stated the application of a paradigm for multiple purposes. On the other hand, some authors did not explicitly state the purpose of the Agents and \acp{DT} they implemented. In such cases, the categories of purposes were identified based on general statements in the reports.

In addition to identifying the capabilities of different use cases, a classification of the two paradigms was made in accordance with the different dimensions of \ac{rami}. The objective was to identify commonalities and differences. Since \ac{rami} is a widely recognized reference architecture for Industry 4.0, in which both paradigms are increasingly used, it provided an appropriate basis for the analysis. Furthermore, \ac{rami} was considered particularly suitable because it allows multidimensional classification along relevant dimensions. This multidimensional classification facilitated a more comprehensive understanding of these paradigms in relation to Industry 4.0. The use of existing standards to characterize the dimensions promoted the establishment of a uniform and consistent understanding of the paradigms. However, it should be noted that a certain bias is inherent in the fact that the analysis was limited to the dimensions prescribed by \ac{rami}. This means that other relevant aspects may have been overlooked. Other architectural models and classifications could have also enabled the classification of the paradigms. \citet{AXZ+21} present an architectural model for the \ac{DT} based on \ac{rami}. Another architectural model worth mentioning is the "China Intelligent Manufacturing System Architecture", which is very similar to \ac{rami}. It consists of dimensions such as "System Hierarchy," "Lifecycle," and "Intelligent Functions," and efforts are being made to align both architectures \citep{FederalMinistryforEconomicAffairsandEnergy.2018}. It can be assumed that using this alternative reference architecture, which shares strong similarities with \ac{rami}, would have yielded similar results.

Generally, it would have been desirable to compare the results of this systematic comparison of Agents and \acp{DT} with previous, similar works on the topic. However, as stated in \autoref{sec:relatedworks}, there exist no other similar comparison-based analyses to which the results could be compared. Despite the lack of comparable analyses, the methodology applied in this work is based on \ac{PRISMA} and leads to reproducible results.

\subsection{Discussion of Results}
The results revealed both differences and similarities between Agents and \acp{DT}. Regarding the capabilities of the two paradigms, it was found that \emph{simulation}, \emph{condition \emph{monitoring}}, \emph{real time capability}, and \emph{visualization} are \ac{DT}-specific capabilities that imply a high degree of \emph{fidelity}. On the Agent side, the dominant capabilities are \emph{encapsulation}, \emph{internal system interaction}, \emph{context awareness}, \emph{decision making}, \emph{external system control}, \emph{coordination}, \emph{cooperation}, and \emph{flexibility}. This means that Agents tend to achieve medium to high scores in \emph{intelligence}, \emph{autonomy}, and \emph{sociability}.

While both Agents and \acp{DT} exhibited paradigm-specific capabilities, all capabilities except for \emph{mobility} occurred in both paradigms. Moreover, more than half of both Agents' and \acp{DT}' applications possessed capabilities that are characteristic of the other paradigm. This indicates that there is no clear boundary between the two paradigms.

The distribution of capabilities further demonstrated that the typical capability profile of Agents was more uniform than that of \acp{DT}. This uniformity was also evident in the analysis of the purposes for which Agents are applied. In the industrial context, Agents are predominantly utilized during the operational phase for \emph{planning}, \emph{scheduling}, \emph{dispatching}, \emph{control}, and \emph{diagnosis and fault management} of production operations. This observation aligns with the allocation of Agents' applications in the \ac{rami} architecture, which shows their primary deployment on the \emph{functional} layer, specifically on computers near the production process (i.e., at the \emph{station} and \emph{work center} levels). This uniformity can be attributed to the existence of established tools and de facto standards in the field of Agents, such as the  \emph{Java Agent Development Framework} and the \emph{Agent Management Specification}. This allows different \acp{MAS} to exhibit a similar structure which also requires a similar capability set within \acp{MAS}.

A different pattern emerged in applications of \acp{DT}. The purposes served by \acp{DT} were more diverse, ranging from the engineering phase (\emph{process optimization}) to real-time \emph{control} of production operations. The allocation of \ac{DT} applications within \ac{rami} also demonstrated greater diversity, with a focus on \emph{information} processing at the \emph{work center} level. Furthermore, the set of capabilities found in \acp{DT} was less uniform compared to Agents, as less than half of the analyzed \ac{DT} applications exhibited over 50~\% of the \ac{DT}-specific capabilities. All of these findings indicated comparatively less standardization and uniformity in \ac{DT} applications within the industrial sector. This observation is consistent with the conclusions of other authors: \emph{There lacks, [...], a clear, encompassing architecture covering necessary components of a \ac{DT} to realize various use cases in an intelligent automation system.} \citep{AJL+19}. Similar to the Software Agents paradigm, there is a standard for \ac{DT} implementations, the \acp{AAS}. However, this work revealed that implementations of \acp{AAS} are rare compared to custom \acp{DT} implementations. 

Summarizing the results and discussion presented earlier, the most significant differences between Agents and \acp{DT} align with their definitions provided in \autoref{sec:agents} and \ref{sec:dt}. It was observed that \acp{DT} are primarily applied to manage \emph{information} related to production processes and present it to the user, while Agents are mostly employed to plan and manage the operation of production processes. This is evident from the purposes served by the paradigms (\autoref{fig:goalCategories}) and the capabilities they possess (\autoref{fig-results-capabilities}). However, a clear similarity exists in that both Agents and \acp{DT} are software programs that run closely to and interact closely with production resources  (\autoref{fig-placementramiagents} and~\ref{fig-placementramidt}). On a more detailed level, it was observed that \acp{DT} primarily possess capabilities that enable them to accurately represent production resources, while Agents possess capabilities for cooperative \emph{planning} and operation of production resources (\autoref{sec:capmasdt}). This is also reflected in the properties observed in Agents and \acp{DT} (\autoref{sec:capdt}). Furthermore, it is noted that there are almost no capabilities exclusive to either Agents or \acp{DT}, and they showed strong resemblance in terms of their ability to interact with other technical systems, interact with human operators, and learn about their environment. Additionally, Agents and \acp{DT} are applied for similar purposes, such as \emph{controlling} production resources, \emph{planning}, \emph{monitoring}, \emph{diagnosing} production operations, and \emph{managing faults}.

Finally, several synergies that represent potential for future research activities can be highlighted. The concept of Industry 4.0 envisions the emergence of \emph{autonomous}, interconnected, and \emph{intelligent} production systems that access, exchange, and process large amounts of data (\autoref{sec:introduction}) \citep{Monostori.2014}. Analyzing this vision and the findings of this work, it is evident that aspects related to the \emph{autonomy}, \emph{intelligence}, and interconnection of production systems are primarily addressed by Agents. This includes capabilities encompassing these properties, such as \emph{cooperation}, \emph{decision making}, \emph{reactivity}, and \emph{flexibility} (\autoref{sec:capandprop}). It is worth noting that interconnection, which refers to the degree of interaction between (sub)systems, is part of the \emph{sociability} property as interpreted in this work. On the other hand, other aspects of envisioned future production systems, such as providing real-time data, documenting the evolution of production (sub)systems, and processing \emph{information}, are typically addressed by \acp{DT}. \acp{DT} commonly employ capabilities such as \emph{simulation}, \emph{real-time capability}, \emph{visualization}, and \emph{condition monitoring}. Based on the analysis presented in this work, it can be concluded that the capabilities and properties of both Agents and \acp{DT} should be combined to realize the current vision of future production systems. Agents and their corresponding methods, tools, and standards should be utilized for the \emph{autonomy}-related and cooperative aspects of managing production systems, while \acp{DT} provide the necessary informational background and real time connections to their corresponding assets. The analysis of Agents' and \acp{DT}' properties supports this conclusion, as a combination of Agents and \acp{DT} promises high degrees of \emph{intelligence}, \emph{fidelity}, \emph{autonomy}, and \emph{sociability}. Therefore, the existing knowledge on Agents and \acp{DT} should be integrated to apply the two paradigms synergistically. Despite the potential of concurrent utilization of Agents and \acp{DT}, relatively little research has been conducted on this topic, as demonstrated in \autoref{sec:motivation}. However, some works have already leveraged the advantages of both Agents and \acp{DT}, similar to the description above \citep{VOS21}. To promote this vision, the connection and collaboration of \acp{DT} and Agents need to be standardized.

Further research opportunities and potential synergies between the research communities of Agents and \acp{DT} can be explored in areas where both paradigms are applied. This encompasses investigating the distinct advantages of Agents and \acp{DT} in purposes such as \emph{planning}, \emph{controlling} production processes, \emph{diagnosing, and managing faults}.  Further, the kind of \emph{learning} algorithms that are applied in either paradigm and their advantages should be investigated. Finally, more research and standardization efforts are needed for the interaction of software entities, such as Agents and \acp{DT}, with other technical systems and humans.

%% file: abbr_table_results.tex
\autoref{tab-resultsoverview} shows the results of the analysis of the reports included during the systematic literature review described in \autoref{sec:method-comp} and~\ref{sec:review-results}. \autoref{tab-resultsoverview} first lists all reports which describe an application of Agents and then all reports with describe an application of \acp{DT}. If both paradigms occur in one report, it is reported twice. The dataset related to this article can be found at \url{https://doi.org/10.5281/zenodo.8120624        } \citep{dataset}.

Due to limited space, the following abbreviations are used in \autoref{tab-resultsoverview}: 

\paragraph{Capabilities (Cap.)} Adaptability (Ad), Condition Monitoring (CM), Context-Awareness (CA), Cooperation (Cop), Coordination (Cor), Decision-Making (DM), Encapsulation (Enc), External-System-Interaction (ESI), Flexibility (Fl), Human-Interaction (HI), Internal-System-Interaction (IS), Learning (Lea), Mobility (Mo), Negotiation (Ne), Prediction (Pre), Prioritization (Pri), Proactivity (Pro), Reacitvity and External-System-Control (RE), Real-time Capability (RT), Robustness and Resilience (Rob), Simulation (Si), Synchronisation (Sy), Uncertainty-handling (Un), Visualisation (Vis)

\paragraph{Properties}
Fidelity (Fid.), 	Intelligence (In.), 	Autonomy (Auto.), 	Sociability (Soc.)

\paragraph{\ac{rami} Layers}
Focus elements are marked by "(F)" directly following the respective abbreviation. 
	Asset (A),	Integration (Int),	Communication (C),	Information (Inf),	Functional (Fun),	Business (Bus)

\paragraph{\ac{rami} Hierarchy Levels}
Similar to the \ac{rami} layers, Hierarchy Levels are reported: 	Product (Pro),	Field Device (FD),	Control Device (CD),	Station (St),	Work Centers (WC),	Enterprise (Ent),	Connected World (CW)

\paragraph{Purposes (Purp.)}
User Assistance (UA), Process Optimization (PO), Virtual Commissioning (VC), Planning (Pl), Scheduling (Sc), Dispatching (Di), Control (Co), Monitoring (Mo), Diagnosis (Di)

%% file: tab_properties.tex
\renewcommand{\arraystretch}{.75}
\scriptsize{
\begin{longtable}{
p{1cm}	p{2.3cm}|	p{.6cm}	p{.6cm}	p{.6cm}	p{.6cm}	||p{1cm}	|p{1cm} ||p{.5cm}
}
\caption{Overview of included reports and corresponding characteristics \label{tab-resultsoverview}} \\
\rot{Reference}& \rot{Capabilities}& \rot{Fidelity}& \rot{Intelligence}& \rot{Autonomy}& \rot{Sociability}& \rot{Layers}& \rot{Levels} & \rot{Purpose} \\
\toprule 
\endfirsthead
\caption*{\textbf{Table~\ref{tab-resultsoverview}} (continued) \\ Overview of Included Reports and its Characteristics} \\
Ref.& Cap.& Fid.& In.& Auto.& Soc.& Layers& Levels & Purp. \\
\toprule 
\endhead 
\multicolumn{8}{l}{\textbf{Reports with focus on Agents}} \\
\\
\input{tab_results_agents} \\
\midrule
\multicolumn{8}{l}{\textbf{Reports with focus on Digital Twins}} \\
\\
\input{tab_results_dt} \\
\bottomrule
\end{longtable}
}

%% file: tab_results_agents.tex
\citet{CGQ+18}& Ad, CM, CA, Cop, Cor, DM, ESI, Fl, HI, IS, Lea, Pri, RE, Si, Vis& 44\%& 65\%& 54\%& 73\%& A, Int(F), C, Inf, Fun& St, WC(F)& Mo\\ 
\citet{MWH+17}& Ad, CA, Cop, Cor, DM, Enc, ESI, Fl, HI, IS, Lea, Ne, Pri, Si, Un, Vis& 33\%& 76\%& 50\%& 100\%& C, Inf(F), Fun& FD, CD, St(F)& Pl\\ 
\citet{VRO+11}& CM, Cor, DM, Enc, ESI, Fl, HI, IS, Ne& 11\%& 40\%& 42\%& 73\%& C, Inf(F), Fun, Bus& St(F), WC(F)& Sc\\ 
\citet{HVN+21}& CM, CA, Cop, Cor, DM, Enc, ESI, IS, RE, Sy& 28\%& 32\%& 27\%& 64\%& A, Int, C, Inf, Fun(F), Bus(F)& Pro, FD, CD, St(F)& Di\\ 
\citet{CiMa11}& Ad, CA, Cop, DM, Enc, HI, IS, Ne, Pro, RE& 0\%& 46\%& 54\%& 73\%& Inf, Fun(F)& FD, St, WC(F)& Mo\\ 
\citet{DAS+21}& Cop, Cor, IS, Sy& 17\%& 17\%& 8\%& 55\%& C(F)& St(F)& Di\\ 
\citet{KML+11b}& Ad, CA, Cop, Cor, ESI, Fl, IS, Pri, RE, Rob& 0\%& 54\%& 58\%& 64\%& Fun(F)& CD(F), WC(F)& Pl\\ 
\citet{KKU+19}& CM, DM, Enc, IS, Pri, RE, Sy& 28\%& 17\%& 23\%& 9\%& Inf(F), Fun(F)& St(F)& Di\\ 
\citet{LLS14}& CM, CA, Cop, Cor, ESI, Fl, IS, RE, Si& 39\%& 33\%& 27\%& 64\%& Fun(F)& St(F)& Pl\\ 
\citet{MMS+17}& CA, Cop, Enc, ESI, Fl, HI, IS, Ne, Pri, RE& 0\%& 43\%& 31\%& 82\%& Fun(F)& St(F)& Di\\ 
\citet{MMV+11}& Cop, Cor, Enc, ESI, Fl, IS, Ne, RE, Rob, Vis& 6\%& 43\%& 46\%& 91\%& C, Inf, Fun(F)& St(F)& Di\\ 
\citet{CAS+19}& DM, Enc, Fl, IS, Ne& 0\%& 22\%& 31\%& 36\%& C(F), Inf, Fun(F)& St(F)& Sc\\ 
\citet{SSF+11}& CM, CA, Cor, Enc, ESI, Fl, IS, Ne, Pre, RE, RT, Rob, Sy, Un& 67\%& 54\%& 46\%& 64\%& C(F), Inf, Fun(F)& CD(F), WC& Pl, Co\\ 
\citet{USV12}& CA, Cor, DM, Enc, Fl, IS, Lea, Pre, RE, RT, Rob& 39\%& 48\%& 54\%& 27\%& C(F), Inf, Fun(F)& CD(F), WC(F)& Co\\ 
\citet{VDP+14}& CA, Cor, DM, Enc, ESI, Fl, HI, IS, Ne, RE& 0\%& 41\%& 42\%& 73\%& Inf(F), Fun& CD, St(F)& Pl\\ 
\citet{ROL18}& Ad, CM, CA, Cor, DM, Enc, Fl, IS, Lea, RE& 11\%& 43\%& 54\%& 27\%& Int, C, Inf, Fun(F)& CD(F), St(F)& Pl\\ 
\citet{RoMa12}& Ad, CM, CA, Cor, DM, Enc, ESI, Fl, IS, Ne, Pro, RE, Rob& 11\%& 62\%& 88\%& 64\%& Int, C, Inf, Fun(F)& WC(F)& Sc\\ 
\citet{LRB+15}& Ad, CM, CA, Cop, Cor, DM, Enc, ESI, RE, Sy& 28\%& 35\%& 42\%& 55\%& Int, C, Inf, Fun(F)& St(F), WC& Pl\\ 
\citet{KBT+22}& Ad, CA, Cop, DM, Enc, ESI, Fl, HI, Ne, Pri, Pro, RE, Rob& 0\%& 67\%& 88\%& 73\%& Int, C(F), Inf, Fun(F)& St(F), WC(F)& Pl\\ 
\citet{LZV+11}& Ad, CA, Cop, Cor, DM, Enc, ESI, Fl, IS, Pre, RE, RT, Rob, Si& 67\%& 59\%& 73\%& 64\%& C(F), Inf, Fun(F)& St(F), WC& Co\\ 
\citet{BGB+22}& Ad, DM, Enc, ESI, Fl, IS, Ne, RE, Vis& 6\%& 35\%& 58\%& 45\%& A, Int(F), C, Inf(F), Fun& CD(F)& Sc\\ 
\citet{Dee22}& CA, ESI, Lea, Si& 28\%& 16\%& 4\%& 9\%& Fun(F)& St(F)& Di\\ 
\citet{KSZ+14}& DM, ESI, Fl, Mo, RE, Rob& 0\%& 24\%& 65\%& 9\%& C, Fun(F)& WC(F)& Sc\\ 
\citet{KrBa18}& Cor, DM, Enc, ESI, Fl, IS, RE& 0\%& 27\%& 42\%& 36\%& A, Int(F), C(F), Inf, Fun& WC(F)& Co\\ 
\citet{MBT+20}& CA, Cop, DM, Enc, ESI, Fl, IS, RE& 0\%& 30\%& 42\%& 45\%& Int, C(F), Inf, Fun(F)& St(F)& Sc\\ 
\citet{OLB+12}& Cop, Cor, DM, Enc, ESI, HI, IS, Ne, RE, RT, Rob& 17\%& 46\%& 42\%& 100\%& Int, C(F), Inf, Fun(F)& St(F)& Pl\\ 
\citet{RPE+21}& CM, CA, Cor, DM, Enc, ESI, Lea, Ne, Pre, RE, Si& 61\%& 41\%& 27\%& 55\%& A, Int, C, Inf(F), Fun(F)& CD(F), WC(F)& Pl\\ 
\citet{SKM19}& CM, CA, Cor, DM, Enc, ESI, Fl, Lea, Ne, Pre, Pri, Pro, RE& 33\%& 62\%& 58\%& 55\%& Int, C(F), Inf, Fun(F)& CD(F), WC(F)& Pl\\ 
\citet{WLM13}& CM, CA, Cop, Cor, DM, Enc, ESI, Fl, HI, IS, Lea, Ne, Pri& 11\%& 63\%& 42\%& 100\%& C(F), Inf(F), Fun& FD(F), CD(F)& Pl\\ 
\citet{SLL+22}& CM, CA, Cop, Cor, DM, ESI, HI, IS, Pre, Pro, RE, Rob& 33\%& 59\%& 54\%& 73\%& A, Int(F), C(F), Inf(F), Fun, Bus& FD(F), CD(F)& Mo\\ 
\citet{AAd19}& CA, Cop, Cor, DM, Enc, ESI, IS, Ne& 0\%& 37\%& 27\%& 91\%& Fun(F), Bus& St(F)& Di\\ 
\citet{CBS+15}& Ad, CM, CA, Cop, DM, Enc, ESI, HI, IS, Pro, RE, Rob, Vis& 17\%& 54\%& 73\%& 55\%& C, Inf(F)& CD(F)& Di\\ 
\citet{CNS18}& CA, DM, Enc, ESI, IS, RE, RT, Si& 44\%& 17\%& 27\%& 18\%& Int(F), C, Fun& CD(F)& Sc\\ 
\citet{CMS+18}& CA, Cop, Cor, DM, Enc, ESI, IS, RE& 0\%& 30\%& 27\%& 64\%& C, Fun(F)& CD, St(F)& Di\\ 
\citet{GrPo17}& CA, Cop, Cor, DM, Enc, ESI, IS, Ne& 0\%& 37\%& 19\%& 91\%& Fun(F)& WC(F)& Sc\\ 
\citet{HTS+16}& CA, Cor, DM, Enc, ESI, IS, Pro, RE, RT, Rob& 17\%& 40\%& 58\%& 36\%& C(F)& WC(F)& Di\\ 
\citet{KQZ22}& CM, CA, Cop, Cor, DM, Enc, ESI, Fl, HI, IS, Ne, Pri, RE, RT, Rob& 28\%& 63\%& 58\%& 100\%& Fun(F)& WC(F)& Pl\\ 
\citet{KDP18}& Ad, CA, Cop, Cor, DM, ESI, Fl, IS, Pri, RE, Si, Sy& 44\%& 51\%& 54\%& 64\%& Fun(F)& CD(F)& Co\\ 
\citet{KBT17}& CM, CA, Cor, DM, Enc, ESI, Fl, IS, Pri, Pro, RE, Rob& 11\%& 54\%& 73\%& 36\%& Fun(F)& St(F)& Pl\\ 
\citet{LZW+15}& CA, DM, Enc, ESI, IS, Pre, Pri, RE, Si, Vis& 56\%& 30\%& 27\%& 18\%& Fun(F)& WC(F)& Pl\\ 
\citet{MHW+18}& Cop, Cor, DM, Enc, ESI, HI, IS, Ne, RE, Vis& 6\%& 38\%& 27\%& 100\%& Fun(F)& Ent(F)& Pl\\ 
\citet{NaSr14}& Ad, CM, CA, Cop, Cor, DM, Enc, ESI, HI, IS, Mo, RE, Rob, Sy, Vis& 33\%& 52\%& 69\%& 73\%& Fun(F)& WC(F)& Di\\ 
\citet{VML+11}& Ad, CA, Cop, Cor, DM, Enc, ESI, IS, Pre, Pri, RE, RT, Rob, Sy, Un& 56\%& 65\%& 58\%& 64\%& Fun(F)& CD(F)& Pl, Di\\ 
\citet{NME+22}& Ad, CA, Cop, Cor, Fl, Pri, RE, Rob& 0\%& 44\%& 46\%& 45\%& A, Fun(F)& St, WC(F)& Sc\\ 
\citet{FPS11}& CM, Enc, IS, RE& 11\%& 6\%& 12\%& 9\%& Inf(F)& CD(F)& Di\\ 
\citet{GLA+21}& Ad, Cor, DM, Enc, ESI, Fl, Ne& 0\%& 37\%& 50\%& 55\%& C(F), Inf, Fun, Bus& CD, St, WC(F)& Pl\\ 
\citet{JJK+18}& Cop, Cor, IS, Mo, RE, Si& 28\%& 17\%& 19\%& 55\%& Int(F), C, Inf, Fun& FD, CD, St(F)& PO\\ 
\citet{KLJ+18}& CM, CA, Cop, Cor, Enc, IS, Ne& 11\%& 29\%& 4\%& 82\%& C(F)& WC(F)& PO\\ 
\citet{KML+11}& Ad, CA, ESI, Sy& 17\%& 16\%& 19\%& 9\%& A, Int, C(F)& FD(F), CD(F)& Di\\ 
\citet{NCS12}& CA, Cop, Cor, DM, Enc, HI, Ne, RE, RT, Rob& 17\%& 40\%& 38\%& 82\%& Int, C, Inf(F)& FD(F), CD(F), St& Co\\ 
\citet{RCK+18}& CA, Cop, Enc, ESI, Lea, RE& 0\%& 22\%& 15\%& 36\%& Int, C(F)& FD(F), CD& Co\\ 
\citet{SSJ11}& Ad, CM, CA, Cop, Enc, Fl, IS, Ne, RE, Sy& 28\%& 37\%& 42\%& 64\%& C(F)& Pro(F)& Pl\\ 
\citet{TAC17}& Cop, DM, Enc, Fl, HI, IS, Lea, Ne, Pre& 22\%& 48\%& 31\%& 73\%& C(F), Inf& FD(F)& Pl\\ 
\citet{WJD17}& CM, CA, Enc, ESI, Fl, RE& 11\%& 16\%& 31\%& 9\%& C, Inf(F)& Pro(F)& Mo\\ 
\citet{WDP+21}& CA, Cop, Cor, DM, ESI, IS, RE, Rob& 0\%& 38\%& 31\%& 64\%& Fun(F)& WC(F)& Co, Mo\\ 
\citet{BDP+18}& Ad, DM, ESI, Lea, Si, Sy& 44\%& 25\%& 38\%& 9\%& Int(F)& St(F)& Pl\\ 
\citet{CrVo22}& CM, CA, Cop, Cor, DM, Enc, ESI, Fl, HI, IS, Lea, Ne, Pro, RE, Sy, Vis& 33\%& 65\%& 50\%& 100\%& A, Int, C, Inf(F), Fun& Pro, FD, CD, St(F)& Mo\\ 
\citet{Giacomo.2023}& CM, Cor, DM, Fl, IS, Sy, Un& 28\%& 30\%& 35\%& 27\%& C, Inf(F), Fun& WC(F)& Co, Mo\\ 
\citet{XSK+21}& ESI, Lea, Si& 0\%& 13\%& 4\%& 9\%& Inf(F)& WC(F)& Co\\ 

%% file: tab_results_dt.tex
\citet{LPZ+22}& CM, CA, Fl, Lea, RE, RT, Si, Vis& 61\%& 19\%& 23\%& 0\%& Int, C, Inf(F)& Pro(F), St& PO\\ 
\citet{GBC+20}& CM, ESI, HI, RE, RT, Si, Vis& 61\%& 11\%& 12\%& 18\%& Int(F)& St(F)& PO\\ 
\citet{HLL+19}& Ad, CM, DM, ESI, Pre, Si& 61\%& 25\%& 31\%& 9\%& Int(F)& St(F)& Mo\\ 
\citet{KJM+18}& CM, Si, Sy& 56\%& 2\%& 0\%& 0\%& A(F), Int(F), C(F), Bus& FD, CD, St, WC(F)& Di\\ 
\citet{WZM18}& ESI, IS& 0\%& 10\%& 4\%& 18\%& A, Int, C(F), Inf(F), Fun& CD, St(F)& UA\\ 
\citet{CaMa19}& Ad, ESI, Pre, Si, Sy& 67\%& 19\%& 19\%& 9\%& Int, C, Inf(F), Fun(F)& FD, CD, St(F)& PO\\ 
\citet{FEE19}& Vis& 6\%& 0\%& 8\%& 0\%& A, Int, C, Inf, Fun(F)& CD(F), St& Mo\\ 
\citet{Park.2020}& CM, CA, DM, ESI, HI, Pre, RE, RT, Rob, Si, Sy, Vis& 100\%& 33\%& 31\%& 18\%& A, Int, C, Inf, Fun(F), Bus(F)& Pro(F), FD, WC& Pl\\ 
\citet{Pires.2020}& CM, Pre, RT, Vis& 56\%& 8\%& 0\%& 0\%& Int, C, Inf, Fun(F)& WC(F)& VC, Mo, Di\\ 
\citet{Jazdi.2021}& ESI, Sy& 17\%& 5\%& 4\%& 9\%& A, Int, C, Inf, Fun(F)& WC(F)& PO\\ 
\citet{Redelinghuys.2020}& CM, ESI, HI, Pre, RE, RT, Si, Sy, Vis& 100\%& 17\%& 12\%& 18\%& A, Int, C(F), Inf(F), Fun(F)& St(F)& PO\\ 
\citet{Santos.2022}& DM, ESI, RE, Si, Sy, Vis& 50\%& 10\%& 23\%& 9\%& C, Inf, Fun(F)& WC(F)& VC\\ 
\citet{Bao.2019}& CM, ESI, RE, RT, Si, Sy& 72\%& 6\%& 12\%& 9\%& C, Inf, Fun(F), Bus& St(F)\\ 
\citet{Martins.2020}& CM, Si& 39\%& 2\%& 8\%& 0\%& Int, C(F)& St(F)& UA\\ 
\citet{ABM+21}& Enc, ESI, Fl, HI, IS, Ne, RE, RT, Si, Vis& 50\%& 27\%& 31\%& 55\%& C(F), Inf, Fun& WC(F)& Di\\ 
\citet{CWB20}& CM, ESI, HI, RT, Si, Vis& 61\%& 11\%& 4\%& 18\%& Int, C(F)& St(F)& Di\\ 
\citet{OOB20}& CM, ESI, HI, RT& 28\%& 11\%& 4\%& 18\%& Int, C, Inf(F)& St(F)& UA\\ 
\citet{ZPP+20bb}& CM, ESI, RT, Si, Sy, Vis& 78\%& 6\%& 4\%& 9\%& C(F), Bus& Pro(F), St& PO\\ 
\citet{PJM+20}& CM, Cop, ESI, Fl, HI, RE, RT, Si, Vis& 61\%& 24\%& 27\%& 45\%& C(F), Inf& St(F)& PO\\ 
\citet{She.21}& CM, DM, ESI, Lea, Pre, RT, Si, Un, Vis& 83\%& 32\%& 23\%& 9\%& Int, C(F)& WC(F), Ent& Di\\ 
\citet{GGG+20}& CM, CA, DM, ESI, HI, IS, Lea, Pre, RE, RT, Vis& 56\%& 38\%& 23\%& 27\%& A, Int(F), C(F)& St(F), WC& PO\\ 
\citet{SZZ+19}& CM, DM, ESI, HI, Pre, Si, Sy& 78\%& 22\%& 15\%& 18\%& Inf(F)& WC(F), Ent(F)& Pl\\ 
\citet{XYX+22}& Ad, CM, ESI, IS, RE, RT, Si, Vis& 61\%& 19\%& 27\%& 18\%& A, Int, C(F)& CD(F)& PO\\ 
\citet{AsWe20}& Ad, CM, ESI, IS, RT, Si, Sy, Vis& 78\%& 19\%& 19\%& 18\%& A, Int, C(F)& WC(F), CW& Co\\ 
\citet{JeSc20}& CM, DM, ESI, HI, IS, Pre, RE, RT, Si, Vis& 83\%& 27\%& 23\%& 27\%& C, Inf(F), Fun& WC(F)& VC\\ 
\citet{MaHa21}& CM, Cor, DM, Enc, ESI, IS, Lea, RE, Si, Vis& 44\%& 30\%& 27\%& 36\%& C, Inf(F), Fun& St(F)& Mo\\ 
\citet{Schweizer.2021}& CA, ESI, IS, Vis& 6\%& 13\%& 4\%& 18\%& C(F)& St(F), WC(F)& UA\\ 
\citet{RMG21}& CM, ESI, IS, RE, RT, Si, Vis& 61\%& 11\%& 12\%& 18\%& C(F), Inf& WC(F)& Pl\\ 
\citet{MSS+22}& CM, Enc, ESI, RE, RT, Si, Vis& 61\%& 6\%& 15\%& 9\%& C(F), Inf& St(F)& Mo\\ 
\citet{ZZH+20}& CM, CA, DM, ESI, IS, Lea, Pre, RT, Si, Vis& 83\%& 33\%& 15\%& 18\%& C, Inf(F)& WC(F)& VC\\ 
\citet{ASY+21}& CM, ESI, IS, Lea, Pre, Si, Sy, Vis& 83\%& 25\%& 4\%& 18\%& C, Inf(F)& WC(F)& Mo\\ 
\citet{AndreMartins.2019}& ESI, Si, Vis& 33\%& 5\%& 4\%& 9\%& C(F)& WC(F)& Mo\\ 
\citet{RSS20}& Fl& 0\%& 6\%& 15\%& 0\%& Bus(F)& WC(F)& PO\\ 
\citet{CHW+20}& Ad, Fl, Si, Vis& 33\%& 14\%& 31\%& 0\%& Inf(F)& St(F)& Pl\\ 
\citet{APM+21}& CM, RT, Si& 56\%& 2\%& 0\%& 0\%& Bus(F)& WC(F)& Pl\\ 
\citet{GaSa19}& CM, Cor, DM, Si& 39\%& 13\%& 12\%& 18\%& C(F)& WC(F)& PO\\ 
\citet{STA19}& Cop, Cor& 0\%& 13\%& 0\%& 45\%& A(F)& FD(F), St(F)& Sc\\ 
\citet{BGJ+21}& Ad, CM, ESI, HI, Si, Vis& 44\%& 19\%& 19\%& 18\%& Int(F)& St(F)& PO\\ 
\citet{QBM+19}& CM, RT, Sy, Vis& 50\%& 2\%& 0\%& 0\%& Int, Inf(F)& St, WC(F)& PO\\ 
\citet{TvHa22}& ESI, Pre, RT, Si& 67\%& 11\%& 4\%& 9\%& Int(F)& St(F)& Mo\\ 
\citet{WWC+22}& CM, ESI, Pre, RT, Si, Vis& 83\%& 13\%& 4\%& 9\%& Int(F)& St(F)& Mo\\ 
\citet{LGG+19}& ESI, HI, Si, Vis& 33\%& 10\%& 4\%& 18\%& Int(F), C& St(F)& Mo\\ 
\citet{LuMa21}& CM, HI, Pre, Si, Sy& 78\%& 13\%& 0\%& 9\%& C, Fun(F)& WC(F)& Mo\\ 
\citet{YYS+21}& ESI, RT& 17\%& 5\%& 4\%& 9\%& C(F), Inf(F), Fun, Bus& WC(F), Ent(F)& PO\\ 
\citet{DAF+22}& CM, Pre, RT& 50\%& 8\%& 8\%& 0\%& Int(F), C(F), Inf(F), Fun(F)& FD(F), WC(F)& Pl\\ 
\citet{GiEs22}& CM, ESI, Lea, Pre, RE& 33\%& 21\%& 4\%& 9\%& C, Fun(F)& St(F)& Mo\\ 
\citet{CSP+19}& ESI, HI, Si& 28\%& 10\%& 4\%& 18\%& C, Inf, Fun(F)& WC(F)& Co\\ 
\citet{SSP+16}& CM, ESI, HI, Si, Vis& 44\%& 11\%& 4\%& 18\%& A, Int, C(F), Inf(F), Fun& FD(F), CD(F)& Co\\ 
\citet{Ward.2021}& CM, ESI, Pre, RE, RT, Si, Sy, Vis& 100\%& 13\%& 12\%& 9\%& A, Int(F), C, Inf, Fun(F)& FD, CD, St(F)& Pl\\ 
\citet{Sierla.2020b}& Si, Vis& 33\%& 0\%& 0\%& 0\%& Inf(F), Fun(F)& WC(F)& UA, PO\\ 
\citet{Schnicke.}& Ad, CM, Cor, Fl, HI, RE& 11\%& 27\%& 38\%& 27\%& Int, C, Inf(F), Fun(F), Bus& CD, WC(F)& VC\\ 
\citet{Bibow.2020}& CM, DM, ESI, HI, Lea, Pre, RT, Sy& 67\%& 30\%& 15\%& 18\%& Int, C, Inf(F), Fun(F)& St(F)& Pl\\ 
\citet{DarioOrive.2019}& HI, Pre, Si, Sy, Vis& 72\%& 11\%& 0\%& 9\%& Fun(F)& FD, St(F), WC& Pl\\ 
\citet{GuerraZubiaga.2021}& ESI, Pre, RE, RT, Si, Sy, Vis& 89\%& 11\%& 12\%& 9\%& A, Int(F), C(F), Inf(F), Fun(F)& CD, St(F), WC(F)& Pl\\ 
\citet{Yang.2022}& Pre, RT, Si, Sy& 83\%& 6\%& 0\%& 0\%& Inf(F), Fun(F), Bus& WC(F)& Co\\ 
\citet{Biesinger.2019}& Pre, RT, Sy& 56\%& 6\%& 0\%& 0\%& A, C(F), Inf(F), Fun(F), Bus& FD, CD(F), St, WC& Co\\ 
\citet{Negri.}& CM, Pre, RT, Si, Sy& 94\%& 8\%& 0\%& 0\%& C(F), Inf(F), Fun(F), Bus& FD(F), CD(F), St, WC& Mo\\ 
\citet{M.Redeker.2021}& Lea, Pre, RT& 39\%& 14\%& 0\%& 0\%& A, C(F), Inf(F), Fun& WC(F)& PO\\ 
\citet{Yang.2020}& CM, HI, Lea, Pre, RT, Si, Sy, Vis& 100\%& 21\%& 0\%& 9\%& Int, C, Inf(F), Fun(F), Bus& St(F), WC(F), Ent(F), CW(F)& Co\\ 
\citet{Kassen.2021}& Cop, Cor, ESI, HI, RT, Si, Vis& 50\%& 22\%& 4\%& 64\%& C, Inf(F), Fun(F), Bus& Pro, CD, St(F), WC(F)& Co\\ 
\citet{Eckhart.2018}& HI, Pre, Sy& 39\%& 11\%& 0\%& 9\%& A, Int(F), C(F), Inf(F)& Pro, FD(F), CD(F), St, WC& Mo\\ 
\citet{He.2019}& CM, ESI, Pre, Si& 61\%& 13\%& 4\%& 9\%& Fun(F)& FD(F), CD(F)& UA\\ 
\citet{BKW20}& Ad, CM, CA, ESI, IS, Lea, RT, Si, Sy& 72\%& 30\%& 19\%& 18\%& C, Inf(F)& CD(F), St(F), WC(F)& Mo\\ 
\citet{OTS+22}& CM, CA, Cop, DM, ESI, Fl, RE, RT, Sy, Vis& 50\%& 27\%& 38\%& 36\%& A, Int, C(F), Inf, Bus& Pro(F), St(F), WC(F), Ent(F), CW(F)& PO\\ 
\citet{PRR+20}& CM, Cop, ESI, RT, Si, Sy, Vis& 78\%& 13\%& 4\%& 36\%& Int(F), C& FD, CD, St(F)& Mo\\ 
\citet{RPU21}& CM, ESI, RT, Si& 56\%& 6\%& 4\%& 9\%& Int(F), C, Inf, Bus& FD, WC(F)& PO\\ 
\citet{Zip21}& CM, ESI, RT, Si, Sy, Vis& 78\%& 6\%& 4\%& 9\%& Inf(F)& CD(F)& UA\\ 
\citet{SPA+18}& CM, CA, Cor, ESI, Fl, HI, IS, Pre, Pro, RE, Si& 61\%& 46\%& 42\%& 45\%& Int(F), C, Inf(F)& FD(F), St(F)& Mo\\ 
\citet{LLY+20}& Ad, CM, CA, DM, ESI, Lea, RT, Si, Sy, Vis& 78\%& 30\%& 31\%& 9\%& Int(F), C, Inf, Fun& FD(F), CD, WC(F)& VC\\ 
\citet{SLR+21}& CM, CA, DM, ESI, Lea, Pro, RT, Si, Vis& 61\%& 30\%& 31\%& 9\%& A(F), Int, C, Inf& WC(F)& Mo\\ 
\citet{IPK+20}& CM& 11\%& 2\%& 0\%& 0\%& A(F)& FD(F)& Mo\\ 
\citet{KSO20}& CM, Fl, RT, Vis& 33\%& 8\%& 15\%& 0\%& Int(F), Fun& FD(F)& PO\\ 
\citet{KaRo21}& CM, CA, DM, Enc, ESI, Fl, Lea, RE, RT, Rob, Si, Vis& 61\%& 37\%& 58\%& 9\%& A(F), Int, C& Ent(F)& VC\\ 
\citet{HaDi21}& CM, CA, Fl, IS, Lea, Pre, RT, Rob, Vis& 56\%& 38\%& 31\%& 9\%& Int, C, Inf(F)& St, WC(F), Ent& Mo, Di\\ 
\citet{TiEr21}& Ad, CA, Fl, Lea, Rob, Si, Sy, Un& 44\%& 40\%& 54\%& 0\%& Int, C, Inf, Fun(F)& St(F)& Di\\ 
\citet{AJL+19}& Ad, CM, ESI, HI, IS, Lea, Pre, RE, Si, Sy& 78\%& 38\%& 19\%& 27\%& Int, C, Inf(F), Fun& CD, St(F)& UA\\ 
\citet{TYS22}& CM, CA, IS, Lea, Pre, Si& 61\%& 24\%& 0\%& 9\%& Int(F), C& St(F)& Mo\\ 
\citet{IBW22}& DM, HI, Lea, Pre, RT, Si& 67\%& 24\%& 12\%& 9\%& A, C(F)& FD(F)& UA\\ 
\citet{JKS+22}& CM, DM, Enc, ESI, HI, Lea, Mo, Pre, RT, Si& 78\%& 30\%& 31\%& 18\%& A(F), Int, C& Pro(F)& Mo\\ 
\citet{HDB+22}& CM, ESI, Lea, Pre, Sy& 50\%& 21\%& 12\%& 9\%& A(F)& Pro, FD(F), CD& Di\\ 
\citet{YZS+23}& CA, ESI, Lea, RE, RT, Si, Vis& 50\%& 16\%& 12\%& 9\%& Int, C(F)& St(F)& Mo\\ 
\citet{Czimmermann.2022}& CM, Cop, Cor, HI, IS, RE, RT, Sy, Vis& 50\%& 24\%& 8\%& 64\%& A, Int, C, Inf, Fun(F)& Pro, FD, CD, St(F)& Di\\ 
\citet{SZM+22}& Ad, CA, ESI, RE, Si& 28\%& 16\%& 19\%& 9\%& A(F), Int& FD(F), CD& Mo\\ 
\citet{SKH23}& CM, ESI, Lea, Pre, RT& 50\%& 21\%& 4\%& 9\%& Int(F), C, Inf& FD(F)& Mo\\ 
\citet{ZXL+23}& CM, ESI, HI, IS, Pre, RT, Si, Sy, Vis& 100\%& 22\%& 4\%& 27\%& A, Int, C, Inf(F)& WC(F)& PO\\ 
\citet{CaSa22}& CM, ESI, HI, RE, Si, Vis& 44\%& 11\%& 12\%& 18\%& A, Int(F)& St(F)& Mo\\ 
\citet{XSK+21}& ESI, Lea, Si& 28\%& 13\%& 4\%& 9\%& Inf(F)& WC(F)& Co\\ 

%% file: _paper-springer.bbl
\begin{thebibliography}{222}
\providecommand{\natexlab}[1]{#1}
\providecommand{\url}[1]{{#1}}
\providecommand{\urlprefix}{URL }
\providecommand{\doi}[1]{\url{https://doi.org/#1}}
\providecommand{\eprint}[2][]{\url{#2}}
 \bibcommenthead

\bibitem[{Adediran et~al(2019)Adediran, Al-Bazi, and {dos Santos}}]{AAd19}
Adediran TV, Al-Bazi A, {dos Santos} LE (2019) {Agent-based modelling and
  heuristic approach for solving complex OEM flow-shop productions under
  customer disruptions}. {Computers {\&} Industrial Engineering} 133:29--41.
  \doi{10.1016/j. cie.2019.04.054}

\bibitem[{Aheleroff et~al(2021)Aheleroff, Xu, Zhong, and Lu}]{AXZ+21}
Aheleroff S, Xu X, Zhong RY, et~al (2021) Digital twin as a service (dtaas) in
  industry 4.0: An architecture reference model. Advanced Engineering
  Informatics 47:101,225. \doi{10.1016/j.aei.2020.101225}

\bibitem[{Aheleroff et~al(2022)Aheleroff, Huang, Xu, and Zhong}]{AHX+22}
Aheleroff S, Huang H, Xu X, et~al (2022) Toward sustainability and resilience
  with industry 4.0 and industry 5.0. Frontiers in Manufacturing Technology 2.
  \doi{10.3389/fmtec.2022.951643}

\bibitem[{Antti et~al(2021)Antti, Paul, Maria, and Lasse}]{APM+21}
Antti K, Paul M, Maria BA, et~al (2021) {Digitalization in the Carbon Area as a
  Means to Improve Productivity}. In: Perander L (ed) {Light Metals 2021}. {The
  Minerals, Metals {\&} Materials Series}, {Springer International Publishing},
  Cham, p 931--939, \doi{10.1007/978-3-030-65396-5_123}

\bibitem[{Arm et~al(2021)Arm, Benesl, Marcon, Bradac, Schr{\"o}der, Belyaev,
  Werner, Braun, Kamensky, Zezulka, Diedrich, and Dohnal}]{ABM+21}
Arm J, Benesl T, Marcon P, et~al (2021) {Automated Design and Integration of
  Asset Administration Shells in Components of Industry 4.0}. {Sensors (Basel,
  Switzerland)} 21(6). \doi{10.3390/s21062004}

\bibitem[{{Ashtari Talkhestani} and Weyrich(2020)}]{AsWe20}
{Ashtari Talkhestani} B, Weyrich M (2020) {Digital Twin of manufacturing
  systems: a case study on increasing the efficiency of reconfiguration}. {at -
  Automatisierungstechnik} 68(6):435--444. \doi{10.1515/auto-2020-0003}

\bibitem[{{Ashtari Talkhestani} et~al(2018){Ashtari Talkhestani}, Jazdi,
  Schloegl, and Weyrich}]{Talkhestani.2018}
{Ashtari Talkhestani} B, Jazdi N, Schloegl W, et~al (2018) {Consistency check
  to synchronize the Digital Twin of manufacturing automation based on anchor
  points}. {Procedia CIRP} 72:159--164. \doi{10.1016/j. procir.2018.03.166}

\bibitem[{{Ashtari Talkhestani} et~al(2019){Ashtari Talkhestani}, Jung,
  Lindemann, Sahlab, Jazdi, Schloegl, and Weyrich}]{AJL+19}
{Ashtari Talkhestani} B, Jung T, Lindemann B, et~al (2019) {An architecture of
  an Intelligent Digital Twin in a Cyber-Physical Production System}. {at -
  Automatisierungstechnik} 67(9):762--782. \doi{10.1515/auto-2019-0039}

\bibitem[{Azangoo et~al(2021)Azangoo, Salmi, Yrjola, Bensky, Santillan,
  Papakonstantinou, Sierla, and Vyatkin}]{ASY+21}
Azangoo M, Salmi J, Yrjola I, et~al (2021) {Hybrid Digital Twin for process
  industry using Apros simulation environment}. In: {2021 26th IEEE
  International Conference on Emerging Technologies and Factory Automation
  (ETFA)}. IEEE, Vasteras, Sweden, pp 01--04,
  \doi{10.1109/ETFA45728.2021.9613416}

\bibitem[{Bakliwal et~al(2018)Bakliwal, Dhada, Palau, Parlikad, and
  Lad}]{BDP+18}
Bakliwal K, Dhada MH, Palau AS, et~al (2018) {A Multi Agent System architecture
  to implement Collaborative Learning for social industrial assets}.
  {IFAC-PapersOnLine} 51(11):1237--1242. \doi{10.1016/j.ifacol.2018.08.421}

\bibitem[{Bamunuarachchi et~al(2021)Bamunuarachchi, Georgakopoulos, Jayaraman,
  and Banerjee}]{BGJ+21}
Bamunuarachchi D, Georgakopoulos D, Jayaraman PP, et~al (2021) {A Framework for
  Enabling Cyber-Twins based Industry 4.0 Application Development}. In: {2021
  IEEE International Conference on Services Computing (SCC)}. IEEE, Chicago,
  IL, USA, pp 340--350, \doi{10.1109/SCC53864.2021.00047}

\bibitem[{Bao et~al(2019)Bao, Guo, Li, and Zhang}]{Bao.2019}
Bao J, Guo D, Li J, et~al (2019) {The modelling and operations for the digital
  twin in the context of manufacturing}. {Enterprise Information Systems}
  13(4):534--556. \doi{10.1080/17517575.2018.1526324}

\bibitem[{Beer et~al(2014)Beer, Fisk, and Rogers}]{BFR14}
Beer JM, Fisk AD, Rogers WA (2014) {Toward a framework for levels of robot
  autonomy in human-robot interaction}. {Journal of human-robot interaction}
  3(2):74--99. \doi{10.5898/JHRI.3.2.Beer}

\bibitem[{Bendjelloul et~al(2022)Bendjelloul, Gaham, Bouzouia, Moufid, and
  Mihoubi}]{BGB+22}
Bendjelloul A, Gaham M, Bouzouia B, et~al (2022) {Multi Agent Systems Based
  CPPS -- An Industry 4.0 Test Case}. In: Lejdel B, Clementini E, Alarabi L
  (eds) {Artificial Intelligence and Its Applications}, {Lecture Notes in
  Networks and Systems}, vol 413. {Springer International Publishing}, Cham, p
  187--196, \doi{10.1007/978-3-030-96311-8_18}

\bibitem[{Bibow et~al(2020)Bibow, Dalibor, Hopmann, Mainz, Rumpe, Schmalzing,
  Schmitz, and Wortmann}]{Bibow.2020}
Bibow P, Dalibor M, Hopmann C, et~al (2020) Model-driven development of a
  digital twin for injection molding. In: Dustdar S, Yu E, Salinesi C, et~al
  (eds) Advanced Information Systems Engineering. Springer International
  Publishing, Cham, pp 85--100

\bibitem[{Biesinger et~al(2019)Biesinger, Meike, Kra{\ss}, and
  Weyrich}]{Biesinger.2019}
Biesinger F, Meike D, Kra{\ss} B, et~al (2019) {A digital twin for production
  planning based on cyber-physical systems: A Case Study for a Cyber-Physical
  System-Based Creation of a Digital Twin}. {Procedia CIRP} 79:355--360.
  \doi{10.1016/j. procir.2019.02.087}

\bibitem[{Birk et~al(2022)Birk, Hostettler, Razi, Atta, and Tammia}]{Birk.2022}
Birk W, Hostettler R, Razi M, et~al (2022) {Automatic generation and updating
  of process industrial digital twins for estimation and control - A review}.
  {Frontiers in Control Engineering} 3. \doi{10.3389/fcteg.2022.954858}

\bibitem[{B{\"o}hm et~al(2021)B{\"o}hm, Broy, Klein, Pohl, Rumpe, and
  Schr{\"o}ck}]{BBK+21}
B{\"o}hm W, Broy M, Klein C, et~al (eds)  (2021) {Model-based engineering of
  collaborative embedded systems: Extensions of the SPES methodology}.
  Springer, Cham, \doi{10.1007/978-3-030-62136-0}

\bibitem[{Boussaada et~al(2016)Boussaada, Curea, Camblong, {Bellaaj Mrabet},
  and Hacala}]{BCC+16}
Boussaada Z, Curea O, Camblong H, et~al (2016) {Multi-agent systems for the
  dependability and safety of microgrids}. {International Journal on
  Interactive Design and Manufacturing (IJIDeM)} 10(1):1--13.
  \doi{10.1007/s12008-014-0257-9}

\bibitem[{Braubach et~al(2005)Braubach, Pokahr, Moldt, and Lamersdorf}]{BPM+05}
Braubach L, Pokahr A, Moldt D, et~al (2005) {Goal Representation for BDI Agent
  Systems}. In: Hutchison D, Kanade T, Kittler J, et~al (eds) {Programming
  Multi-Agent Systems}, {Lecture Notes in Computer Science}, vol 3346.
  {Springer Berlin Heidelberg}, Berlin, Heidelberg, p 44--65,
  \doi{10.1007/978-3-540-32260-3_3}

\bibitem[{Brosinsky et~al(2020)Brosinsky, Krebs, and Westermann}]{BKW20}
Brosinsky C, Krebs R, Westermann D (2020) {Embedded Digital Twins in future
  energy management systems: paving the way for automated grid control}. {at -
  Automatisierungstechnik} 68(9):750--764. \doi{10.1515/auto-2020-0086}

\bibitem[{Bussmann et~al(2004)Bussmann, Jennings, and Wooldridge}]{BJW04}
Bussmann S, Jennings NR, Wooldridge M (2004) {Multiagent Systems for
  Manufacturing Control: A Design Methodology}. {Springer Series on Agent
  Technology}, Springer, Berlin and Heidelberg, \doi{10.1007/978-3-662-08872-2}

\bibitem[{Caesar et~al(2020)Caesar, Hanel, Wenkler, Corinth, Ihlenfeldt, and
  Fay}]{CHW+20}
Caesar B, Hanel A, Wenkler E, et~al (2020) {Information Model of a Digital
  Process Twin for Machining Processes}. In: {2020 25th IEEE International
  Conference on Emerging Technologies and Factory Automation (ETFA)}. IEEE,
  Vienna, Austria, pp 1765--1772, \doi{10.1109/ETFA46521.2020.9212085}

\bibitem[{Cagnin et~al(2018)Cagnin, Guilherme, Queiroz, Paulo, and
  Neto}]{CGQ+18}
Cagnin RL, Guilherme IR, Queiroz J, et~al (2018) {A Multi-agent System Approach
  for Management of Industrial IoT Devices in Manufacturing Processes}. In:
  {2018 IEEE 16th International Conference on Industrial Informatics (INDIN)}.
  IEEE, Porto, pp 31--36, \doi{10.1109/INDIN.2018.8471926}

\bibitem[{Cai et~al(2020)Cai, Wang, and Burnett}]{CWB20}
Cai Y, Wang Y, Burnett M (2020) {Using augmented reality to build digital twin
  for reconfigurable additive manufacturing system}. {Journal of Manufacturing
  Systems} 56:598--604. \doi{10.1016/j.jmsy.2020.04.005}

\bibitem[{Caiza and Sanz(2022)}]{CaSa22}
Caiza G, Sanz R (2022) {Digital Twin for Monitoring an Industrial Process Using
  Augmented Reality}. In: {2022 17th Iberian Conference on Information Systems
  and Technologies (CISTI)}. IEEE, Madrid, Spain, pp 1--5,
  \doi{10.23919/CISTI54924.2022.9820356}

\bibitem[{Cardin(2019)}]{Cardin.2019}
Cardin O (2019) {Classification of cyber-physical production systems
  applications: Proposition of an analysis framework}. {Computers in Industry}
  104:11--21. \doi{10.1016/j. compind.2018.10.002}

\bibitem[{Casquero et~al(2019)Casquero, Armentia, Sarachaga, Perez, Orive, and
  Marcos}]{CAS+19}
Casquero O, Armentia A, Sarachaga I, et~al (2019) {Distributed scheduling in
  Kubernetes based on MAS for Fog-in-the-loop applications}. In: {2019 24th
  IEEE International Conference on Emerging Technologies and Factory Automation
  (ETFA)}. IEEE, Zaragoza, Spain,
  \urlprefix\url{https://ieeexplore.ieee.org/servlet/opac?punumber=8851311}

\bibitem[{Cattaneo and Macchi(2019)}]{CaMa19}
Cattaneo L, Macchi M (2019) {A Digital Twin Proof of Concept to Support Machine
  Prognostics with Low Availability of Run-To-Failure Data}.
  {IFAC-PapersOnLine} 52(10):37--42. \doi{10.1016/j. ifacol.2019.10.016}

\bibitem[{Chaplin et~al(2015)Chaplin, Bakker, de~Silva, Sanderson, Kelly,
  Logan, and Ratchev}]{CBS+15}
Chaplin JC, Bakker OJ, de~Silva L, et~al (2015) {Evolvable Assembly Systems: A
  Distributed Architecture for Intelligent Manufacturing}. {IFAC-PapersOnLine}
  48(3):2065--2070. \doi{10.1016/j.ifacol.2015.06.393}

\bibitem[{Chirico et~al(2019)Chirico, Spellini, Panato, Lora, and
  Fummi}]{CSP+19}
Chirico R, Spellini S, Panato M, et~al (2019) {A Contract-based Methodology for
  Production Lines Validation}. In: {2019 IEEE 17th International Conference on
  Industrial Informatics (INDIN)}. IEEE, Helsinki, Finland, pp 695--698,
  \doi{10.1109/INDIN41052.2019.8972100}

\bibitem[{Cicirelli et~al(2018)Cicirelli, Nigro, and Sciammarella}]{CNS18}
Cicirelli F, Nigro L, Sciammarella PF (2018) {Model continuity in
  cyber-physical systems: A control-centered methodology based on agents}.
  {Simulation Modelling Practice and Theory} 83:93--107.
  \doi{10.1016/j.simpat.2017.12.008}

\bibitem[{Cimino and Marcelloni(2011)}]{CiMa11}
Cimino MG, Marcelloni F (2011) {Autonomic tracing of production processes with
  mobile and agent-based computing}. {Information Sciences} 181(5):935--953.
  \doi{10.1016/j.ins.2010.11.015}

\bibitem[{Cohen et~al(2023)Cohen, Aperstein, and Reis}]{CAR23}
Cohen Y, Aperstein Y, Reis J (2023) {A Framework for Integrating Artificial
  Intelligence in Digital Twins of Manufacturing Systems}. In: {Proceedings of
  the 22nd IFAC World Congress}, Yokohama, Japan

\bibitem[{Colombo(1998)}]{Colombo.2020}
Colombo AW (1998) {Development and Implementation of Hierarchical Control
  Structures of Flexible Production Systems Using High Level Petri Nets}. PhD
  thesis, Meisenbach, \doi{10.25593/3-87525-109-1}

\bibitem[{{Cruz Salazar} and Vogel-Heuser(2022)}]{CrVo22}
{Cruz Salazar} LA, Vogel-Heuser B (2022) {A CPPS-architecture and workflow for
  bringing agent-based technologies as a form of artificial intelligence into
  practice}. {at - Automatisierungstechnik} 70(6):580--598.
  \doi{10.1515/auto-2022-0008}

\bibitem[{{Cruz Salazar} et~al(2018){Cruz Salazar}, Mayer, Sch{\"u}tz, and
  Vogel-Heuser}]{CMS+18}
{Cruz Salazar} LA, Mayer F, Sch{\"u}tz D, et~al (2018) {Platform Independent
  Multi-Agent System for Robust Networks of Production Systems}.
  {IFAC-PapersOnLine} 51(11):1261--1268. \doi{10.1016/j.ifacol.2018.08.359}

\bibitem[{Czimmermann et~al(2022)Czimmermann, Chiurazzi, Milazzo, Roccella,
  Barbieri, Dario, Oddo, and Ciuti}]{Czimmermann.2022}
Czimmermann T, Chiurazzi M, Milazzo M, et~al (2022) {An Autonomous Robotic
  Platform for Manipulation and Inspection of Metallic Surfaces in Industry
  4.0}. {IEEE Transactions on Automation Science and Engineering}
  19(3):1691--1706. \doi{10.1109/TASE.2021.3122820}

\bibitem[{DallaOra et~al(2022)DallaOra, Alamin, Fraccaroli, Poncino, Quaglia,
  and Vinco}]{DAF+22}
DallaOra N, Alamin K, Fraccaroli E, et~al (2022) {Digital Transformation of a
  Production Line: Network Design, Online Data Collection and Energy
  Monitoring}. {IEEE Transactions on Emerging Topics in Computing}
  10(1):46--59. \doi{10.1109/TETC.2021.3132432}

\bibitem[{{Dario Orive} et~al(2019){Dario Orive}, {Nagore Iriondo}, {Arantza
  Burgos}, {Isabel Sarachaga}, {Maria Luz Alvarez}, and {Marga
  Marcos}}]{DarioOrive.2019}
{Dario Orive}, {Nagore Iriondo}, {Arantza Burgos}, et~al (2019) {Fault
  injection in Digital Twin as a means to test the response to process faults
  at virtual commissioning: Paraninfo Building, University of Zaragoza,
  Zaragoza, Spain, 10-13 September, 2019 : proceedings}. {2019 24th IEEE
  International Conference on Emerging Technologies and Factory Automation
  (ETFA)}

\bibitem[{Deeks(2022)}]{Dee22}
Deeks C (2022) {Adapting Multi-agent Swarm Robotics to Achieve Synchronised
  Behaviour from Production Line Automata}. In: Holderbaum W, Selig JM (eds)
  {2nd IMA Conference on Mathematics of Robotics}, {Springer Proceedings in
  Advanced Robotics}, vol~21. {Springer International Publishing}, Cham, p
  13--24, \doi{10.1007/978-3-030-91352-6_2}

\bibitem[{Dehghanpour and Nehrir(2018)}]{DeNe18}
Dehghanpour K, Nehrir H (2018) {A Market-Based Resilient Power Management
  Technique for Distribution Systems with Multiple Microgrids Using a
  Multi-Agent System Approach}. {Electric Power Components and Systems}
  46(16-17):1744--1755. \doi{10.1080/15325008.2018.1527869}

\bibitem[{Dey(2001)}]{Dey01}
Dey AK (2001) {Understanding and Using Context}. {Personal and Ubiquitous
  Computing} 5(1):4--7. \doi{10.1007/s007790170019}

\bibitem[{Dimolitsas et~al(2021)Dimolitsas, Avgeris, Spatharakis, Dechouniotis,
  and Papavassiliou}]{DAS+21}
Dimolitsas I, Avgeris M, Spatharakis D, et~al (2021) {Enabling Industrial
  Network Slicing Orchestration: A Collaborative Edge Robotics Use Case}. In:
  {2021 IEEE International Mediterranean Conference on Communications and
  Networking (MeditCom)}. IEEE, Athens, Greece, pp 215--220,
  \doi{10.1109/MeditCom49071.2021.9647615}

\bibitem[{DIN(2016)}]{91345}
DIN (2016) {DIN SPEC 91345:2016-04, Referenzarchitekturmodell Industrie 4.0
  (RAMI4.0): Reference Architecture Model Industrie 4.0 (RAMI4.0)}.
  \doi{10.31030/2436156}

\bibitem[{Eckhart and Ekelhart(2018)}]{Eckhart.2018}
Eckhart M, Ekelhart A (2018) {A Specification-based State Replication Approach
  for Digital Twins}. In: Lie D (ed) {Proceedings of the 2018 Workshop on
  Cyber-Physical Systems Security and PrivaCy}. ACM, Toronto Canada, {ACM
  Conferences}, pp 36--47, \doi{10.1145/3264888.3264892}

\bibitem[{{Elsevier B.V.}(2022)}]{Els22}
{Elsevier B.V.} (2022) {Scopus}.
  \urlprefix\url{https://www.scopus.com/search/form.uri?display=advanced}

\bibitem[{Etukudor et~al(2020)Etukudor, Couraud, Robu, Fr{\"u}h, Flynn, and
  Okereke}]{ECR+20}
Etukudor C, Couraud B, Robu V, et~al (2020) {Automated Negotiation for
  Peer-to-Peer Electricity Trading in Local Energy Markets}. {Energies}
  13(4):920. \doi{10.3390/en13040920}

\bibitem[{Fay et~al(2019)Fay, Gehlhoff, Seitz, and Vogel-Heuser}]{FGS+19}
Fay A, Gehlhoff F, Seitz M, et~al (2019) {Agenten zur Realisierung von
  Industrie 4.0: VDI Statusreport}.
  \urlprefix\url{https://www.vdi.de/ueber-uns/presse/publikationen/details/agenten-zur-realisierung-von-industrie-40}

\bibitem[{{Federal Ministry for Economic Affairs and
  Energy}(2018)}]{FederalMinistryforEconomicAffairsandEnergy.2018}
{Federal Ministry for Economic Affairs and Energy} (2018) {Alignment Report for
  Reference Architectural Model for Industrie 4.0 / Intelligent Manufacturing
  Architecture: Sino-German Industrie 4.0 / Intelligent Manufacturing
  Standardisation Sub-Working Group}.
  \urlprefix\url{https://www.plattform-i40.de/IP/Redaktion/EN/Downloads/Publikation/hm-2018-manufacturing.pdf?__blob=publicationFile&v=1}

\bibitem[{{Federal Ministry for Economic Affairs and Energy}(2019)}]{Fed19}
{Federal Ministry for Economic Affairs and Energy} (2019) {Technology Scenario
  ‘Artificial Intelligence in Industrie 4.0’}.
  \urlprefix\url{https://www.plattform-i40.de/IP/Redaktion/EN/Downloads/Publikation/AI-in-Industrie4.0.pdf?__blob=publicationFile&v=1}

\bibitem[{Fernandez et~al(2019)Fernandez, Eguia, and Echeverria}]{FEE19}
Fernandez IA, Eguia MA, Echeverria LE (2019) {Virtual commissioning of a
  robotic cell: an educational case study}. In: {2019 24th IEEE International
  Conference on Emerging Technologies and Factory Automation (ETFA)}. IEEE,
  Zaragoza, Spain, pp 820--825, \doi{10.1109/ETFA.2019.8869373}

\bibitem[{Frank et~al(2011)Frank, Papenfort, and Sch{\"u}tz}]{FPS11}
Frank U, Papenfort J, Sch{\"u}tz D (2011) {Real-time capable software agents on
  IEC 61131 systems -- Developing a tool supported method}. {IFAC Proceedings
  Volumes} 44(1):9164--9169. \doi{10.3182/20110828-6-IT-1002.01390}

\bibitem[{Froese and Kleinblotekamp(2019)}]{FrKl19}
Froese A, Kleinblotekamp J (2019) {SEPDiT - Handbuch Lehrmaterial}. Froese,
  Andreas

\bibitem[{Gangoiti et~al(2021)Gangoiti, L{\'o}pez, Armentia, Est{\'e}vez, and
  Marcos}]{GLA+21}
Gangoiti U, L{\'o}pez A, Armentia A, et~al (2021) {Model-Driven Design and
  Development of Flexible Automated Production Control Configurations for
  Industry 4.0}. {Applied Sciences} 11(5):2319. \doi{10.3390/app11052319}

\bibitem[{Garrido and S{\'a}ez(2019)}]{GaSa19}
Garrido J, S{\'a}ez J (2019) {Integration of automatic generated simulation
  models, machine control projects and management tools to support whole life
  cycle of industrial digital twins}. {IFAC-PapersOnLine} 52(13):1814--1819.
  \doi{10.1016/j. ifacol.2019.11.465}

\bibitem[{Geisberger et~al(2011)Geisberger, Cengarle, Keil, Niehaus, Thiel, and
  Th{\"o}nni{\ss}en-Fries}]{GCK+11}
Geisberger E, Cengarle MV, Keil P, et~al (2011) {Cyber-physical systems:
  Driving force for innovation in mobility, health, energy and production:
  acatech POSITION PAPER}.
  \urlprefix\url{https://en.acatech.de/publication/cyber-physical-systems-driving-force-for-innovation-in-mobility-health-energy-and-production/download-pdf?lang=en_CPS_Englisch_WEB-1.pdf}

\bibitem[{de~Giacomo et~al(2023)de~Giacomo, Favorito, Leotta, Mecella, and
  Silo}]{Giacomo.2023}
de~Giacomo G, Favorito M, Leotta F, et~al (2023) {Digital twin composition in
  smart manufacturing via Markov decision processes}. {Computers in Industry}
  149:103,916. \doi{10.1016/j.compind.2023.103916}

\bibitem[{Giannetti and Essien(2022)}]{GiEs22}
Giannetti C, Essien A (2022) {Towards scalable and reusable predictive models
  for cyber twins in manufacturing systems}. {Journal of Intelligent
  Manufacturing} 33(2):441--455. \doi{10.1007/s10845-021-01804-0}

\bibitem[{Gichane et~al(2020)Gichane, Byiringiro, Chesang, Nyaga, Langat,
  Smajic, and Kiiru}]{GBC+20}
Gichane MM, Byiringiro JB, Chesang AK, et~al (2020) {Digital Triplet Approach
  for Real-Time Monitoring and Control of an Elevator Security System}.
  {Designs} 4(2):9. \doi{10.3390/designs4020009}

\bibitem[{Girletti et~al(2020)Girletti, Groshev, Guimaraes, Bernardos, and
  de~{La Oliva}}]{GGG+20}
Girletti L, Groshev M, Guimaraes C, et~al (2020) {An Intelligent Edge-based
  Digital Twin for Robotics}. In: {2020 IEEE Globecom workshops (GC Wkshps)}.
  IEEE, Taipei, Taiwan, pp 1--6, \doi{10.1109/GCWkshps50303.2020.9367549}

\bibitem[{Gonz{\'a}lez-Briones et~al(2018)Gonz{\'a}lez-Briones, {La Prieta},
  Mohamad, Omatu, and Corchado}]{GLM+18}
Gonz{\'a}lez-Briones A, {La Prieta} FD, Mohamad M, et~al (2018) {Multi-Agent
  Systems Applications in Energy Optimization Problems: A State-of-the-Art
  Review}. {Energies} 11(8):1928. \doi{10.3390/en11081928},
  \urlprefix\url{https://www.mdpi.com/1996-1073/11/8/1928}

\bibitem[{Gorodetskii(2012)}]{Gor12}
Gorodetskii VI (2012) {Self-organization and multiagent systems: I. Models of
  multiagent self-organization}. {Journal of Computer and Systems Sciences
  International} 51(2):256--281. \doi{10.1134/S106423071201008X}

\bibitem[{Gr{\"a}{\ss}ler and P{\"o}hler(2017)}]{GrPo17}
Gr{\"a}{\ss}ler I, P{\"o}hler A (2017) {Implementation of an Adapted Holonic
  Production Architecture}. {Procedia CIRP} 63:138--143.
  \doi{10.1016/j.procir.2017.03.176}

\bibitem[{Grieves(2014)}]{Gri14}
Grieves M (2014) {Digital Twin: Manufacturing Excellence through Virtual
  Factory Replication}

\bibitem[{Groover(2016)}]{Gro16}
Groover MP (2016) {Automation, production systems, and computer-integrated
  manufacturing}, fourth edition, global edition edn. Pearson, Boston and
  Columbus and Indianapolis

\bibitem[{Guerra-Zubiaga et~al(2021)Guerra-Zubiaga, Kuts, Mahmood, Bondar,
  Nasajpour-Esfahani, and Otto}]{GuerraZubiaga.2021}
Guerra-Zubiaga D, Kuts V, Mahmood K, et~al (2021) {An approach to develop a
  digital twin for industry 4.0 systems: manufacturing automation case
  studies}. {International Journal of Computer Integrated Manufacturing}
  34(9):933--949. \doi{10.1080/0951192X.2021.1946857}

\bibitem[{Haben et~al(2021)Haben, Vogel-Heuser, Najjari, Seitz, Trunzer, and
  Salazar}]{HVN+21}
Haben F, Vogel-Heuser B, Najjari H, et~al (2021) {Low-entry Barrier Multi-Agent
  System for Small- and Middle-sized Enterprises in the Sector of Automated
  Production Systems}. In: {2021 IEEE International Conference on Industrial
  Engineering and Engineering Management (IEEM)}. IEEE, Singapore, Singapore,
  pp 1351--1357, \doi{10.1109/IEEM50564.2021.9672973}

\bibitem[{Haddod and Dingli(2021)}]{HaDi21}
Haddod F, Dingli A (2021) {Intelligent Digital Twin System in the
  Semiconductors Manufacturing}. In: Dingli A, Haddod F, Kl{\"u}ver C (eds)
  {Artificial Intelligence in Industry 4.0}, {Studies in Computational
  Intelligence}, vol 928. {Springer International Publishing}, Cham, p 99--113,
  \doi{10.1007/978-3-030-61045-6_8}

\bibitem[{Han(2012)}]{Han12}
Han Z (2012) {Game theory in wireless and communication networks: Theory,
  models, and applications}. {Cambridge University Press}, Cambridge, UK and
  New York, \doi{10.1017/CBO9780511895043}

\bibitem[{He et~al(2019)He, Chen, Dong, Sun, and Shen}]{He.2019}
He R, Chen G, Dong C, et~al (2019) {Data-driven digital twin technology for
  optimized control in process systems}. {ISA transactions} 95:221--234.
  \doi{10.1016/j.isatra.2019.05.011}

\bibitem[{Hichri et~al(2022)Hichri, Driate, Borghesi, and Giovannini}]{HDB+22}
Hichri B, Driate A, Borghesi A, et~al (2022) {Predictive Maintenance Based on
  Machine Learning Model}. In: Maglogiannis I, Iliadis L, Macintyre J, et~al
  (eds) {Artificial Intelligence Applications and Innovations}, {IFIP Advances
  in Information and Communication Technology}, vol 647. {Springer
  International Publishing}, Cham, p 250--261,
  \doi{10.1007/978-3-031-08337-2_21}

\bibitem[{Hoebert et~al(2019)Hoebert, Lepuschitz, List, and Merdan}]{HLL+19}
Hoebert T, Lepuschitz W, List E, et~al (2019) {Cloud-Based Digital Twin for
  Industrial Robotics}. In: Ma{\v{r}}{\'i}k V, Kadera P, Rzevski G, et~al (eds)
  {Industrial Applications of Holonic and Multi-Agent Systems}, {Springer
  eBooks Computer Science}, vol 11710. Springer, Cham, p 105--116,
  \doi{10.1007/978-3-030-27878-6_9}

\bibitem[{Hoffmann et~al(2016)Hoffmann, Thomas, Schutz, Vogel-Heuser, Meisen,
  and Jeschke}]{HTS+16}
Hoffmann M, Thomas P, Schutz D, et~al (2016) {Semantic integration of
  multi-agent systems using an OPC UA information modeling approach}. In: {2016
  IEEE 14th International Conference on Industrial Informatics (INDIN)}. IEEE,
  Poitiers, France, pp 744--747, \doi{10.1109/INDIN.2016.7819258}

\bibitem[{Huang et~al(2007)Huang, Pavek, Novak, Albus, and Messina}]{HPN+07}
Huang HM, Pavek K, Novak B, et~al (2007) {Autonomy Levels for Unmanned Systems
  (ALFUS) Framework}. In: {Proceedings of the AUVSI{\grq}s Unmanned Systems
  North America}. Association for Computing Machinery, Baltimore, Maryland,
  \doi{10.1145/1660877.1660883}

\bibitem[{Ioshchikhes et~al(2022)Ioshchikhes, Borst, and Weigold}]{IBW22}
Ioshchikhes B, Borst F, Weigold M (2022) {Assessing Energy Efficiency Measures
  for Hydraulic Systems using a Digital Twin}. {Procedia CIRP} 107:1232--1237.
  \doi{10.1016/j.procir.2022.05.137}

\bibitem[{Iñigo et~al(2020)Iñigo, Porto, Kremer, Perez, Larrinaga, and
  Cuenca}]{IPK+20}
Iñigo MA, Porto A, Kremer B, et~al (2020) Towards an asset administration
  shell scenario: a use case for interoperability and standardization in
  industry 4.0. In: NOMS 2020 - 2020 IEEE/IFIP Network Operations and
  Management Symposium, pp 1--6, \doi{10.1109/NOMS47738.2020.9110410}

\bibitem[{Jarvis et~al(2018)Jarvis, Jarvis, Kalachev, Zhabelova, and
  Vyatkin}]{JJK+18}
Jarvis D, Jarvis J, Kalachev A, et~al (2018) {PROSA/G: An architecture for
  agent-based manufacturing execution}. In: {2018 IEEE 23rd International
  Conference on Emerging Technologies and Factory Automation (ETFA)}. IEEE,
  Turin, pp 155--160, \doi{10.1109/ETFA.2018.8502598}

\bibitem[{Jazdi et~al(2021)Jazdi, {Ashtari Talkhestani}, Maschler, and
  Weyrich}]{Jazdi.2021}
Jazdi N, {Ashtari Talkhestani} B, Maschler B, et~al (2021) {Realization of
  AI-enhanced industrial automation systems using intelligent Digital Twins}.
  {Procedia CIRP} 97:396--400. \doi{10.1016/j. procir.2020.05.257}

\bibitem[{Jennings and Bussmann(2003)}]{JeBu03}
Jennings N, Bussmann S (2003) {Agent-based control systems: Why are they suited
  to engineering complex systems?} {IEEE Control Systems} 23(3):61--73.
  \doi{10.1109/MCS.2003.1200249}

\bibitem[{Jennings et~al(1998)Jennings, Sycara, and Wooldridge}]{JSW98}
Jennings NR, Sycara K, Wooldridge M (1998) {A Roadmap of Agent Research and
  Development. Autonomous Agents and Multi-Agent Systems}. {Autonomous Agents
  and Multi-Agent Systems} 1(1):7--38. \doi{10.1023/A:1010090405266}

\bibitem[{Jeon and Schuesslbauer(2020)}]{JeSc20}
Jeon SM, Schuesslbauer S (2020) {Digital Twin Application for Production
  Optimization}. In: {2020 IEEE International Conference on Industrial
  Engineering and Engineering Management (IEEM)}. IEEE, Singapore, Singapore,
  pp 542--545, \doi{10.1109/IEEM45057.2020.9309874}

\bibitem[{Jones et~al(2020)Jones, Snider, Nassehi, Yon, and Hicks}]{JSN+20}
Jones D, Snider C, Nassehi A, et~al (2020) {Characterising the Digital Twin: A
  systematic literature review}. {CIRP Journal of Manufacturing Science and
  Technology} 29:36--52. \doi{10.1016/j.cirpj.2020.02.002}

\bibitem[{Juhlin et~al(2022)Juhlin, Karaagac, Schlake, Gruner, and
  Ruckert}]{JKS+22}
Juhlin P, Karaagac A, Schlake JC, et~al (2022) {Cloud-enabled Drive-Motor-Load
  Simulation Platform using Asset Administration Shell and Functional Mockup
  Units}. In: {2022 IEEE 27th International Conference on Emerging Technologies
  and Factory Automation (ETFA)}. IEEE, Stuttgart, Germany, pp 1--8,
  \doi{10.1109/ETFA52439.2022.9921678}

\bibitem[{Julian and Botti(2004)}]{JuBo04}
Julian V, Botti V (2004) {Developing real-time multi-agent systems}.
  {Integrated Computer-Aided Engineering} 11(2):135--149.
  \doi{10.3233/ICA-2004-11204}

\bibitem[{Kaigom and Rosmann(2021)}]{KaRo21}
Kaigom EG, Rosmann J (2021) {Value-Driven Robotic Digital Twins in
  Cyber--Physical Applications}. {IEEE Transactions on Industrial Informatics}
  17(5):3609--3619. \doi{10.1109/TII.2020.3011062}

\bibitem[{Kang et~al(2022)Kang, {Qing Tan}, and Zhong}]{KQZ22}
Kang K, {Qing Tan} B, Zhong RY (2022) {Multi-attribute negotiation mechanism
  for manufacturing service allocation in smart manufacturing}. {Advanced
  Engineering Informatics} 51:101,523. \doi{10.1016/j. aei.2021.101523}

\bibitem[{Karanjkar et~al(2018)Karanjkar, Joglekar, Mohanty, Prabhu, Raghunath,
  and Sundaresan}]{KJM+18}
Karanjkar N, Joglekar A, Mohanty S, et~al (2018) {Digital Twin for Energy
  Optimization in an SMT-PCB Assembly Line}. In: {2018 IEEE International
  Conference on Internet of Things and Intelligence System (IOTAIS)}. IEEE,
  Bali, pp 85--89, \doi{10.1109/IOTAIS.2018.8600830},
  \urlprefix\url{http://ieeexplore.ieee.org/servlet/opac?punumber=8589237}

\bibitem[{Karnouskos et~al(2019)Karnouskos, Ribeiro, Leitao, Luder, and
  Vogel-Heuser}]{KRL+19}
Karnouskos S, Ribeiro L, Leitao P, et~al (2019) {Key Directions for Industrial
  Agent Based Cyber-Physical Production Systems}. In: {2019 IEEE International
  Conference on Industrial Cyber Physical Systems (ICPS)}. IEEE, Taipei,
  Taiwan, pp 17--22, \doi{10.1109/ICPHYS.2019.8780360}

\bibitem[{Karnouskos et~al(2020)Karnouskos, Leitao, Ribeiro, and
  Colombo}]{KLR+20}
Karnouskos S, Leitao P, Ribeiro L, et~al (2020) Industrial agents as a key
  enabler for realizing industrial cyber-physical systems: Multiagent systems
  entering industry 4.0. IEEE Industrial Electronics Magazine 14(3):18--32.
  \doi{10.1109/MIE.2019.2962225}

\bibitem[{Kasprzak et~al(2014)Kasprzak, Szynkiewicz, Zlatanov, and
  Zieli{\'n}ska}]{KSZ+14}
Kasprzak W, Szynkiewicz W, Zlatanov D, et~al (2014) {A hierarchical CSP search
  for path planning of cooperating self-reconfigurable mobile fixtures}.
  {Engineering Applications of Artificial Intelligence} 34:85--98.
  \doi{10.1016/j.engappai.2014.05.013}

\bibitem[{Kassen et~al(2021)Kassen, Tammen, Zarte, and Pechmann}]{Kassen.2021}
Kassen S, Tammen H, Zarte M, et~al (2021) {Concept and Case Study for a Generic
  Simulation as a Digital Shadow to Be Used for Production Optimisation}.
  {Processes} 9(8):1362. \doi{10.3390/pr9081362},
  \urlprefix\url{https://www.mdpi.com/2227-9717/9/8/1362}

\bibitem[{Kattepur et~al(2018)Kattepur, Dey, and P.}]{KDP18}
Kattepur A, Dey S, P. B (2018) {Knowledge Based Hierarchical Decomposition of
  Industry 4.0 Robotic Automation Tasks}. In: {IECON 2018 - 44th Annual
  Conference of the IEEE Industrial Electronics Society}. IEEE, Washington, DC,
  pp 3665--3672, \doi{10.1109/IECON.2018.8592800}

\bibitem[{Khalgui et~al(2011{\natexlab{a}})Khalgui, Mosbahi, Li, and
  Hanisch}]{KML+11}
Khalgui M, Mosbahi O, Li Z, et~al (2011{\natexlab{a}}) {Reconfigurable
  Multiagent Embedded Control Systems: From Modeling to Implementation}. {IEEE
  Transactions on Computers} 60(4):538--551. \doi{10.1109/TC.2010.96}

\bibitem[{Khalgui et~al(2011{\natexlab{b}})Khalgui, Mosbahi, Li, and
  Hanisch}]{KML+11b}
Khalgui M, Mosbahi O, Li Z, et~al (2011{\natexlab{b}}) {Reconfiguration of
  Distributed Embedded-Control Systems}. {IEEE/ASME Transactions on
  Mechatronics} 16(4):684--694. \doi{10.1109/TMECH.2010.2050697}

\bibitem[{Khan et~al(2019)Khan, Khan, {Ul Haq}, and Shah}]{KKU+19}
Khan ZA, Khan MT, {Ul Haq} I, et~al (2019) {Agent-based fault tolerant
  framework for manufacturing process automation}. {International Journal of
  Computer Integrated Manufacturing} 32(3):268--277.
  \doi{10.1080/0951192X.2019.1571235}

\bibitem[{Klein et~al(2018)Klein, L{\"o}cklin, Jazdi, and Weyrich}]{KLJ+18}
Klein M, L{\"o}cklin A, Jazdi N, et~al (2018) {A negotiation based approach for
  agent based production scheduling}. {Procedia Manufacturing} 17:334--341.
  \doi{10.1016/j.promfg.2018.10.054}

\bibitem[{Kovalenko et~al(2017)Kovalenko, Barton, and Tilbury}]{KBT17}
Kovalenko I, Barton K, Tilbury D (2017) {Design and implementation of an
  intelligent product agent architecture in manufacturing systems}. In: {2017
  22nd IEEE International Conference on Emerging Technologies and Factory
  Automation (ETFA)}. IEEE, Limassol, pp 1--8, \doi{10.1109/ETFA.2017.8247652}

\bibitem[{Kovalenko et~al(2022)Kovalenko, Balta, Tilbury, and Barton}]{KBT+22}
Kovalenko I, Balta EC, Tilbury DM, et~al (2022) {Cooperative Product Agents to
  Improve Manufacturing System Flexibility: A Model-Based Decision Framework}.
  {IEEE Transactions on Automation Science and Engineering} pp 1--18.
  \doi{10.1109/TASE.2022.3156384}

\bibitem[{KrishnaKumar(2003)}]{Kri}
KrishnaKumar K (2003) {Intelligent Systems For Aerospace Engineering: An
  Overview}

\bibitem[{Kritzinger et~al(2018)Kritzinger, Karner, Traar, Henjes, and
  Sihn}]{Kritzinger.2018}
Kritzinger W, Karner M, Traar G, et~al (2018) {Digital Twin in manufacturing: A
  categorical literature review and classification}. {IFAC-PapersOnLine}
  51(11):1016--1022. \doi{10.1016/j. ifacol.2018.08.474}

\bibitem[{Kruger and Basson(2018)}]{KrBa18}
Kruger K, Basson A (2018) {Erlang-based holonic controller for a palletized
  conveyor material handling system}. {Computers in Industry} 101:120--126.
  \doi{10.1016/j. compind.2018.07.003}

\bibitem[{Kruse et~al(2022)Kruse, Mostaghim, Borgelt, Braune, and
  Steinbrecher}]{KMB+22}
Kruse R, Mostaghim S, Borgelt C, et~al (2022) {Introduction}. In: Kruse R,
  Mostaghim S, Borgelt C, et~al (eds) {Computational Intelligence}. {Texts in
  Computer Science}, {Springer International Publishing}, Cham, p 1--3,
  \doi{10.1007/978-3-030-42227-1_1}

\bibitem[{Kuhn et~al(2020)Kuhn, Schnicke, and {Oliveira Antonino}}]{KSO20}
Kuhn T, Schnicke F, {Oliveira Antonino} P (2020) {Service-Based Architectures
  in Production Systems: Challenges, Solutions {\&} Experiences}. In: {2020 ITU
  Kaleidoscope: Industry-Driven Digital Transformation (ITU K)}. IEEE, Ha Noi,
  Vietnam, pp 1--7, \doi{10.23919/ITUK50268.2020.9303207}

\bibitem[{Laux et~al(2018)Laux, Gillenkirch, and Schenk-Mathes}]{LGS18}
Laux H, Gillenkirch RM, Schenk-Mathes HY (2018) {Entscheidungstheorie:
  (Decision Theory)}, 10th edn. {Lehrbuch}, {Springer Gabler}, Berlin and
  Heidelberg, \doi{10.1007/978-3-662-57818-6}

\bibitem[{Lee and Park(2014)}]{Lee.2014}
Lee CG, Park SC (2014) {Survey on the virtual commissioning of manufacturing
  systems}. {Journal of Computational Design and Engineering} 1(3):213--222.
  \doi{10.7315/JCDE.2014.021}

\bibitem[{Lehmann et~al(2023)Lehmann, Lober, Häußermann, Rache, Ollinger,
  Baumgärtel, and Reichwald}]{LLH+23}
Lehmann J, Lober A, Häußermann T, et~al (2023) The anatomy of the internet of
  digital twins: A symbiosis of agent and digital twin paradigms enhancing
  resilience (not only) in manufacturing environments. Machines 11(5).
  \doi{10.3390/machines11050504}

\bibitem[{Leit{\~a}o(2009)}]{Lei09}
Leit{\~a}o P (2009) {Agent-based distributed manufacturing control: A
  state-of-the-art survey}. {Engineering Applications of Artificial
  Intelligence} 22(7):979--991. \doi{10.1016/j. engappai.2008.09.005}

\bibitem[{Leit{\~a}o et~al(2015)Leit{\~a}o, Rodrigues, Barbosa, Turrin, and
  Pagani}]{LRB+15}
Leit{\~a}o P, Rodrigues N, Barbosa J, et~al (2015) {Intelligent products: The
  grace experience}. {Control Engineering Practice} 42:95--105. \doi{10.1016/j.
  conengprac.2015.05.001}

\bibitem[{Leng et~al(2020)Leng, Liu, Ye, Jing, Wang, Zhang, Zhang, and
  Chen}]{LLY+20}
Leng J, Liu Q, Ye S, et~al (2020) {Digital twin-driven rapid reconfiguration of
  the automated manufacturing system via an open architecture model}. {Robotics
  and Computer-Integrated Manufacturing} 63:101,895.
  \doi{10.1016/j.rcim.2019.101895}

\bibitem[{Lepuschitz et~al(2011)Lepuschitz, Zoitl, Vall{\'e}e, and
  Merdan}]{LZV+11}
Lepuschitz W, Zoitl A, Vall{\'e}e M, et~al (2011) {Toward Self-Reconfiguration
  of Manufacturing Systems Using Automation Agents}. {IEEE Transactions on
  Systems, Man, and Cybernetics, Part C (Applications and Reviews)}
  41(1):52--69. \doi{10.1109/TSMCC.2010.2059012}

\bibitem[{Li et~al(2022)Li, Pang, Zheng, Guan, and Le}]{LPZ+22}
Li J, Pang D, Zheng Y, et~al (2022) {A flexible manufacturing assembly system
  with deep reinforcement learning}. {Control Engineering Practice}
  118:104,957. \doi{10.1016/j. conengprac.2021.104957}

\bibitem[{Liu et~al(2014)Liu, Li, and Shen}]{LLS14}
Liu C, Li Y, Shen W (2014) {Integrated manufacturing process planning and
  control based on intelligent agents and multi-dimension features}. {The
  International Journal of Advanced Manufacturing Technology}
  75(9-12):1457--1471. \doi{10.1007/s00170-014-6246-0}

\bibitem[{L{\'o}pez et~al(2023)L{\'o}pez, Casquero, Est{\'e}vez, Armentia,
  Orive, and Marcos}]{LCE+23}
L{\'o}pez A, Casquero O, Est{\'e}vez E, et~al (2023) {An industrial agent-based
  customizable platform for I4.0 manufacturing systems}. {Computers in
  Industry} 146:103,859. \doi{10.1016/j.compind.2023.103859}

\bibitem[{Lou et~al(2019)Lou, Guo, Gao, Waedt, and Parekh}]{LGG+19}
Lou X, Guo Y, Gao Y, et~al (2019) {An idea of using Digital Twin to platform
  the functional safety and cybersecurity analysis}.
  \doi{10.18420/inf2019_ws32}

\bibitem[{Lugaresi and Matta(2021)}]{LuMa21}
Lugaresi G, Matta A (2021) {Discovery and digital model generation for
  manufacturing systems with assembly operations}. In: {2021 IEEE 17th
  International Conference on Automation Science and Engineering (CASE)}. IEEE,
  Lyon, France, pp 752--757, \doi{10.1109/CASE49439.2021.9551479}

\bibitem[{Luo et~al(2015)Luo, Zhong, Wan, Ye, and Qian}]{LZW+15}
Luo N, Zhong W, Wan F, et~al (2015) {An agent-based service-oriented
  integration architecture for chemical process automation}. {Chinese Journal
  of Chemical Engineering} 23(1):173--180. \doi{10.1016/j.cjche.2014.09.047}

\bibitem[{{M. Redeker} et~al(2021){M. Redeker}, {C. Klarhorst}, {D.
  G{\"o}llner}, {D. Quirin}, {P. Wi{\ss}brock}, {S. Althoff}, and {M.
  Hesse}}]{M.Redeker.2021}
{M. Redeker}, {C. Klarhorst}, {D. G{\"o}llner}, et~al (2021) {Towards an
  Autonomous Application of Smart Services in Industry 4.0}. In: {2021 26th
  IEEE International Conference on Emerging Technologies and Factory Automation
  (ETFA)}, pp 1--4, \doi{10.1109/ETFA45728.2021.9613369}

\bibitem[{Maier et~al(2017)Maier, Eckert, and Clarkson}]{Maier.2017}
Maier JF, Eckert CM, Clarkson PJ (2017) Model granularity in engineering design
  – concepts and framework. Design Science 3. \doi{10.1017/dsj.2016.16}

\bibitem[{Marcus et~al(2003)Marcus, Feng, and Maletic}]{Marcus.06112003}
Marcus A, Feng L, Maletic JI (2003) {3D representations for software
  visualization}. In: Diehl S, Stasko JT (eds) {Proceedings of the 2003 ACM
  symposium on Software visualization}. ACM, San Diego California, p~27,
  \doi{10.1145/774833.774837}

\bibitem[{Marks et~al(2017)Marks, Weyrich, Hoang, and Fay}]{MWH+17}
Marks P, Weyrich M, Hoang XL, et~al (2017) {Agent-based adaptation of automated
  manufacturing machines}. In: {2017 22nd IEEE International Conference on
  Emerging Technologies and Factory Automation (ETFA)}. IEEE, Limassol, pp
  1--8, \doi{10.1109/ETFA.2017.8247572}

\bibitem[{Marks et~al(2018)Marks, Hoang, Weyrich, and Fay}]{MHW+18}
Marks P, Hoang XL, Weyrich M, et~al (2018) {A systematic approach for
  supporting the adaptation process of discrete manufacturing machines}.
  {Research in Engineering Design} 29(4):621--641.
  \doi{10.1007/s00163-018-0296-5}

\bibitem[{Martins et~al(2019)Martins, Costelha, and Neves}]{AndreMartins.2019}
Martins A, Costelha H, Neves C (2019) {Shop Floor Virtualization and Industry
  4.0}. In: Almeida L (ed) {19th IEEE International Conference on Autonomous
  Robot Systems and Competitions (ICARSC 2019)}. IEEE, Piscataway, NJ, p 1--6,
  \doi{10.1109/ICARSC.2019.8733657}

\bibitem[{Martins et~al(2020)Martins, Costelha, and Neves}]{Martins.2020}
Martins A, Costelha H, Neves C (2020) {Supporting the Design, Commissioning and
  Supervision of Smart Factory Components through their Digital Twin}. In: Lau
  N (ed) {2020 IEEE International Conference on Autonomous Robot Systems and
  Competitions (ICARSC)}. IEEE, Ponta Delgada, Portugal, pp 114--119,
  \doi{10.1109/ICARSC49921.2020.9096072}

\bibitem[{Mashaly(2021)}]{Mashaly.2021}
Mashaly M (2021) {Connecting the Twins: A Review on Digital Twin Technology
  {\&} its Networking Requirements}. {Procedia Computer Science} 184:299--305.
  \doi{10.1016/j.procs.2021.03.039}

\bibitem[{Mathiesen et~al(2022)Mathiesen, S{\o}rensen, Sartori, Lindvig, Waspe,
  and Schlette}]{MSS+22}
Mathiesen S, S{\o}rensen LC, Sartori A, et~al (2022) {Applying Robotics
  Centered Digital Twins in a Smart Factory for Facilitating Integration and
  Improved Process Monitoring}. In: Andersen AL, Andersen R, Brunoe TD, et~al
  (eds) {Towards Sustainable Customization: Bridging Smart Products and
  Manufacturing Systems}. {Springer eBook Collection}, {Springer International
  Publishing} and {Imprint Springer}, Cham, p 305--313,
  \doi{10.1007/978-3-030-90700-6_34}

\bibitem[{Matulis and Harvey(2021)}]{MaHa21}
Matulis M, Harvey C (2021) {A robot arm digital twin utilising reinforcement
  learning}. {Computers {\&} Graphics} 95:106--114.
  \doi{10.1016/j.cag.2021.01.011}

\bibitem[{Melo et~al(2023)Melo, {La Prieta}, and Leit{\~a}o}]{MLL23}
Melo V, {La Prieta} FD, Leit{\~a}o P (2023) {Alignment of Digital Twin Systems
  with the RAMI 4.0 Model Using Multi-agent Systems}. In: Borangiu T,
  Trentesaux D, Leit{\~a}o P (eds) {Service Oriented, Holonic and Multi-Agent
  Manufacturing Systems for Industry of the Future}, {Studies in Computational
  Intelligence}, vol 1083. {Springer Cham}, pp 23--35,
  \doi{10.1007/978-3-031-24291-5_2}

\bibitem[{Merdan et~al(2011)Merdan, Moser, Vrba, and Biffl}]{MMV+11}
Merdan M, Moser T, Vrba P, et~al (2011) {Investigating the robustness of
  re-scheduling policies with multi-agent system simulation}. {The
  International Journal of Advanced Manufacturing Technology} 55(1-4):355--367.
  \doi{10.1007/s00170-010-3049-9}

\bibitem[{Merdan et~al(2017)Merdan, Moser, Sunindyo, Biffl, and Vrba}]{MMS+17}
Merdan M, Moser T, Sunindyo W, et~al (2017) {Workflow scheduling using
  multi-agent systems in a dynamically changing environment}. {Journal of
  Simulation} \doi{10.1057/jos.2012.15}

\bibitem[{Meyer et~al(2009)Meyer, Fr{\"a}mling, and Holmstr{\"o}m}]{MFH09}
Meyer GG, Fr{\"a}mling K, Holmstr{\"o}m J (2009) {Intelligent Products: A
  survey}. {Computers in Industry} 60(3):137--148.
  \doi{10.1016/j.compind.2008.12.005}

\bibitem[{Mihoubi et~al(2020)Mihoubi, Bouzouia, Tebani, and Gaham}]{MBT+20}
Mihoubi B, Bouzouia B, Tebani K, et~al (2020) {Hardware in the loop simulation
  for product driven control of a cyber-physical manufacturing system}.
  {Production Engineering} 14(3):329--343. \doi{10.1007/s11740-020-00957-w}

\bibitem[{Monostori(2014)}]{Monostori.2014}
Monostori L (2014) {Cyber-physical Production Systems: Roots, Expectations and
  R{\&}D Challenges}. {Procedia CIRP} 17:9--13.
  \doi{10.1016/j.procir.2014.03.115}

\bibitem[{M{\"u}ller et~al(2021)M{\"u}ller, M{\"u}ller, {Ashtari Talkhestani},
  Marks, Jazdi, and Weyrich}]{MMA+21}
M{\"u}ller M, M{\"u}ller T, {Ashtari Talkhestani} B, et~al (2021) {Industrial
  autonomous systems: a survey on definitions, characteristics and abilities}.
  {at - Automatisierungstechnik} 69(1):3--13. \doi{10.1515/auto-2020-0131}

\bibitem[{Nagorny et~al(2012{\natexlab{a}})Nagorny, Colombo, and
  Schmidtmann}]{NCS12}
Nagorny K, Colombo AW, Schmidtmann U (2012{\natexlab{a}}) {A service- and
  multi-agent-oriented manufacturing automation architecture}. {Computers in
  Industry} 63(8):813--823. \doi{10.1016/j.compind.2012.08.003}

\bibitem[{Nagorny et~al(2012{\natexlab{b}})Nagorny, Colombo, and
  Schmidtmann}]{Negri.}
Nagorny K, Colombo AW, Schmidtmann U (2012{\natexlab{b}}) A service- and
  multi-agent-oriented manufacturing automation architecture. Comput Ind
  63(8):813–823. \doi{10.1016/j.compind.2012.08.003}

\bibitem[{Natarajan and Srinivasan(2014)}]{NaSr14}
Natarajan S, Srinivasan R (2014) {Implementation of multi agents based system
  for process supervision in large-scale chemical plants}. {Computers {\&}
  Chemical Engineering} 60:182--196. \doi{10.1016/j.compchemeng.2013.08.012}

\bibitem[{Notomista et~al(2022)Notomista, Mayya, Emam, Kroninger, Bohannon,
  Hutchinson, and Egerstedt}]{NME+22}
Notomista G, Mayya S, Emam Y, et~al (2022) {A Resilient and Energy-Aware Task
  Allocation Framework for Heterogeneous Multirobot Systems}. {IEEE
  Transactions on Robotics} 38(1):159--179. \doi{10.1109/TRO.2021.3102379}

\bibitem[{Onaji et~al(2022)Onaji, Tiwari, Soulatiantork, Song, and
  Tiwari}]{OTS+22}
Onaji I, Tiwari D, Soulatiantork P, et~al (2022) {Digital twin in
  manufacturing: conceptual framework and case studies}. {International Journal
  of Computer Integrated Manufacturing} 35(8):831--858.
  \doi{10.1080/0951192X.2022.2027014}

\bibitem[{Onori et~al(2012)Onori, Lohse, Barata, and Hanisch}]{OLB+12}
Onori M, Lohse N, Barata J, et~al (2012) {The IDEAS project: plug {\&} produce
  at shop--floor level}. {Assembly Automation} 32(2):124--134.
  \doi{10.1108/01445151211212280}

\bibitem[{O'Sullivan et~al(2020)O'Sullivan, O'Sullivan, and Bruton}]{OOB20}
O'Sullivan J, O'Sullivan D, Bruton K (2020) {A case-study in the introduction
  of a digital twin in a large-scale smart manufacturing facility}. {Procedia
  Manufacturing} 51:1523--1530. \doi{10.1016/j.promfg.2020.10.212}

\bibitem[{Padgham and Winikoff(2004)}]{PaWi04}
Padgham L, Winikoff M (2004) {Developing Intelligent Agent Systems}. Wiley,
  \doi{10.1002/0470861223}

\bibitem[{Page et~al(2021)Page, McKenzie, Bossuyt, Boutron, Hoffmann, Mulrow,
  Shamseer, Tetzlaff, Akl, Brennan, Chou, Glanville, Grimshaw,
  Hr{\'o}bjartsson, Lalu, Li, Loder, Mayo-Wilson, McDonald, McGuinness,
  Stewart, Thomas, Tricco, Welch, Whiting, and Moher}]{PMB+21}
Page MJ, McKenzie JE, Bossuyt PM, et~al (2021) {The PRISMA 2020 statement: an
  updated guideline for reporting systematic reviews}. {BMJ (Clinical research
  ed)} 372:n71. \doi{10.1136/bmj.n71}

\bibitem[{Park et~al(2020)Park, Lee, Kim, and Noh}]{Park.2020}
Park KT, Lee J, Kim HJ, et~al (2020) {Digital twin-based cyber physical
  production system architectural framework for personalized production}. {The
  International Journal of Advanced Manufacturing Technology}
  106(5-6):1787--1810. \doi{10.1007/s00170-019-04653-7}

\bibitem[{de~{Paula Ferreira} et~al(2020)de~{Paula Ferreira}, Armellini, and
  de~Santa-Eulalia}]{PAS20}
de~{Paula Ferreira} W, Armellini F, de~Santa-Eulalia LA (2020) {Simulation in
  industry 4.0: A state-of-the-art review}. {Computers {\&} Industrial
  Engineering} 149:106,868. \doi{10.1016/j.cie.2020.106868}

\bibitem[{P{\'e}rez et~al(2020)P{\'e}rez, Rodr{\'i}guez-Jim{\'e}nez,
  Rodr{\'i}guez, Usamentiaga, and Garc{\'i}a}]{PRR+20}
P{\'e}rez L, Rodr{\'i}guez-Jim{\'e}nez S, Rodr{\'i}guez N, et~al (2020)
  {Digital Twin and Virtual Reality Based Methodology for Multi-Robot
  Manufacturing Cell Commissioning}. {Applied Sciences} 10(10):3633.
  \doi{10.3390/app10103633}

\bibitem[{Pires et~al(2020)Pires, Melo, Almeida, and Leitao}]{Pires.2020}
Pires F, Melo V, Almeida J, et~al (2020) {Digital Twin Experiments Focusing
  Virtualisation, Connectivity and Real-time Monitoring}. In: {2020 IEEE
  Conference on Industrial Cyberphysical Systems (ICPS)}. IEEE, Tampere,
  Finland, pp 309--314, \doi{10.1109/ICPS48405.2020.9274739}

\bibitem[{Pires et~al(2021)Pires, Ahmad, Moreira, and Leitao}]{Pires.2021b}
Pires F, Ahmad B, Moreira AP, et~al (2021) {Recommendation System using
  Reinforcement Learning for What-If Simulation in Digital Twin}. In: {IEEE
  International Conference on Industrial Informatics, INDIN'21},
  \doi{https://doi.org/10.1109/INDIN45523.2021.9557372}

\bibitem[{Protic et~al(2020)Protic, Jin, Marian, Abd, Campbell, and
  Chahl}]{PJM+20}
Protic A, Jin Z, Marian R, et~al (2020) {Implementation of a Bi-Directional
  Digital Twin for Industry 4 Labs in Academia: A Solution Based on OPC UA}.
  In: {2020 IEEE International Conference on Industrial Engineering and
  Engineering Management (IEEM)}. IEEE, Singapore, Singapore, pp 979--983,
  \doi{10.1109/IEEM45057.2020.9309953}

\bibitem[{Qamsane et~al(2019)Qamsane, Balta, Moyne, Tilbury, and
  Barton}]{QBM+19}
Qamsane Y, Balta EC, Moyne J, et~al (2019) {Dynamic Rerouting of Cyber-Physical
  Production Systems in Response to Disruptions Based on SDC Framework}. In:
  {2019 American Control Conference (ACC)}. IEEE, Philadelphia, PA, USA, pp
  3650--3657, \doi{10.23919/acc.2019.8814412}

\bibitem[{Ralph et~al(2020)Ralph, Schwarz, and Stockinger}]{RSS20}
Ralph BJ, Schwarz A, Stockinger M (2020) An implementation approach for an
  academic learning factory for the metal forming industry with special focus
  on digital twins and finite element analysis. Procedia Manufacturing
  45:253--258. \doi{https://doi.org/10.1016/j.promfg.2020.04.103}, learning
  Factories across the value chain – from innovation to service – The 10th
  Conference on Learning Factories 2020

\bibitem[{Redelinghuys et~al(2020)Redelinghuys, Basson, and
  Kruger}]{Redelinghuys.2020}
Redelinghuys AJH, Basson AH, Kruger K (2020) {A six-layer architecture for the
  digital twin: a manufacturing case study implementation}. {Journal of
  Intelligent Manufacturing} 31(6):1383--1402. \doi{10.1007/s10845-019-01516-6}

\bibitem[{Rehman et~al(2021)Rehman, Pulikottil, Estrada-Jimenez, Mo, Chaplin,
  Barata, and Ratchev}]{RPE+21}
Rehman HU, Pulikottil T, Estrada-Jimenez LA, et~al (2021) {Cloud Based Decision
  Making for Multi-agent Production Systems}. In: Marreiros G, Melo FS, Lau N,
  et~al (eds) {Progress in Artificial Intelligence}, {Lecture Notes in Computer
  Science}, vol 12981. {Springer International Publishing}, Cham, p 673--686,
  \doi{10.1007/978-3-030-86230-5_53}

\bibitem[{Reiche et~al(2021)Reiche, Gundlach, Mewes, and Fay}]{Reiche.2021}
Reiche LT, Gundlach CS, Mewes GF, et~al (2021) {The Digital Twin of a System: A
  Structure for Networks of Digital Twins}. In: {2021 26th IEEE International
  Conference on Emerging Technologies and Factory Automation (ETFA)}. IEEE, pp
  1--8, \doi{10.1109/ETFA45728.2021.9613594}

\bibitem[{Reinpold et~al(2023)Reinpold, Wagner, Gehlhoff, Ramonat, Kilthau,
  Gill, Reif, Henkel, Scholz, and Fay}]{dataset}
Reinpold LM, Wagner LP, Gehlhoff F, et~al (2023) {Systematic Comparison of
  Software Agents and Digital Twins: Differences, Similarities, and Synergies
  in Industrial Production: A Dataset}. \doi{10.5281/zenodo.8120623}

\bibitem[{Ricondo et~al(2021)Ricondo, Porto, and Ugarte}]{RPU21}
Ricondo I, Porto A, Ugarte M (2021) {A digital twin framework for the
  simulation and optimization of production systems}. {CIRP Conference on
  Manufacturing Systems} 104:762--767. \doi{10.1016/j.procir.2021.11.128}

\bibitem[{Rodrigues et~al(2018)Rodrigues, Oliveira, and Leit{\~a}o}]{ROL18}
Rodrigues N, Oliveira E, Leit{\~a}o P (2018) {Decentralized and on-the-fly
  agent-based service reconfiguration in manufacturing systems}. {Computers in
  Industry} 101:81--90. \doi{10.1016/j.compind.2018.06.003}

\bibitem[{Rolle et~al(2021)Rolle, Martucci, and Godoy}]{RMG21}
Rolle RP, Martucci VdO, Godoy EP (2021) {Modular Framework for Digital Twins:
  Development and Performance Analysis}. {Journal of Control, Automation and
  Electrical Systems} 32(6):1485--1497. \doi{10.1007/s40313-021-00830-w}

\bibitem[{Rol{\'o}n and Mart{\'i}nez(2012)}]{RoMa12}
Rol{\'o}n M, Mart{\'i}nez E (2012) {Agent-based modeling and simulation of an
  autonomic manufacturing execution system}. {Computers in Industry}
  63(1):53--78. \doi{10.1016/j.compind.2011.10.005}

\bibitem[{Rosenberger and Gerhard(2018)}]{RoGe18}
Rosenberger P, Gerhard D (2018) {Context-awareness in industrial applications:
  definition, classification and use case}. {Procedia CIRP} 72:1172--1177.
  \doi{10.1016/j.procir.2018.03.242}

\bibitem[{Rosendahl et~al(2018)Rosendahl, Cal{\`a}, Kirchheim, L{\"u}der, and
  D'Agostino}]{RCK+18}
Rosendahl R, Cal{\`a} A, Kirchheim K, et~al (2018) {Towards Smart Factory:
  Multi-Agent Integration on Industrial Standards for Service-oriented
  Communication and Semantic Data Exchange}. In: {WOA 2018 19th Workshop ``From
  Objects to Agents'}, Palermo

\bibitem[{Sahlab et~al(2022)Sahlab, Braun, K{\"o}hler, Jazdi, and
  Weyrich}]{Sahlab.}
Sahlab N, Braun D, K{\"o}hler C, et~al (2022) {Extending the Intelligent
  Digital Twin with a context modeling service: A decision support use case:
  PREPRINT}. In: {Proceedings of CIRP 2022}, pp 463--468, \doi{10.1016/j.
  procir.2022.05.009}

\bibitem[{Sakurada and Leit{\~a}o(2020)}]{SaLe20}
Sakurada L, Leit{\~a}o P (2020) {Multi-Agent Systems to Implement Industry 4.0
  Components}. In: {2020 IEEE Conference on Industrial Cyberphysical Systems
  (ICPS)}. IEEE, Tampere, Finland, pp 21--26,
  \doi{10.1109/ICPS48405.2020.9274745}

\bibitem[{Sakurada et~al(2022{\natexlab{a}})Sakurada, Leitao, and {La
  Prieta}}]{SLL22}
Sakurada L, Leitao P, {La Prieta} FD (2022{\natexlab{a}}) {Agent-Based Asset
  Administration Shell Approach for Digitizing Industrial Assets}.
  {IFAC-PapersOnLine} 55(2):193--198. \doi{10.1016/j.ifacol.2022.04.192}

\bibitem[{Sakurada et~al(2022{\natexlab{b}})Sakurada, Leit{\~a}o, {La Prieta},
  and Corchado}]{SLL+22}
Sakurada L, Leit{\~a}o P, {La Prieta} FD, et~al (2022{\natexlab{b}})
  {Multi-Agent Systems to Realize Intelligent Asset Administration Shells}. In:
  {III Workshop on Disruptive Information and Communication Technologies for
  Innovation and Digital Transformation}. {Ediciones Universidad de Salamanca},
  pp 43--58, \doi{10.14201/0AQ03114358}

\bibitem[{Santos et~al(2022)Santos, de~Queiroz, Leal, and
  Montevechi}]{Santos.2022}
Santos CHd, de~Queiroz JA, Leal F, et~al (2022) {Use of simulation in the
  industry 4.0 context: Creation of a Digital Twin to optimise decision making
  on non-automated process}. {Journal of Simulation} 16(3):284--297.
  \doi{10.1080/17477778.2020.1811172}

\bibitem[{Schluse et~al(2018)Schluse, Priggemeyer, Atorf, and
  Rossmann}]{SPA+18}
Schluse M, Priggemeyer M, Atorf L, et~al (2018) {Experimentable Digital
  Twins---Streamlining Simulation-Based Systems Engineering for Industry 4.0}.
  {IEEE Transactions on Industrial Informatics} 14(4):1722--1731.
  \doi{10.1109/TII.2018.2804917}

\bibitem[{Schnicke et~al(2020)Schnicke, Kuhn, and Antonino}]{Schnicke.}
Schnicke F, Kuhn T, Antonino PO (2020) Enabling industry 4.0 service-oriented
  architecture through digital twins. In: Muccini H, Avgeriou P, Buhnova B,
  et~al (eds) Software Architecture. Springer International Publishing, Cham,
  pp 490--503

\bibitem[{Schoenwald et~al(2004)Schoenwald, Barton, and Ehlen}]{SBE04}
Schoenwald DA, Barton DC, Ehlen MA (2004) {An agent-based simulation laboratory
  for economics and infrastructure interdependency}. In: {Proceedings of the
  2004 American Control Conference}. IEEE, Boston, MA, USA, pp 1295--1300
  vol.2, \doi{10.23919/ACC.2004.1386752}

\bibitem[{Schroeder et~al(2016)Schroeder, Steinmetz, Pereira, and
  Espindola}]{SSP+16}
Schroeder GN, Steinmetz C, Pereira CE, et~al (2016) {Digital Twin Data Modeling
  with AutomationML and a Communication Methodology for Data Exchange}.
  {IFAC-PapersOnLine} 49(30):12--17. \doi{10.1016/j.ifacol.2016.11.115}

\bibitem[{Schutz et~al(2011)Schutz, Schraufstetter, Folmer, Vogel-Heuser,
  Gmeiner, and Shea}]{SSF+11}
Schutz D, Schraufstetter M, Folmer J, et~al (2011) {Highly reconfigurable
  production systems controlled by real-time agents}. In: {ETFA2011}. IEEE,
  Toulouse, France, pp 1--8, \doi{10.1109/ETFA.2011.6058991}

\bibitem[{Schweizer et~al(2021)Schweizer, Braunisch, Alt, Schmitz, and
  Wollschlaeger}]{Schweizer.2021}
Schweizer H, Braunisch N, Alt R, et~al (2021) {Prozesskomposition in verteilten
  Automatisierungssystemen}. {at - Automatisierungstechnik} 69(3):242--255.
  \doi{10.1515/auto-2020-0118}

\bibitem[{Seif et~al(2019)Seif, Toro, and Akhtar}]{STA19}
Seif A, Toro C, Akhtar H (2019) {Implementing Industry 4.0 Asset Administrative
  Shells in Mini Factories}. {Procedia Computer Science} 159:495--504.
  \doi{10.1016/j.procs.2019.09.204}

\bibitem[{{Seng Ng} and Srinivasan(2010)}]{SeSr10}
{Seng Ng} Y, Srinivasan R (2010) {Multi-agent based collaborative fault
  detection and identification in chemical processes}. {Engineering
  Applications of Artificial Intelligence} 23(6):934--949.
  \doi{10.1016/j.engappai.2010.01.026}

\bibitem[{Serik et~al(2022)Serik, Zhetpissov, Mosadeghzad, and
  Alizadeh}]{SZM+22}
Serik A, Zhetpissov Y, Mosadeghzad M, et~al (2022) {Digital Twins Development
  of Automatic Storage and Retrieval Station in a Production Line and an
  Integrated Robotic Manipulator}. In: Kim J, Englot B, Park HW, et~al (eds)
  {Robot Intelligence Technology and Applications 6}, {Lecture Notes in
  Networks and Systems}, vol 429. {Springer International Publishing}, Cham, p
  210--223, \doi{10.1007/978-3-030-97672-9_19}

\bibitem[{She(2021)}]{She.21}
She M (2021) {Deep Reinforcement Learning-Based Smart Manufacturing Plants with
  a Novel Digital Twin Training Model}. {Wireless Personal Communications}
  \doi{10.1007/s11277-021-09072-0}

\bibitem[{Shi et~al(2020)Shi, Gasser, Seeck, and Auerswald}]{SGS+20}
Shi E, Gasser TM, Seeck A, et~al (2020) {The Principles of Operation Framework:
  A Comprehensive Classification Concept for Automated Driving Functions}. {SAE
  International Journal of Connected and Automated Vehicles} 3(1).
  \doi{10.4271/12-03-01-0003}

\bibitem[{Shin et~al(2019)Shin, Kim, and Meilanitasari}]{SKM19}
Shin SJ, Kim YM, Meilanitasari P (2019) {A Holonic-Based Self-Learning
  Mechanism for Energy-Predictive Planning in Machining Processes}. {Processes}
  7(10):739. \doi{10.3390/pr7100739}

\bibitem[{Siddiqui et~al(2023)Siddiqui, Kahandawa, and Hewawasam}]{SKH23}
Siddiqui M, Kahandawa G, Hewawasam H (2023) {Artificial Intelligence Enabled
  Digital Twin For Predictive Maintenance in Industrial Automation System: A
  Novel Framework and Case Study}. In: {2023 IEEE International Conference on
  Mechatronics (ICM)}. IEEE, Loughborough, United Kingdom, pp 1--6,
  \doi{10.1109/ICM54990.2023.10101971}

\bibitem[{Sierla et~al(2020)Sierla, Sorsam{\"a}ki, Azangoo, Villberg,
  Hyt{\"o}nen, and Vyatkin}]{Sierla.2020b}
Sierla S, Sorsam{\"a}ki L, Azangoo M, et~al (2020) {Towards Semi-Automatic
  Generation of a Steady State Digital Twin of a Brownfield Process Plant}.
  {Applied Sciences} 10(19):6959. \doi{10.3390/app10196959}

\bibitem[{Sjarov et~al(2020)Sjarov, Lechler, Fuchs, Brossog, Selmaier, Faltus,
  Donhauser, and Franke}]{SLF+20}
Sjarov M, Lechler T, Fuchs J, et~al (2020) {The Digital Twin Concept in
  Industry -- A Review and Systematization}. In: {2020 25th IEEE International
  Conference on Emerging Technologies and Factory Automation (ETFA)}. IEEE,
  Vienna, Austria, pp 1789--1796, \doi{10.1109/ETFA46521.2020.9212089}

\bibitem[{Sjarov et~al(2021)Sjarov, Lechler, Russwurm, Fuchs, Faltus,
  Sch{\"a}ffer, Brossog, and Franke}]{SLR+21}
Sjarov M, Lechler T, Russwurm E, et~al (2021) {Life Cycle of a Digital Resource
  Twin: Meta-Modeling and Application Example}. {CIRP Conference on
  Manufacturing Systems} 104:1644--1649. \doi{10.1016/j.procir.2021.11.277}

\bibitem[{Steringer et~al(2019)Steringer, Z{\"o}rrer, Zambal, and
  Eitzinger}]{SZZ+19}
Steringer R, Z{\"o}rrer H, Zambal S, et~al (2019) {Using Discrete Event
  Simulation in multiple System Life Cycles to support Zero-Defect Composite
  Manufacturing in Aerospace Industry}. {IFAC-PapersOnLine} 52(13):1467--1472.
  \doi{10.1016/j. ifacol.2019.11.406}

\bibitem[{{\v{S}}vaco et~al(2011){\v{S}}vaco, {\v{S}}ekoranja, and
  Jerbic}]{SSJ11}
{\v{S}}vaco M, {\v{S}}ekoranja B, Jerbic B (2011) {A multiagent framework for
  industrial robotic applications}. {Procedia Computer Science} 6:291--296.
  \doi{10.1016/j.procs.2011.08.054}

\bibitem[{Taha et~al(2022)Taha, Yacout, and Shaban}]{TYS22}
Taha HA, Yacout S, Shaban Y (2022) {Deep Reinforcement Learning for autonomous
  pre-failure tool life improvement}. {The International Journal of Advanced
  Manufacturing Technology} 121(9-10):6169--6192.
  \doi{10.1007/s00170-022-09700-4}

\bibitem[{Tao et~al(2018)Tao, Cheng, Qi, Zhang, Zhang, and Sui}]{Tao.2018}
Tao F, Cheng J, Qi Q, et~al (2018) {Digital twin-driven product design,
  manufacturing and service with big data}. {The International Journal of
  Advanced Manufacturing Technology} 94(9-12):3563--3576.
  \doi{10.1007/s00170-017-0233-1}

\bibitem[{Tao et~al(2019)Tao, Zhang, Liu, and Nee}]{Tao.2019}
Tao F, Zhang H, Liu A, et~al (2019) {Digital Twin in Industry:
  State-of-the-Art}. {IEEE Transactions on Industrial Informatics}
  15(4):2405--2415. \doi{10.1109/TII.2018.2873186}

\bibitem[{Tao et~al(2023)Tao, Qiu, Chen, Stojanovic, and Cheng}]{TQC+23}
Tao H, Qiu J, Chen Y, et~al (2023) Unsupervised cross-domain rolling bearing
  fault diagnosis based on time-frequency information fusion. Journal of the
  Franklin Institute 360(2):1454--1477. \doi{10.1016/j.jfranklin.2022.11.004}

\bibitem[{Ter{\'a}n et~al(2017)Ter{\'a}n, Aguilar, and Cerrada}]{TAC17}
Ter{\'a}n J, Aguilar J, Cerrada M (2017) {Integration in industrial automation
  based on multi-agent systems using cultural algorithms for optimizing the
  coordination mechanisms}. {Computers in Industry} 91:11--23. \doi{10.1016/j.
  compind.2017.05.002}

\bibitem[{Terwiesch and {E. Bohn}(2001)}]{Terwiesch.2001}
Terwiesch C, {E. Bohn} R (2001) {Learning and process improvement during
  production ramp-up}. {International Journal of Production Economics}
  70(1):1--19. \doi{10.1016/S0925-5273(00)00045-1}

\bibitem[{Tipary and {Erd{\H o}s.}(2021)}]{TiEr21}
Tipary B, {Erd{\H o}s.} G (2021) {Generic development methodology for flexible
  robotic pick-and-place workcells based on Digital Twin}. {Robotics and
  Computer-Integrated Manufacturing} 71:102,140.
  \doi{10.1016/j.rcim.2021.102140}

\bibitem[{Tveito and H{\aa}konsen(2022)}]{TvHa22}
Tveito KO, H{\aa}konsen A (2022) {Digital Twin for Design and Optimization of
  DC Casting Lines}. In: Eskin D (ed) {Light Metals 2022}. {The Minerals,
  Metals {\&} Materials Series}, {Springer International Publishing}, Cham, p
  674--680, \doi{10.1007/978-3-030-92529-1_88}

\bibitem[{Ulewicz et~al(2012)Ulewicz, Schutz, and Vogel-Heuser}]{USV12}
Ulewicz S, Schutz D, Vogel-Heuser B (2012) {Design, implementation and
  evaluation of a hybrid approach for software agents in automation}. In: {2012
  International Conference on Collaboration Technologies and Systems (CTS)}.
  IEEE, Denver, CO, USA, pp 1--4, \doi{10.1109/ETFA.2012.6489766}

\bibitem[{Vallee et~al(2011)Vallee, Merdan, Lepuschitz, and
  Koppensteiner}]{VML+11}
Vallee M, Merdan M, Lepuschitz W, et~al (2011) {Decentralized Reconfiguration
  of a Flexible Transportation System}. {IEEE Transactions on Industrial
  Informatics} 7(3):505--516. \doi{10.1109/TII.2011.2158839}

\bibitem[{VDI(2010)}]{vdi12345}
VDI (2010) {VDI 2653 Multi-agent Systems in industrial automation -
  Fundamentals: Sheet 1}.
  \urlprefix\url{https://www.beuth.de/en/technical-rule/vdi-vde-2653-blatt-1/282864028}

\bibitem[{Verbeet and Baumg{\"a}rtel(2020)}]{VeBa20}
Verbeet R, Baumg{\"a}rtel H (2020) {Implementierung von autonomen I4.0-Systemen
  mit BDI-Agenten}. In: ten Hompel M, Vogel-Heuser B, Bauernhansl T (eds)
  {Handbuch Industrie 4.0}. {Springer Reference Technik}, {Springer Berlin
  Heidelberg}, Berlin, Heidelberg, p 1--36,
  \doi{10.1007/978-3-662-45537-1_130-1}

\bibitem[{Villalonga et~al(2021)Villalonga, Negri, Biscardo, Castano, Haber,
  Fumagalli, and Macchi}]{VNB+21}
Villalonga A, Negri E, Biscardo G, et~al (2021) {A decision-making framework
  for dynamic scheduling of cyber-physical production systems based on digital
  twins}. {Annual Reviews in Control} 51:357--373.
  \doi{10.1016/j.arcontrol.2021.04.008}

\bibitem[{Vogel-Heuser et~al(2014)Vogel-Heuser, Diedrich, Pantforder, and
  Gohner}]{VDP+14}
Vogel-Heuser B, Diedrich C, Pantforder D, et~al (2014) {Coupling heterogeneous
  production systems by a multi-agent based cyber-physical production system}.
  In: {2014 12th IEEE International Conference on Industrial Informatics
  (INDIN)}. IEEE, Porto Alegre RS, Brazil, pp 713--719,
  \doi{10.1109/INDIN.2014.6945601}

\bibitem[{Vogel-Heuser et~al(2020)Vogel-Heuser, Seitz, {Cruz Salazar},
  Gehlhoff, Dogan, and Fay}]{VSC+20}
Vogel-Heuser B, Seitz M, {Cruz Salazar} LA, et~al (2020) {Multi-agent systems
  to enable Industry 4.0}. {at - Automatisierungstechnik} 68(6):445--458.
  \doi{10.1515/auto-2020-0004}

\bibitem[{Vogel-Heuser et~al(2021)Vogel-Heuser, Ocker, and Scheuer}]{VOS21}
Vogel-Heuser B, Ocker F, Scheuer T (2021) {An approach for leveraging Digital
  Twins in agent-based production systems}. {at - Automatisierungstechnik}
  69(12):1026--1039. \doi{10.1515/auto-2021-0081}

\bibitem[{Vrba et~al(2011)Vrba, Radakovi{\v{c}}, Obitko, and
  Ma{\v{r}}{\'i}k}]{VRO+11}
Vrba P, Radakovi{\v{c}} M, Obitko M, et~al (2011) {Semantic technologies:
  latest advances in agent-based manufacturing control systems}. {International
  Journal of Production Research} 49(5):1483--1496.
  \doi{10.1080/00207543.2010.518746}

\bibitem[{Wang et~al(2017)Wang, Jiang, and Ding}]{WJD17}
Wang C, Jiang P, Ding K (2017) {A hybrid-data-on-tag--enabled decentralized
  control system for flexible smart workpiece manufacturing shop floors}.
  {Proceedings of the Institution of Mechanical Engineers, Part C: Journal of
  Mechanical Engineering Science} 231(4):764--782.
  \doi{10.1177/0954406215620452}

\bibitem[{Wang et~al(2013)Wang, Li, and Ma}]{WLM13}
Wang W, Li Y, Ma Y (2013) {Towards a Feature-based Agent-driven NC Tool Path
  Generation to Support Design and Process Changes}. {Computer-Aided Design and
  Applications} 10(4):603--618. \doi{10.3722/cadaps.2013.603-618}

\bibitem[{Ward et~al(2021)Ward, Sun, Dominguez-Caballero, Ojo, Ayvar-Soberanis,
  Curtis, and Ozturk}]{Ward.2021}
Ward R, Sun C, Dominguez-Caballero J, et~al (2021) {Machining Digital Twin
  using real-time model-based simulations and lookahead function for closed
  loop machining control}. {The International Journal of Advanced Manufacturing
  Technology} 117(11-12):3615--3629. \doi{10.1007/s00170-021-07867-w}

\bibitem[{Weilkiens(2014)}]{Wei14}
Weilkiens T (2014) {Systems Engineering mit SysML/UML: Anforderungen, Analyse,
  Architektur}, 3rd edn. dpunkt.verl., Heidelberg

\bibitem[{Wein et~al(2021)Wein, Dassen, Pallasch, Miny, Storms, and
  Brecher}]{WDP+21}
Wein S, Dassen Y, Pallasch C, et~al (2021) {Konzept und Anwendung Autonomer
  Industrie 4.0-Komponenten auf Basis Agenten-basierter Ans{\"a}tze}. {at -
  Automatisierungstechnik} 69(6):430--441. \doi{10.1515/auto-2020-0117}

\bibitem[{Wenger et~al(2018)Wenger, Zoitl, and Muller}]{WZM18}
Wenger M, Zoitl A, Muller T (2018) {Connecting PLCs With Their Asset
  Administration Shell For Automatic Device Configuration}. In: {2018 IEEE 16th
  International Conference on Industrial Informatics (INDIN)}. IEEE, Porto, pp
  74--79, \doi{10.1109/INDIN.2018.8472022}

\bibitem[{Wenna et~al(2022)Wenna, Weili, Changchun, Heng, Haibing, and
  Yao}]{WWC+22}
Wenna W, Weili D, Changchun H, et~al (2022) {A digital twin for 3D path
  planning of large-span curved-arm gantry robot}. {Robotics and
  Computer-Integrated Manufacturing} 76:102,330.
  \doi{10.1016/j.rcim.2022.102330}

\bibitem[{Wooldridge and Jennings(1995)}]{WoJe95}
Wooldridge M, Jennings NR (1995) {Intelligent agents: theory and practice}.
  {The Knowledge Engineering Review} 10(2):115--152.
  \doi{10.1017/S0269888900008122}

\bibitem[{Wooldridge(2002)}]{Woo02}
Wooldridge MJ (2002) {An introduction to multiagent systems}, 2nd edn. Wiley,
  Chichester, West Essex, United Kingdom

\bibitem[{Xia et~al(2021)Xia, Sacco, Kirkpatrick, Saidy, Nguyen, Kircaliali,
  and Harik}]{XSK+21}
Xia K, Sacco C, Kirkpatrick M, et~al (2021) {A digital twin to train deep
  reinforcement learning agent for smart manufacturing plants: Environment,
  interfaces and intelligence}. {Journal of Manufacturing Systems} 58:210--230.
  \doi{10.1016/j.jmsy.2020.06.012}

\bibitem[{Xie et~al(2022)Xie, Yang, Xu, Li, and Hu}]{XYX+22}
Xie G, Yang K, Xu C, et~al (2022) {Digital Twinning Based Adaptive Development
  Environment for Automotive Cyber-Physical Systems}. {IEEE Transactions on
  Industrial Informatics} 18(2):1387--1396. \doi{10.1109/TII.2021.3064364}

\bibitem[{Xu et~al(2019)Xu, Sun, Liu, and Zheng}]{XSL+19}
Xu Y, Sun Y, Liu X, et~al (2019) {A Digital-Twin-Assisted Fault Diagnosis Using
  Deep Transfer Learning}. {IEEE Access} 7:19,990--19,999.
  \doi{10.1109/ACCESS.2018.2890566}

\bibitem[{Yang et~al(2020)Yang, Lee, Kang, Do~Noh, Choi, Jung, Lee, Kang, Lee,
  and Kim}]{Yang.2020}
Yang J, Lee S, Kang YS, et~al (2020) Integrated platform and digital twin
  application for global automotive part suppliers. In: Lalic B, Majstorovic V,
  Marjanovic U, et~al (eds) Advances in Production Management Systems. Towards
  Smart and Digital Manufacturing. Springer International Publishing, Cham, pp
  230--237

\bibitem[{Yang et~al(2022)Yang, Son, Lee, and Noh}]{Yang.2022}
Yang J, Son YH, Lee D, et~al (2022) {Digital Twin-Based Integrated Assessment
  of Flexible and Reconfigurable Automotive Part Production Lines}. {Machines}
  10(2):75. \doi{10.3390/machines10020075}

\bibitem[{Yang et~al(2023)Yang, Zhou, S{\o}rensen, Christensen, {\"U}nalan, and
  Zhang}]{YZS+23}
Yang X, Zhou Z, S{\o}rensen JH, et~al (2023) {Automation of SME production with
  a Cobot system powered by learning-based vision}. {Robotics and
  Computer-Integrated Manufacturing} 83:102,564. \doi{10.1016/j.
  rcim.2023.102564}

\bibitem[{Ye et~al(2021)Ye, Yu, Song, and Hong}]{YYS+21}
Ye X, Yu M, Song WS, et~al (2021) {An Asset Administration Shell Method for
  Data Exchange Between Manufacturing Software Applications}. {IEEE Access}
  9:144,171--144,178. \doi{10.1109/ACCESS.2021.3122175}

\bibitem[{Zhang et~al(2020)Zhang, Zhou, Hu, and Li}]{ZZH+20}
Zhang C, Zhou G, Hu J, et~al (2020) {Deep learning-enabled intelligent process
  planning for digital twin manufacturing cell}. {Knowledge-Based Systems}
  191:105,247. \doi{10.1016/j. knosys.2019.105247}

\bibitem[{Zhang et~al(2023{\natexlab{a}})Zhang, Xiao, Liu, Duan, and
  Qin}]{ZXL+23}
Zhang Q, Xiao R, Liu Z, et~al (2023{\natexlab{a}}) {Process Simulation and
  Optimization of Arc Welding Robot Workstation Based on Digital Twin}.
  {Machines} 11(1):53. \doi{10.3390/machines11010053}

\bibitem[{Zhang et~al(2023{\natexlab{b}})Zhang, Song, Sun, and
  Stojanovic}]{ZSS+23}
Zhang Z, Song X, Sun X, et~al (2023{\natexlab{b}}) Hybrid-driven-based fuzzy
  secure filtering for nonlinear parabolic partial differential equation
  systems with cyber attacks. International Journal of Adaptive Control and
  Signal Processing 37(2):380--398. \doi{10.1002/acs.3529}

\bibitem[{Zidek et~al(2020)Zidek, Pitel, Pavlenko, Lazorik, and
  Hosovsky}]{ZPP+20bb}
Zidek K, Pitel J, Pavlenko I, et~al (2020) {Digital Twin of Experimental
  Workplace for Quality Control with Cloud Platform Support}. In: Knapcikova L,
  Balog M, Perakovic D, et~al (eds) {4th EAI International Conference on
  Management of Manufacturing Systems}. {EAI/Springer Innovations in
  Communication and Computing}, {Springer International Publishing}, Cham, p
  135--145, \doi{10.1007/978-3-030-34272-2_13}

\bibitem[{Zipper(2021)}]{Zip21}
Zipper H (2021) {Real-Time-Capable Synchronization of Digital Twins}.
  {IFAC-PapersOnLine} 54(4):147--152. \doi{10.1016/j.ifacol.2021.10.025}

\end{thebibliography}
